\documentclass[a4paper,12pt]{article}
\usepackage[letter paper, margin=1 in]{geometry}
\usepackage[numbers]{natbib}

\usepackage{lmodern} 
\usepackage{amsmath, bm}
\usepackage{authblk}
\usepackage[utf8]{inputenc} 
\usepackage[T1]{fontenc}    
\usepackage[hidelinks,colorlinks=true,linkcolor=blue,citecolor=blue]{hyperref}
\usepackage{url}            
\usepackage{booktabs}
\usepackage[titletoc]{appendix}
\usepackage{algorithm}
\usepackage{algorithmic}
\usepackage[algo2e,ruled,vlined,linesnumbered]{algorithm2e}

\usepackage{amsfonts,amsmath,amssymb,amsthm}      
\usepackage{bbm}
\usepackage{nicefrac}       
\usepackage{microtype}      
\usepackage[capitalize,noabbrev]{cleveref}
\usepackage{graphicx}
\usepackage{wrapfig}
\usepackage{subcaption}
\usepackage{xcolor}
\usepackage{comment}
\usepackage{empheq}
\usepackage[most]{tcolorbox}
\usepackage{framed}

\newcommand{\mbb}{\mathbb}

\newcommand{\mcl}{\mathcal}
\newcommand{\bs}{\boldsymbol}

\newcommand{\T}{\textnormal}

\newcommand{\crls}{\texttt{CRLS}}
\theoremstyle{plain}

\theoremstyle{definition}

\theoremstyle{remark}

\newtcbox{\mymath}[1][]{%
	nobeforeafter, math upper, tcbox raise base,
	enhanced, colframe=gray!30!black,
	colback=gray!30, boxrule=1pt,
	#1}
\tcbset{highlight math style={boxsep=5mm,colback=gray!30!red!30!white}}

\title{\texttt{CRLS}: Convolutional Regularized Least Squares Framework for Reduced Order Modeling of Transonic  Flows}
\author[1]{Muhammad Bilal \footnote{Graduate student, Kevin T. Crofton Department of Aerospace and Ocean Engineering.}}
\author[2]{Ashwin Renganathan\footnote{Assistant professor, Aerospace Engineering and the Institute of Computational and Data Science (ICDS).}}
\affil[1]{Penn State}
\affil[2]{Virginia Tech}

\begin{document}
	\vspace{-4 mm}
	\date{}
	\maketitle
	%
	
	%
	
	\begin{abstract}
		We develop a convolutional regularized least squares (\texttt{CRLS}) framework for reduced-order modeling of transonic flows with shocks. Conventional proper orthogonal decomposition (POD)–based reduced models are attractive because of their optimality and low online cost; however, but they perform poorly when snapshots contain parameter-dependent discontinuities, leading to smeared shocks, stair-stepping, or non-physical oscillations. In \texttt{CRLS}, we first map each full-order snapshot to a smoother representation by applying a one-dimensional Gaussian convolution with reflect padding along the flow field coordinates. The convolution hyperparameters (kernel width and support) are selected automatically by Bayesian optimization on a held-out set of snapshots. POD bases are then extracted from the smoothed data, and the parametric dependence of the POD coefficients is learned via radial basis function interpolation. To recover sharp shock structures, we introduce an efficient deconvolution step formulated as a regularized least squares problem, where the regularization centers the reconstruction around a nearest-neighbor reference snapshot in parameter space. The resulting \texttt{CRLS} surrogate is evaluated on inviscid transonic flow over the RAE2822 airfoil, modeled by the steady compressible Euler equations solved with SU2 over a Latin–hypercube sample of Mach number and angle of attack. Compared with standard POD and smoothed-POD baselines, \texttt{CRLS} yields markedly improved shock location and strength, lower surface-pressure and field-level errors, and a $42$\% reduction in the number of POD modes required to capture a fixed fraction of snapshot energy. These results demonstrate that \texttt{CRLS} provides an accurate, data-efficient, and largely automated route to shock-aware reduced-order models for high-speed aerodynamic design.
	\end{abstract}

	\section{Introduction}
	\label{sec:intro}
	
	In aircraft design, full-scale numerical simulations of aerodynamic fluid flows, particularly flows involving phenomena like shock waves, require extensive computational resources and run times. With the revival of interest in commercial supersonic transportation, the need for accurately capturing the parametric variation of shock strength and location has been unprecedented. While advances in numerical methods, e.g., with shock capturing schemes \cite{Rusanov1962, vanLeer1979, Steger1981, Jiang1996, Cockburn1998, Fu2016}, have enabled this, they are still prohibitive to be used in a design setting where they will have to be queried many times. Emulating such flows with a computationally cheap surrogate model would be more appropriate for a design setting; however, that introduces a unique set of challenges that must still be addressed. This work aims to address the challenges involved in accurately emulating shock-dominated flows with a surrogate model.
	
	Reduced order modeling (ROM) \cite{sirovich1987turbulence} approximates the state of a system of partial differential equations (PDEs) in terms of a finite number of ``basis'' solutions. If the basis solutions are well chosen, then this offers an effective way to reduce the number of degrees of freedom in the PDE system, thereby offering a computationally cheap surrogate model.
	At the core of most ROMs are the methods for determining a suitable reduced basis. Approaches such as proper orthogonal decomposition (POD) \cite{Berkooz1993, Rowley2004}, dynamic mode decomposition (DMD) \cite{Schmid2010} for unsteady flows, balanced POD \cite{Rowley2011}, deep neural autoencoders \cite{Otto2023, Maulik2021, Fu2023}, and Fourier neural operators (FNO) \cite{li2021fno, Dai2023}, offer powerful tools for identifying the most dominant flow structures in the full order model (FOM). By capturing the key features of the underlying system with a handful of dominant modes, these methods preserve the essential physics of fluid flow while significantly reducing the computational cost.
	
	Among all the reduced basis methods, POD based methods are particularly attractive. This is because (i) the POD bases are \emph{optimal}~\cite{Berkooz1993} in the sense that they represent the best choice of basis solutions that capture the dominant energy modes in the flow, and (ii) they offer a linear mapping between the reduced and full order solution, making computation simple; we provide a formal introduction to POD in section \ref{sec:pod}.
	However, POD-based ROM methods face challenges when dealing with convection-dominated and multi-scale flows \cite{Renganathan2020, Renganathan2020_1, Renganathan2018, Nair2019, Welper2017}. There are two perspectives as to why this is true. First, convection-dominated and multiscale flows are endowed with a wide spectrum of scales; POD, which is based on the singular value decomposition (SVD), is ideal when larger integral scales of the flow dominate. Second, a simple linear combination of a basis set might not be adequate to capture sharp gradients; for instance, a linear combination of two step functions cannot lead to another step function for non-zero weights; we provide an illustration in section \ref{sec:pod2}. Shocks are an example of sharp changes in flow features. Projection-based ROMs typically depend on the smooth low-dimensional modal expansions, and these expansions struggle to capture sharp changes without smearing or introducing non-physical oscillations (similar to the Gibbs phenomenon \cite{jerri1998gibbs}). 
	
	Researchers gravitate towards POD because of its low computational cost,  theoretical soundness, and the ability to handle large data sets. But it struggles to capture strong nonlinear effects, like shock waves. Several works exist that address the aforementioned shortcomings of POD-based ROMs for shock-dominated flows. Examples include adaptive h-refinement \cite{Carlberg2015} which adaptively refines reduced-order models a posteriori on the inviscid Burgers equations, projection-based ROM using a Koopman-based approach for high subsonic and transonic flows \cite{Renganathan2018, renganathan2018methodology}, transported snapshot model order reduction \cite{Nair2019} which works by transporting the snapshots to the desired location and solving the minimization problem, combining POD with DMD \cite{Yao2022} to construct an evolution model for POD mode coefficients which uses DMD with specific control to predict unsteady transonic flow. Most of these POD-based ROM models have traditionally treated high-order systems as if they were linear. But transonic aerodynamic flows with strong nonlinear features such as shocks are discontinuous in nature as well. This makes it hard to accurately predict the flow features using POD without manipulating the data to remove these discontinuities. 
	
	Recently, researchers are more focused on developing  data-driven flow models based on deep learning techniques to predict the flow behaviour for highly non-linear systems \cite{Kutz2017}, \cite{Bhatnagar2019}, \cite{Zhang2021}. 
	The combination of complex network architectures and advanced algorithms to train them offer tremendous potential. This made deep-learning approaches well-suited to tackle high-dimensional flow problems that feature nonlinear behaviors. Examples include transonic flow prediction over airfoils using deep learning based multi-layer perceptron neural network \cite{Sun2021}, CNN-based deep learning  predictions of flow  around transonic airfoils \cite{Duru2022}, transonic flow prediction by using vision transformer (ViT)-based architecture \cite{Deng2023}, using transformed-encoded geometric input, transonic flow prediction by fusing deep learning and a reduced-order model (CNN-POD) \cite{Jia2024}, to name a few. Following this, recently, graph neural networks (GNNs) \cite{Sanchez-Gonzalez2020}, \cite{Peng2023} and autoencoders \cite{Eivazi2020}, \cite{Wang2023} are also developed to emulate different kinds of data-driven flow modeling. However, deep learning approaches are data hungry which can be prohibitive when expensive fluid flows are involved; furthermore, their hyperparameter tuning, despite the advances in large scale stochastic optimization, requires additional care.
	
	We are interested in learning surrogate models capturing the parametric variation of shock-dominated fluid flows.
    When the underlying flowfield is smooth (that is, with continuous first derivatives with bounded Lipschitz constant), POD is known to work very well -- we exploit this strength of POD in our proposed method. In short, our approach performs an invertible parametric transformation of flowfields with shocks into a smooth flowfield that is better suited to POD-based reduced basis methods; the parametric transformation is then inverted to recover the original flow field. This way, our method is generic, with just a few learnable parameters, while retaining all the advantages of POD-based ROMs; we provide more details in section \ref{sec:mathod}.
	
	\subsection{Proper orthogonal decomposition (POD)}\label{sec:pod}
	Singular value decomposition (SVD) is a commonly used matrix factorization technique that could be used to decompose the given  matrix ${U} \in \mathbb{R}^{n \times m}$ into left orthogonal singular vectors ${\Phi} \in \mathbb{R}^{n \times n}$, a non-negative monotonically decreasing singular values diagonal matrix ${\Sigma} \in \mathbb{R}^{n \times m}$, and right orthogonal singular vectors ${V} \in \mathbb{R}^{m \times m}$. Consider $U = [\bm{u}_1,...,\bm{u}_m]$ to be our snapshot matrix, where each snapshot $\bm{u}_i$ is a column vector of $n$ elements. Mathematically, the SVD of snapshot matrix $U$ can be written as;
	\begin{equation}\label{1.1.1}
		U = \Phi \Sigma V^\top.
	\end{equation}
	POD \cite{Berkooz1993} is essentially applying SVD to the snapshot matrix to obtain an orthonormal basis set of vectors ${\Phi} = [\bm{\phi}_1,...,\bm{\phi}_n]$, also known as ``POD modes''. These modes are ordered by the magnitude of the corresponding singular values $\{s_1, \ldots, s_m \}$ in $\Sigma$. The singular value $s_i$ represents the “energy” captured by the 
	$i$th POD mode. According to the Eckart–Young theorem \cite{Eckart_Young_1936}, the best rank-$r$ approximation of $U$ is given by the truncated SVD as;
	\begin{equation}\label{1.1.2}
		U \approx \bar{U} = \Phi_r \Sigma_r V_r^\top,
	\end{equation}
	where $\Phi_r \in \mathbb{R}^{n \times r}$ contains  the first $r$ columns of $\Phi$ (dominant modes), $\Sigma_r$ is the $r \times r$ square matrix from the leading diagonal block of ${\Sigma} \in \mathbb{R}^{n \times m}$ and $V_r \in \mathbb{R}^{m \times r}$ contains  the first $r$ columns of $V$. In our problem setting, the matrix $U$ contains a column stack of snapshot solutions for different values of parameter $\mu \in \mcl{M} \subset \mbb{R}^d$. Then, as described above, POD can be used to express snapshot $\bs{u}_j$ via a rank-$r$ approximation given by
	
	\begin{equation}\label{1.1.3}
		\bm{u_j} \approx \sum_{i=1}^{r} a_i^j \bm{\phi_i},
	\end{equation}
	where $a_i^j$ is the $i${th} component of coefficients of the basis expansion. 
	Due to the orthonormality of the POD basis set, the coefficients $\bm{a^j}$ can be expressed as
	\begin{equation}\label{1.1.4}
		\bm{a^j} = \Phi_r^\top \bm{u_j}.
	\end{equation}
	Crucially, every $\bm{a}^j = [{a_1}^j,...,{a_r}^j]^\top$ corresponds to a particular parameter snapshot $\mu_j \in \mcl{M}$. If the parametric map $\mcl{M} \rightarrow \bm{a}^j,~\forall j$ is learned, e.g., via an interpolator $\hat{\bm{a}}$, then an unknown $\bm{u}$ can be approximated as 
	\begin{equation}
		\bm{u} \approx \hat{\bm{u}} =  \Phi_r \hat{\bm{a}}.
	\end{equation}
	
	\subsection{Limitations of POD for nonsmooth data} \label{sec:pod2}
\begin{figure}[htb!]
		\centering
		\begin{subfigure}{0.5\textwidth}
			\includegraphics[width=1\linewidth]{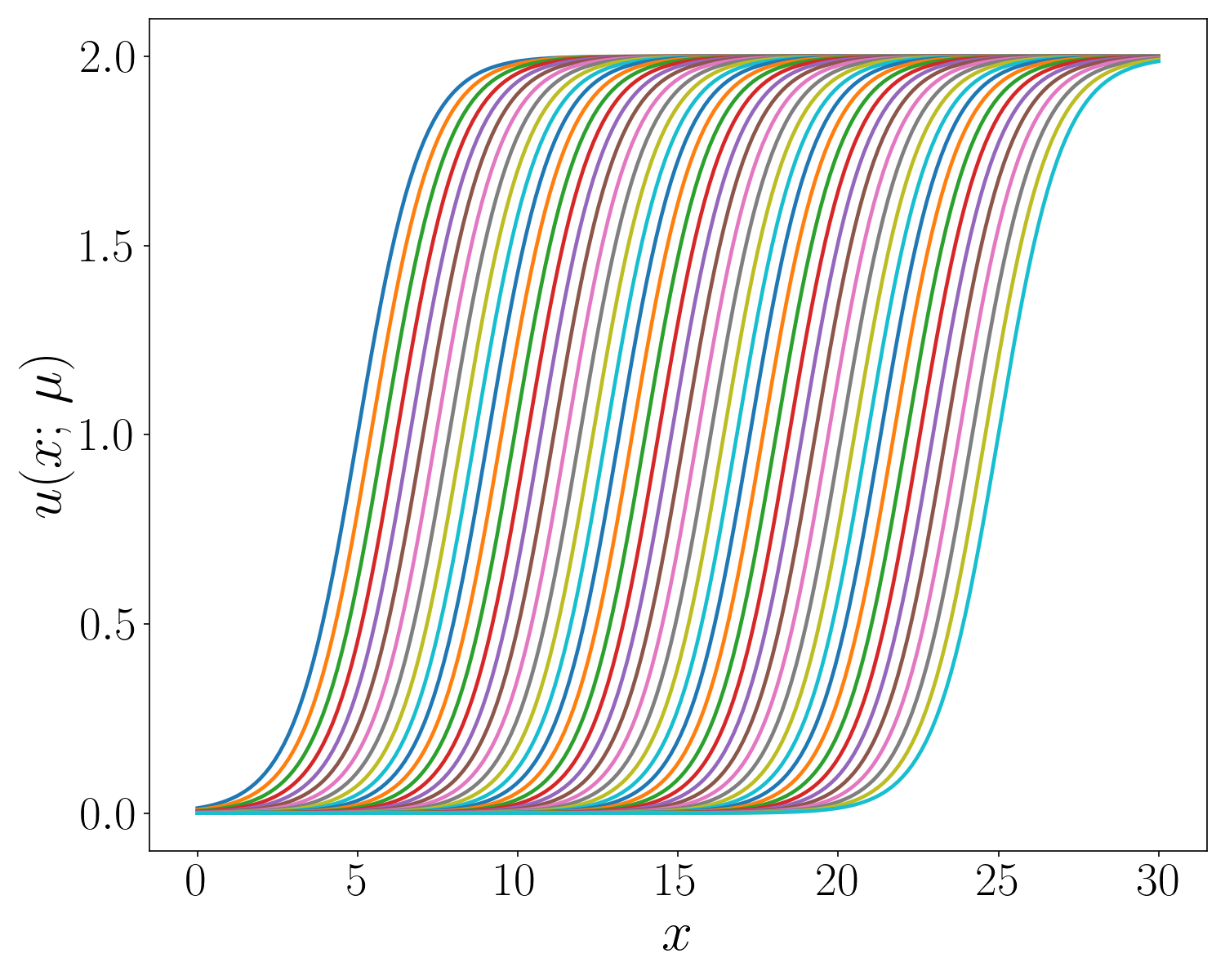}
			\caption{Smoothed step (sigmoid) functions snapshots}
		\end{subfigure}%
		\begin{subfigure}{0.5\textwidth}
			\includegraphics[width=1\linewidth]{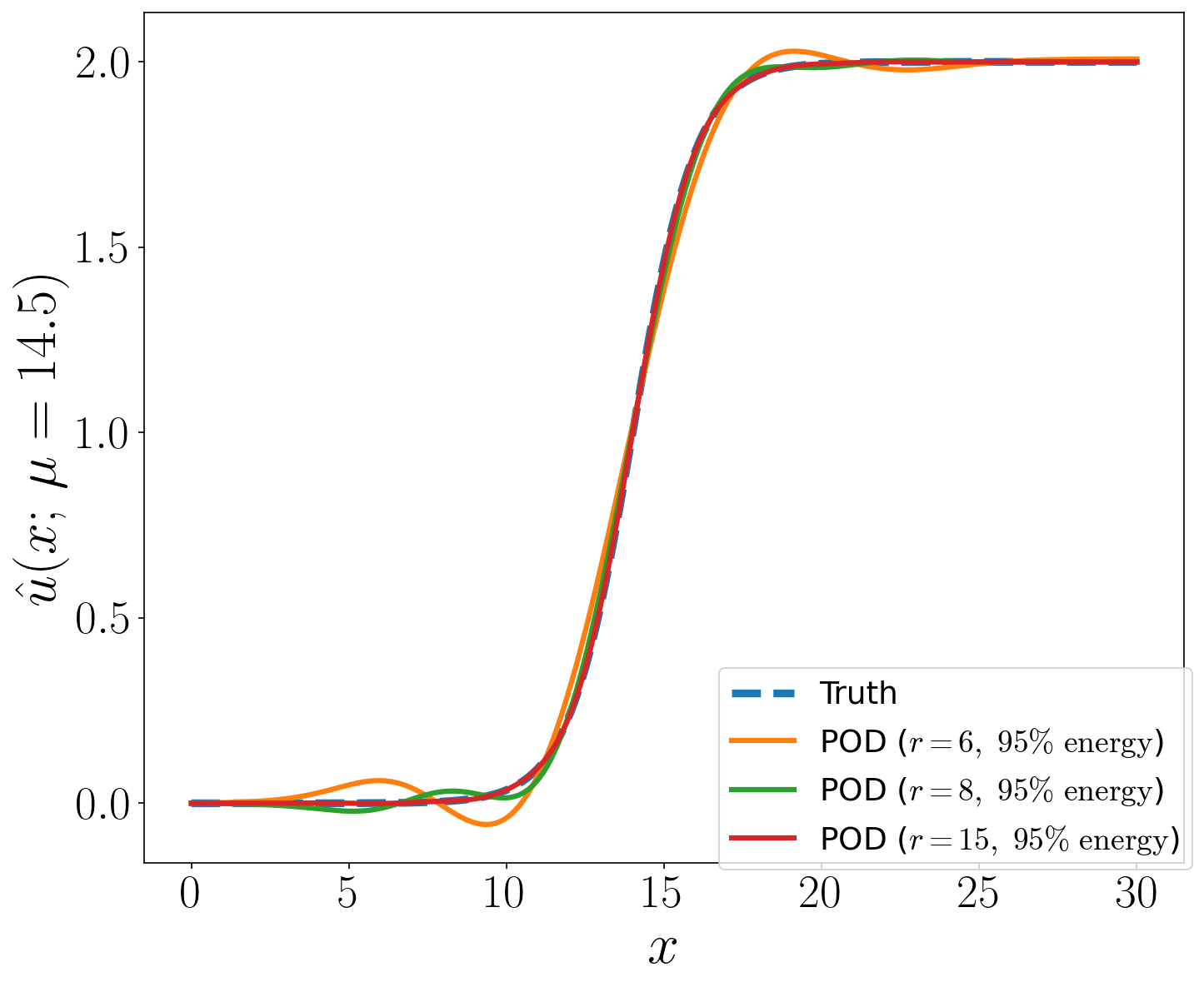}
			\caption{POD reconstruction}
		\end{subfigure}
		\caption{FOM and prediciton of sigmoid function for different \% energy  at  $\mu=14.5$ using POD}
		\label{fig:smooth_func}
	\end{figure}

\begin{figure}[!ht]
		\centering
		\includegraphics[width=0.65\textwidth]{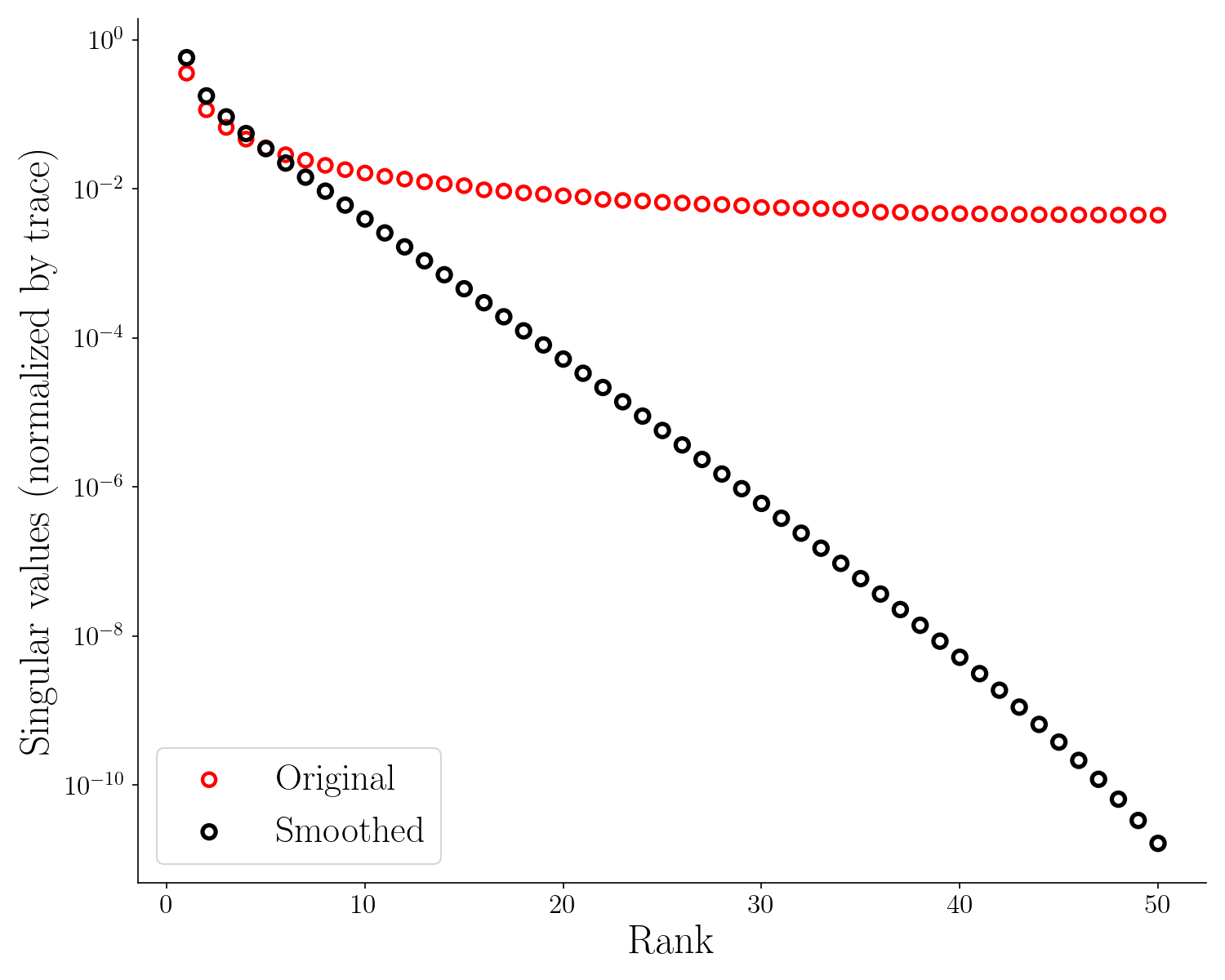}\\[-1pt]
		\includegraphics[width=0.65\textwidth]{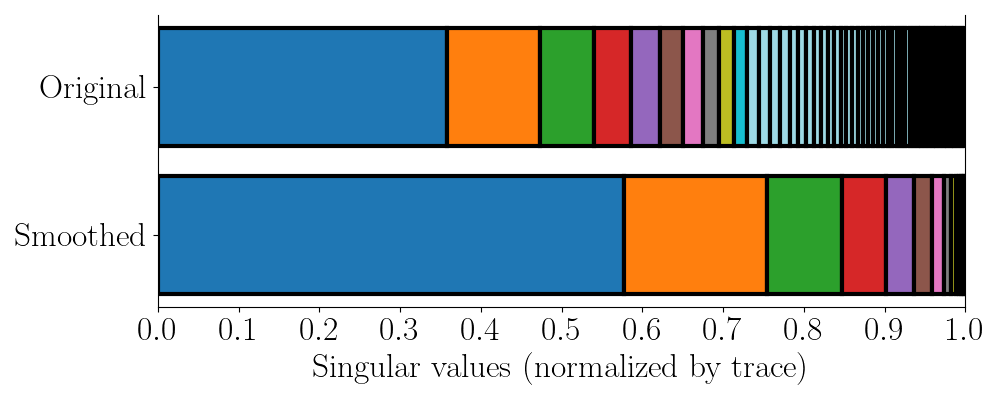}
		\captionsetup{justification=centering, labelfont=bf, font=small}
		\caption{Comparison of normalized singular values and and cumulative contribution for step functions (original) and sigmoid functions (smoothed) }
		\label{fig:RANK}
	\end{figure}

	POD-based reconstruction of flow fields offers two primary benefits: (i) linearity (which eases the inversion from $\hat{\bm{a}}$ to $\hat{u}$) and (ii) the interpretability (in terms of capturing the high-energy coherent structures). However, these benefits can be offset when the flow field has nonsmooth features -- we illustrate this with a simple example. Consider the parametric ``step'' function \eqref{1.2.1} which is evaluated at set of discrete points $x$ to obtain the snapshot matrix $U$ for different values of $\mu$;
	\begin{equation}\label{1.2.1}
		u(x; \mu) = b H(x - \mu)
	\end{equation}
	where $H$ is the Heaviside step function, and $b$ is a scalar that determines the height of the step (we set $b=2$). Using this, we have generated 50 equally spaced (for $\mu \in [5,25]$) snapshots. $x \in [0,30]$ is discretized into $200$ equally spaced points. This results in a snapshot matrix $U$ of size $200 \times 50$. The original data set obtained through this is shown to the left of \Cref{fig:step_func} below. To the right of \Cref{fig:step_func}, we show POD reconstructions at an unseen parameter value $\mu = 14.5$, where we use a radial basis function (RBF) \cite{fasshauer2007meshfree} interpolant to predict the POD coefficients at $\mu = 14.5$. Notice the ``stairstepping'' in the predictions, regardless of the number of modes used.
	\begin{figure}[htb!]
		\centering
		\begin{subfigure}{0.5\textwidth}
			\includegraphics[width=1\linewidth]{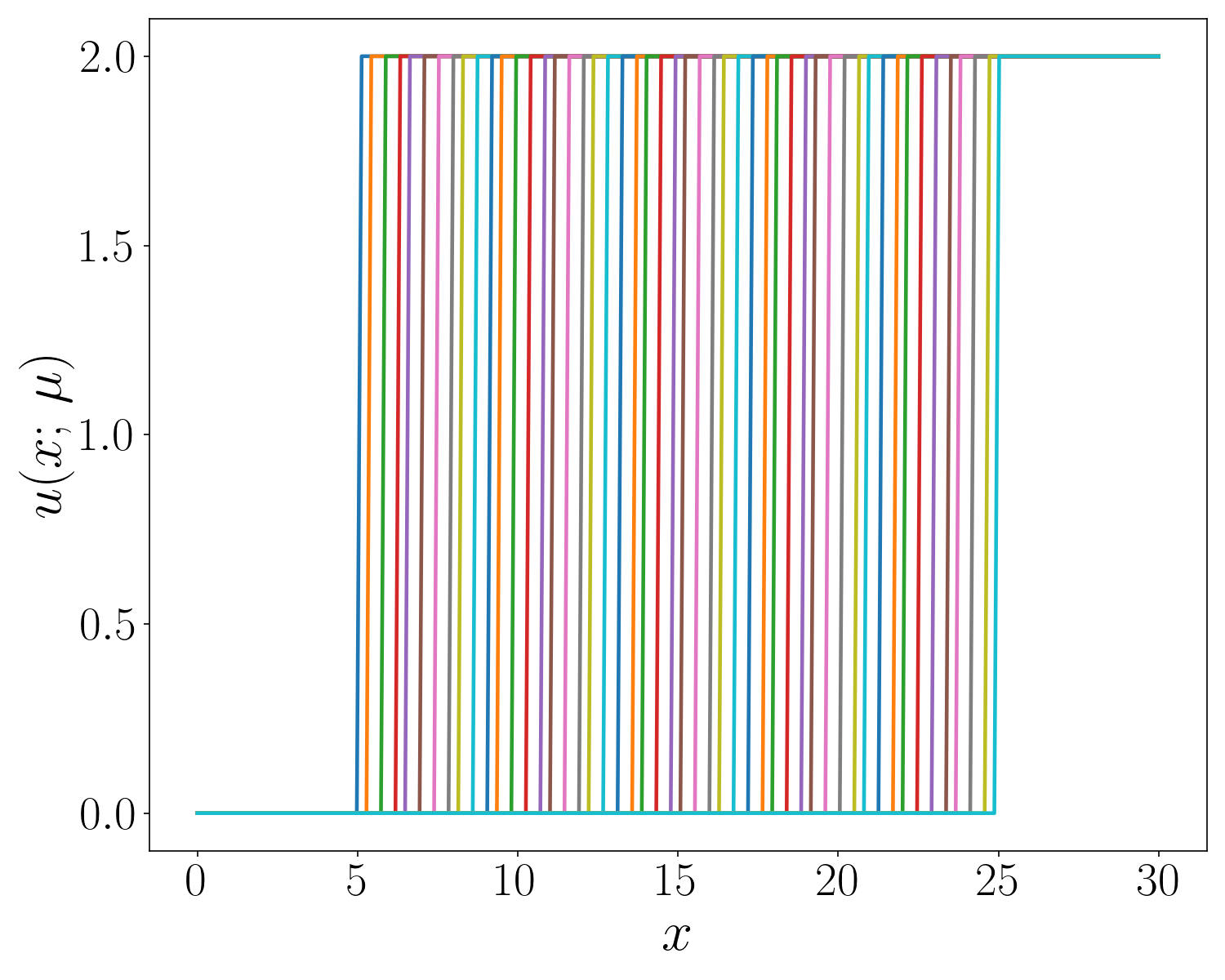}
			\caption{Step functions snapshots}
		\end{subfigure}%
		\begin{subfigure}{0.492\textwidth}
			\includegraphics[width=1\linewidth]{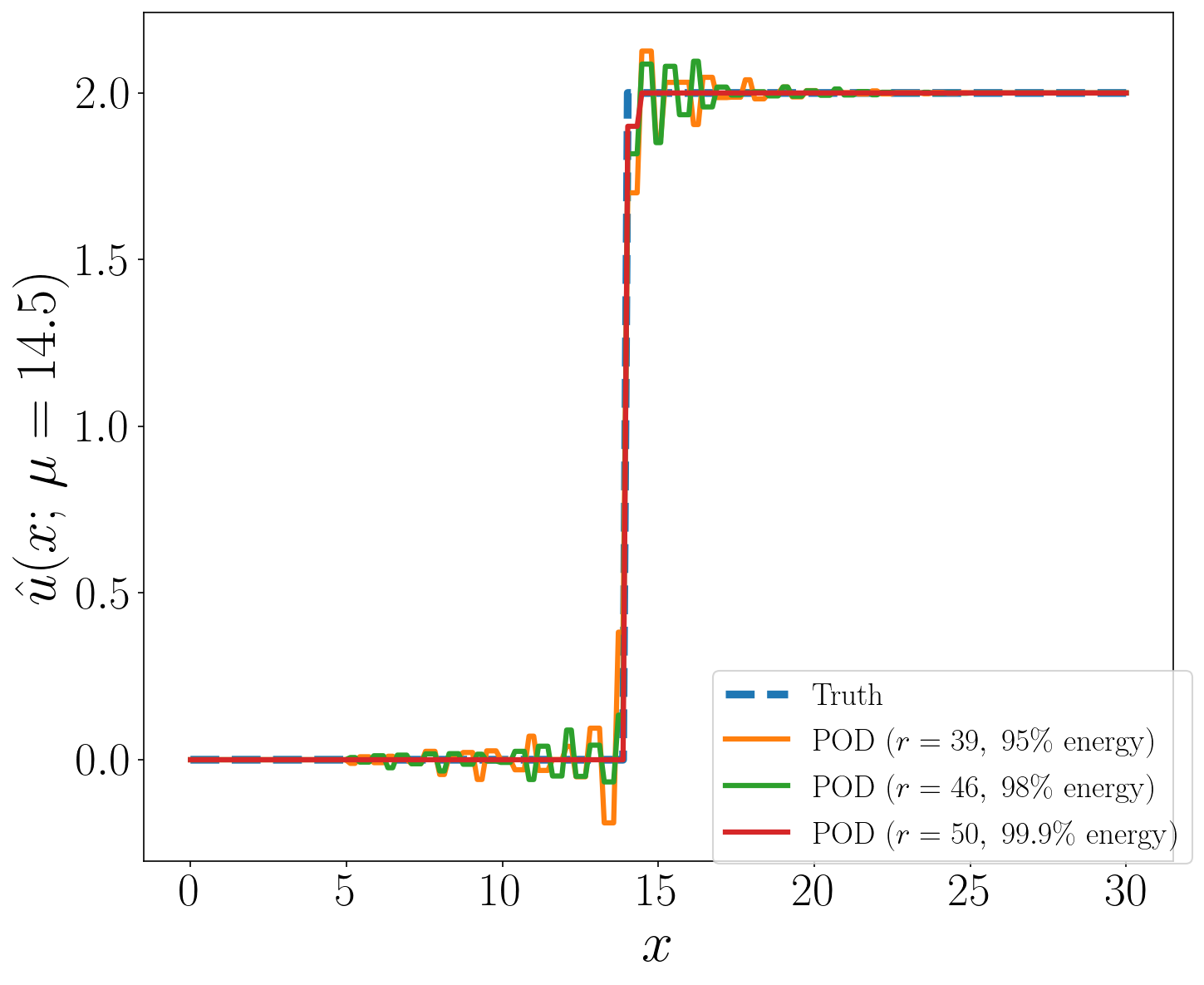}
			\caption{POD prediction reconstruction}
		\end{subfigure}
		\caption{FOM and prediciton of step function for different \% energy  at  $\mu=14.5$ using POD }
		\label{fig:step_func}
	\end{figure}
Now, consider a ``smoothed'' version of the step function given as in which similar to \eqref{1.2.1}, is evaluated at set of discrete points $x$ to obtain the snapshot matrix $U$ for different values of $\mu$;
	\begin{equation}\label{1.2.2}
		u(x; \mu) =  \frac{b}{{1+e^{(\mu - x)}}}.
	\end{equation}
	Using this, we again generated 50 equally spaced snapshots (for $\mu \in [5,25]$). The function snapshots and the POD reconstruction are shown in \Cref{fig:smooth_func}; notice the improvement in the predictions (right) even with as few as $8$ modes.
	This $1D$ example illustrates the dependence on smoothness of the underlying snapshots for accurate POD reconstruction. The singular values plot is particularly illuminating; in \Cref{fig:RANK}, we show the singular values for the original and smoothed snapshot matrices; notice the ``slower'' variation for the original snapshots (step function) compared to the smoothed snapshots. Therefore, smoothing additionally enables capturing more energy in the flow with far fewer modes than the original snapshots -- this is visualized by the bottom part of \Cref{fig:RANK}. Drawing from this, our work proposes a strategy, where inherently nonsmooth flow fields are first smoothed using an appropriate smoothing transformation. Then, POD-based reduced basis methods are constructed on the smoothed flow fields, following which we invert the smoothing to recover the original flow fields. In this regard, we propose \crls: convolutional regularized least squares, which uses a convolution operator to smooth flow fields for better POD reconstruction, followed by a regularized de-convolution to invert the smoothing --  we provide more detail in \Cref{sec:method}. 
	The rest of this paper is organized as follows. We present details of the \crls~method in \Cref{sec:method}. Then, in \Cref{sec:experiments}, we present details of our experiments, followed by results and discussion in \Cref{sec:results}. Finally, we conclude in \Cref{sec:conclusion} with a summary and an outlook for future work.
	
	\section{Methodology} \label{sec:mathod}
	\label{sec:method}
	Our overall method \crls~is an amalgamation of several ingredients -- we present each of them as follows.
	\subsection{Convolution}
	\label{sec:convolution}
	For two continuous functions $u(x)$ and $k(x)$, the convolution operation is defined as:
	\begin{equation}\label{2.1a.1}
		(u * k)(x) = \int_{-\infty}^{\infty} u(\tau) \cdot k(x - \tau) \, d\tau,
	\end{equation}
	where $\tau$ is a dummy integration variable, which affects the width of the convolution window. In the discrete setting, eq (\ref{2.1a.1}) can be expressed with the integral replaced with a summation operator as
	\begin{equation}\label{2.1b.1}
		(u * k)(x) = \sum_{j=-\infty}^{\infty} u[j] \cdot k[x - j]
	\end{equation}
	where analogous to $\tau$ in \eqref{2.1a.1}, $j$ is a dummy summation index.  In discrete terms, the convolution involves sliding the ``filter kernel'' $k$ over the signal $u$. At each position, the overlapping values from the image and the kernel are multiplied element-wise and summed to produce a new value in the output smoothened/convolved snapshot. 	Gaussian kernel \cite{gaussian_filter} is a common choice for the kernel which can be expressed as	
	\begin{equation}\label{2.1.1}
		k(x) = \frac{1}{\sqrt{2\pi}\,\sigma} \exp \left({-\frac{x}{2\sigma^2}} \right),
	\end{equation}
    where $k$ is a scalar-valued function $k(x): \mbb{R} \rightarrow \mbb{R}_+$, $\sigma$ is the standard deviation or the measure of spread of Gaussian. In the discrete setting, the kernel is generally evaluated at odd-numbered discrete locations $x_1, x_2, x_3, \ldots, x_p $, which are symmetric around zero, and normalized such that: 
		\begin{equation}\label{2.1.2}
		\sum_{i = 1}^{p} {k}(x_i) = 1
	\end{equation}
		Consider the vector $\bm{u}_i $ of $n$ elements and let $\bm{k}$ be a kernel vector of $p$ elements obtained by evaluating \eqref{2.1.1} on a set of points $\{x_1, \ldots, x_p\}$ such that $\bm{k} = k(\bm{x}) = [k(x_1),...,k(x_p)]^\top$; note that $p < n$ necessarily. 
	In the context of 1D Gaussian smoothing, convolution is used to apply the Gaussian kernel to the snapshot data matrix sequentially (one snapshot at a time) to produce a smoothed version of the data matrix $U$. For simplicity, the convolved value in the discrete case, i.e., for a sampled snapshot $\bm{u}_i$ at index $j$ can  be expressed  as:
	\begin{equation}\label{2.1.3}
		{{\bm{u}}^s_i} [j] = \sum_{l=1}^{p} {\bm{u}}_i{[(j-1+l)]}\cdot k(x_l)),
	\end{equation}
	where ${{\bm{u}}^s_i}[j]$  is the convolved value of $\bm{u}_i$ at index $j$. This means that each new smoothed value of $\bm{u}_i$ is computed as a weighted sum of a set of $p$ elements of ${{\bm{u}}_i[j]}$, where the weights are determined by the Gaussian kernel $k$. 
		
\begin{figure}[!h]
		\centering
		\includegraphics[width=0.75\textwidth]{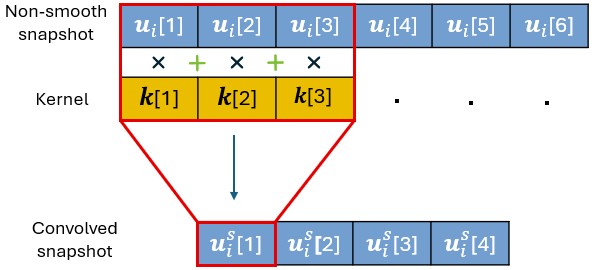}\\[-1pt]
		\captionsetup{justification=centering, labelfont=bf, font=small}
		\caption{Convolution without padding}
		\label{fig:Convolution_1}
	\end{figure}

\begin{figure}[!h]
		\centering
		\includegraphics[width=0.75\textwidth]{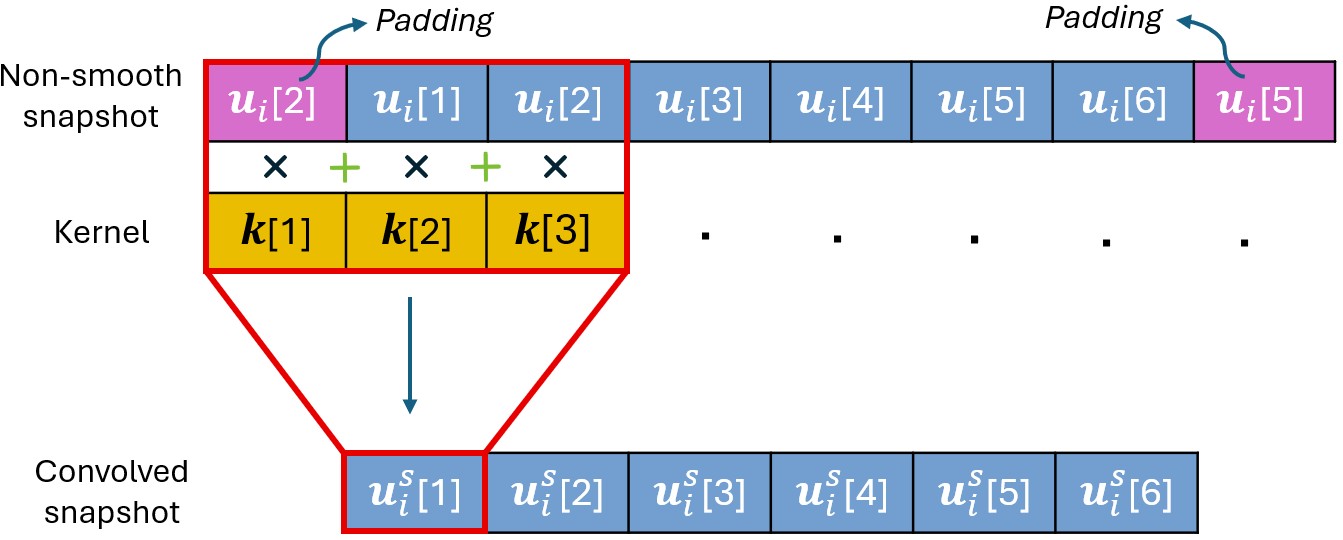}\\[-1pt]
		\captionsetup{justification=centering, labelfont=bf, font=small}
		\caption{Convolution with reflect padding}
		\label{fig:Convolution_2}
	\end{figure}

	This smoothing process, which involves a moving window of $p$ elements, needs special attention to ensure the smoothed snapshot retains the cardinality of the original nonsmooth snapshot. This is illustrated as follows: consider the snapshot $\bm{u}_i$ of six elements and kernel window of three elements, then the convolved output snapshot $\bm{u}^s_i$ would have four elements as shown in the \Cref{fig:Convolution_1}.  
	To overcome this problem, padding is generally used to extend the length of $\bm{u}_i$ before applying the convolution. For a kernel window $\bm{k}$ of $p$ elements, $L$ additional elements are added for the padding. Multiple types of padding such as zero padding, constant padding, and reflect padding can be used. The choice of padding affects how smoothly the edges are handled. To prevent abrupt changes in the convolved values without compromising the quality of data, we have used the reflect padding here. Consider again the same snapshot $\bm{u}_i$ of six elements and kernel window of three elements. The snapshot now has $1$ additional element on each side as a padding (reflect padding). The  convolved output snapshot $\bm{u}^s_i$, thus, would now have six elements as shown in \Cref{fig:Convolution_2}.
	
	\subsection{Reflect-padded 1D convolution in matrix form}
	In this subsection, we will provide the details of mathematical formulation to perform the convolution with the reflect-padded data using the kernel in the matrix form. Consider again the input snapshot data $\bm{u}_i$ of $n$ elements and kernel $\bm{k}$ of $p$ elements, then to have the smoothened output $\bm{u^s_i}$ of same cardinality as input ${\bm{u}_i}$, a padding size $L$ is chosen as:
	\begin{equation}\label{2.2.3}
		L=\frac{(p-1)}{2},
	\end{equation}
	where $p$ is the kernel length, and notice that $p$ must be an odd number for $L$ to be an integer. This padding will extend ${\bm{u}_i}$ with $n$ elements to another vector $\tilde{\bm{u}_i}$ with $n+2L$ total elements. Consider a set of indices $\{\,j\in\mathbb{Z}\mid -L \le j \le n + L - 1\}$, then $\tilde{\bm{u}_i}$ can be written as
	\begin{equation}\label{2.2.4}
		\tilde{\bm{u}_i}
		\;=\;
		\begin{cases}
			\bm{u}_i[-j], 
			& \text{if } -L \,\le j \,< 0,\\[6pt]
			\bm{u}_i[j],  
			& \text{if } 0 \,\le j \,\le n,\\[6pt]
			\bm{u}_i\bigl[2\,n - 2 - j\bigr], 
			& \text{if } n \,< j \,\le n + L.
		\end{cases}
	\end{equation}
    Note that we introduce negative indices $\{-L, -L+1, \ldots, 0\}$ to preserve the fact that indices $\{1, \ldots, n\}$ corresponds to the original snapshot. This way, the front padding ($-L,\ldots,0$) mirrors the first $L$ elements of the snapshot, the rear padding ($n+1, \ldots, n+L$) mirrors that last $L$ elements of the snapshot and the rest are unchanged. The reason for doing this will become clear next: the convolved snapshot ${u^s_i}$ can be written as
	\begin{equation}\label{2.2.5}
		{{\bm{u}}^s_i} [j] = \sum_{l=1}^{p} \tilde{\bm{u}}_i{[(j-1+l)]}\cdot \bs{k}[l]),~\forall j=1,\ldots,n.
	\end{equation}
	It is also possible to write the convolved output $\bm{u^s_i},~\forall i$ in matrix form as  $U^s = BU $ if we define \( {B} \) to be an \( n \times n \) matrix with entries \( B_{i,j} \) given by:
	
	\begin{equation}\label{2.2.6}
		B_{i,j} =
		\begin{cases} 
			k_{q}, & \text{if } (i+q - L < 0 \text{ and } j = -(i+q - L)) \\
			& \quad \text{or } (0 \leq i+q - L < n \text{ and } j = i+q - L) \\
			& \quad \text{or } (i+q - L \geq n \text{ and } j = 2n - 2 - (i+q - L)), \\
			0, & \text{otherwise}.
		\end{cases}
	\end{equation}
    where $q \in \{0, \dots, p-1\} $. We illustrate the construction of the $B$ matrix for a simple case of $p=3$ and $L=5$ in Appendix \ref{Appendix_1}.

	\subsection{Deconvolution}
	The process of inverting a convolution is called deconvolution, which is a lossy operation and is generally difficult to recover the original data from it exactly. With the information about how the convolution is performed, as presented above i.e; $\bm{u^s}_i = {B} \bm{u}_i$, one can perform the deconvolution to invert $\bm{u}_i$ from $\bm{u}$ by minimizing the residual $\frac{1}{2} \| B\bm{u}_i - \bm{u}^s_i \|^2$.
   However, to overcome the ill-conditioning of $B$ (which leads to the lossiness of this inversion), we regularize the residual $\frac{1}{2} \| B\bm{u}_i - \bm{u}^s_i \|^2$ accordingly -- this leads to our convolutional regularized least squares (\texttt{CRLS}) method. 
    We will provide more detail on this regularization momentarily, but before that, we will address how to estimate the hyperparameters of the convolution operation next.
	
	\subsection{Estimation of hyperparameters}
	The extent of smoothing in our convolution operation is affected by the Gaussian width $\sigma$ and the size of the kernel window $\texttt{len}(\bm{k})$. $\texttt{len}(\bm{k})$ is an integer varying from $1$ to $n$ and $\sigma \in [0, \infty)$ is a nonnegative real-valued parameter. At $\sigma = 0$ the Gaussian becomes a Dirac delta function, i.e., no smoothing at all; on the other hand, as $\sigma \rightarrow \infty$, the Gaussian flattens out and becomes constant, and the snapshots will be completely blurred, i.e., no variation at all. In practice, the range of $\sigma$ for Gaussian smoothing is meaningful only within certain bounds. On the other hand, having a smaller kernel window, which, while resulting in a sparse matrix $B$, eases the deconvolution computation, can lead to insufficient smoothing. And, vise versa, larger kernel windows are also undesirable. 
    
    We formulate an optimization problem to optimally estimate these two hyperparameters. Specifically, we choose hyperparameters by minimizing the maximum infinity norm of the prediction error over a ``held out'' test set of snapshots. This is presented as follows
    \newcommand{\argmin}{\text{arg}\min}
	\begin{equation} \label{3.4.1}
		\begin{split}
			\{\sigma^*,~ \texttt{len}(\bm k)^*\} =& \argmin_{\sigma,~ \texttt{len}(\bm k)} \quad  \underset{i \in \mathcal{T}}{\text{max}} \left\| \bm{u}_{i, \text{predict}} - \bm{u}_{i} \right\|_\infty \\
            \text{s.t.} \quad 
			& \; \sigma \in [0, 100] \subset \mathbb{R} \\
			& \; \texttt{len}(\bm k) \in [1, \texttt{len}(\bm u)] \subset \mathbb{N} ,
		\end{split}
	\end{equation}
    where $\mathcal{T}$ is a set of indices corresponding to the held out snapshot set, $\bm{u}_{i, \text{predict}}$ is the  predicted snapshot after deconvolution at an unseen parameter location, and $\bm{u}_{i}$ is an original snapshot in the held out set. The bounds on $\texttt{len}(\bm k)$ are natural, and the bounds on $\sigma$ are chosen based on experiments with a variety of ranges; note that one can still set $0 \leq \sigma < \infty$, and our method works seamlessly, but we recommend a finite range. Solving this minimax problem is hard without access to gradients and due to the fact that the $\max$ operator being non-differentiable. Therefore, we solve it using Bayesian optimization~\cite{frazier2018tutorial,renganathan2025q,renganathan2021enhanced,renganathan2023camera, renganathan2021lookahead} with Gaussian process~\cite{williams2006gaussian} surrogate models. 
	
	\subsection{Least squares regularization}
    Once we have the hyperparameters estimated for the convolution, we are ready to pose and solve our regularized least squares optimization problem to invert the {smoothened} prediction $\bm u^s$ to obtain the prediction with shocks $\bm u$. In this work, we regularize the least squares problem with the Euclidean distance to a reference snapshot $\bm u_\text{ref}$. The \texttt{CRLS} optimization problem is then given as
	\begin{equation} \label{3.5.1}
		\bm{u}_{\T{predict}} = \min_{\bm{u}} \frac{1}{2} \left\| B \bm{u} - \bm{u}^s \right\|_2^2 + \frac{\gamma}{d^2} \left\| \bm{u} - \bm{u}_{\text{ref}} \right\|_2^2,
	\end{equation}
	where $\bm{u}$ is the unknown solution to be determined, $\bm{u}^s$ is the predicted smooth snapshot after convolution, $\gamma$ is the regularization parameter, $d$ is a weighting parameter, and $\bm{u}_{\text{ref}}$ is the reference snapshot chosen to be the nearest neighbor (in the parameter space) to the current parameter of interest $\mu$; we use a KDTree (implemented as \texttt{KDTree} in SciPy~\cite{virtanen2020scipy}) based nearest neighbor algorithm for this purpose. The weighting parameter measures the Euclidean distance, in the parameter space, between $\mu$ and $\mu_\text{ref}$. This inverse squared-distance weighting follows the intuition that for parameters close by in parameter space, the corresponding snapshots are likely to be close by in the snapshot space.
	Finally, the POD coefficients at unseen parameter locations are interpolated via a radial basis function (RBF) interpolant, for which we use the \texttt{Rbf} implementation in SciPy.

\begin{figure*}[!htbp]
		\centering
		\includegraphics[width=1\textwidth]{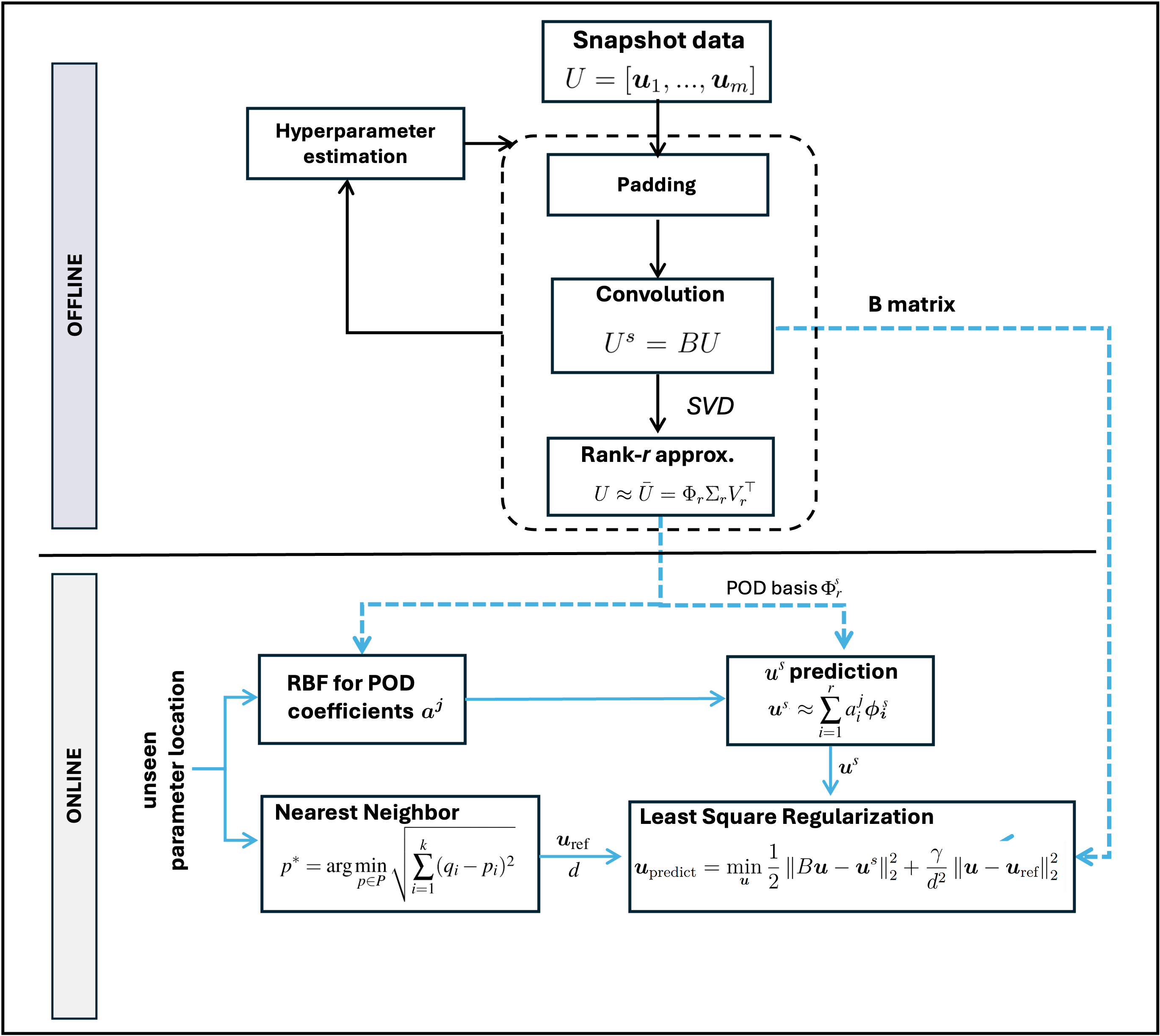}
		\captionsetup{justification=centering, labelfont=bf, font=small}
		\caption{\texttt{CRLS} algorithm overview}
		\label{fig:algo_2}
	\end{figure*}

	The complete algorithm consists of two distinct parts. The offline part, which needs to run once for a unique set of snapshot data, will perform convolution and POD basis extraction, following the hyperparameter estimation. Next, the online part includes prediction and regularized least squares inversion.  The overall methodology is summarized in \Cref{fig:algo_2}.
	
	
	\section{Experiments}
	\label{sec:experiments}
	
	We demonstrate our methodology on the transonic flow past the RAE2822 airfoil.
	For the full order model (FOM), we consider the two-dimensional, steady, compressible Euler equations to model fluid flow. These equations capture the conservation of mass, momentum, and energy and are summarized as follows:
	\begin{equation}\label{3.1.1}
		\begin{gathered}
			(\rho u)_x + (\rho v)_y = 0, \\
			(\rho u^2 + p)_x + (\rho u\,v)_y = 0, \\
			(\rho u\,v)_x + (\rho v^2 + p)_y = 0, \\
			(\rho u H)_x + (\rho v H)_y = 0.
		\end{gathered}
	\end{equation}
	where
	\begin{equation*}\label{3.1.2}
		H = E + \tfrac{p}{\rho}, \, E = \frac{u^2 + v^2}{2} + \frac{p}{\rho (\gamma -1)}.
	\end{equation*}	
	Here, $u$ and $v$ are x and y-components of the velocity, $\rho$ is the density, $p$ is pressure, $E$ is internal energy, $H$ is the enthalpy, and $\gamma$ is the adiabatic index. 
	We use the open-source finite-volume-based CFD solver SU2~\cite{economon2016su2} to solve the FOM and generate the snapshot data. Assuming a quasi-1D, irrotational flow in the boundary-normal direction, the free-stream values at boundary faces are then computed by extrapolating Riemann invariants. The numerical solution uses a second-order spatial discretization and a coupled implicit method with a hybrid Gauss-least-squares method for gradient computations. Lastly, to handle steep gradients (shocks), the Venkatakrishnan limiter is used for the numerical stability of the solution. On the airfoil surface, an adiabatic slip-wall condition is imposed, and by using reconstruction gradients, primitive and thermodynamic variables are extrapolated from the interior of the domain. This provides a robust and accurate framework for capturing shocks in transonic flow over the airfoil to generate FOM snapshots.
	
	The RAE2822 airfoil is a widely used canonical transonic aerodynamics test case. To model the farfield domain, we use a circular region with a radius set to $100$ times the chord length of the airfoil, with an O-grid structured mesh of $32,900$ cells and near-field mesh refinement for accurate capture of the shock wave. We collect snapshots of the pressure coefficient ($C_p$) defined as:
	\begin{equation} \label{3.1.4}
		C_p = \frac{p - p_{\infty}}{\tfrac{1}{2}\,\rho_{\infty}\,\bigl(M_{\infty}\,a_{\infty}\bigr)^2}
	\end{equation}
	where \(p\) is the local pressure, \(p_{\infty}\) and \(\rho_{\infty}\) are 
	the free-stream pressure and density, and \(M_{\infty} \) and $a_{\infty}$ 
	are the free-stream Mach number and speed of sound, respectively. 
	
	\subsection{Training and test data generation}
\begin{figure}[!ht]
    \centering

    \begin{minipage}[c]{0.6\textwidth}
        \centering
        \includegraphics[width=1\textwidth]{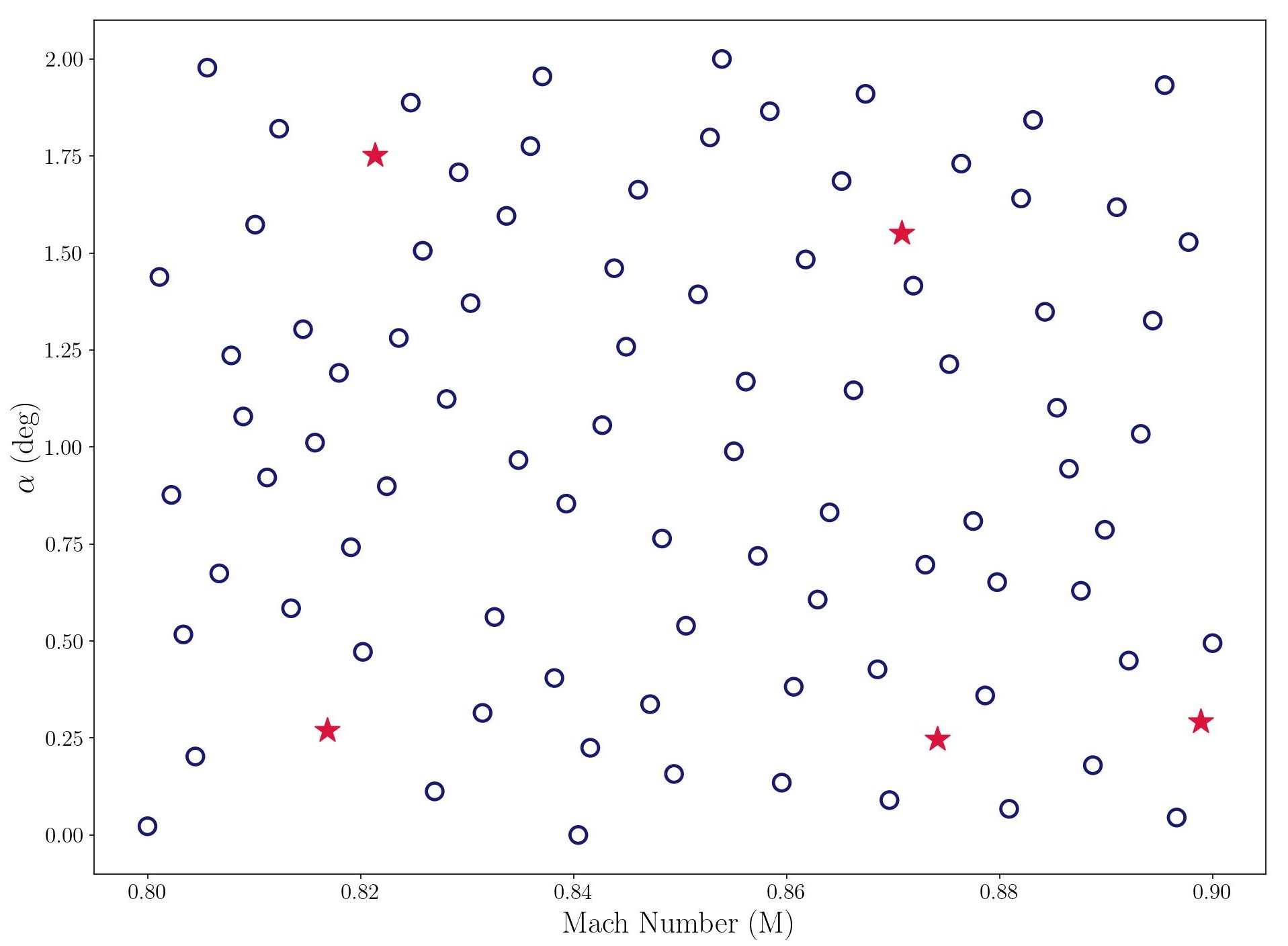}
    \end{minipage}%
    \hfill
    \begin{minipage}[c]{0.4\textwidth}
        \centering
        \includegraphics[width=1\textwidth]{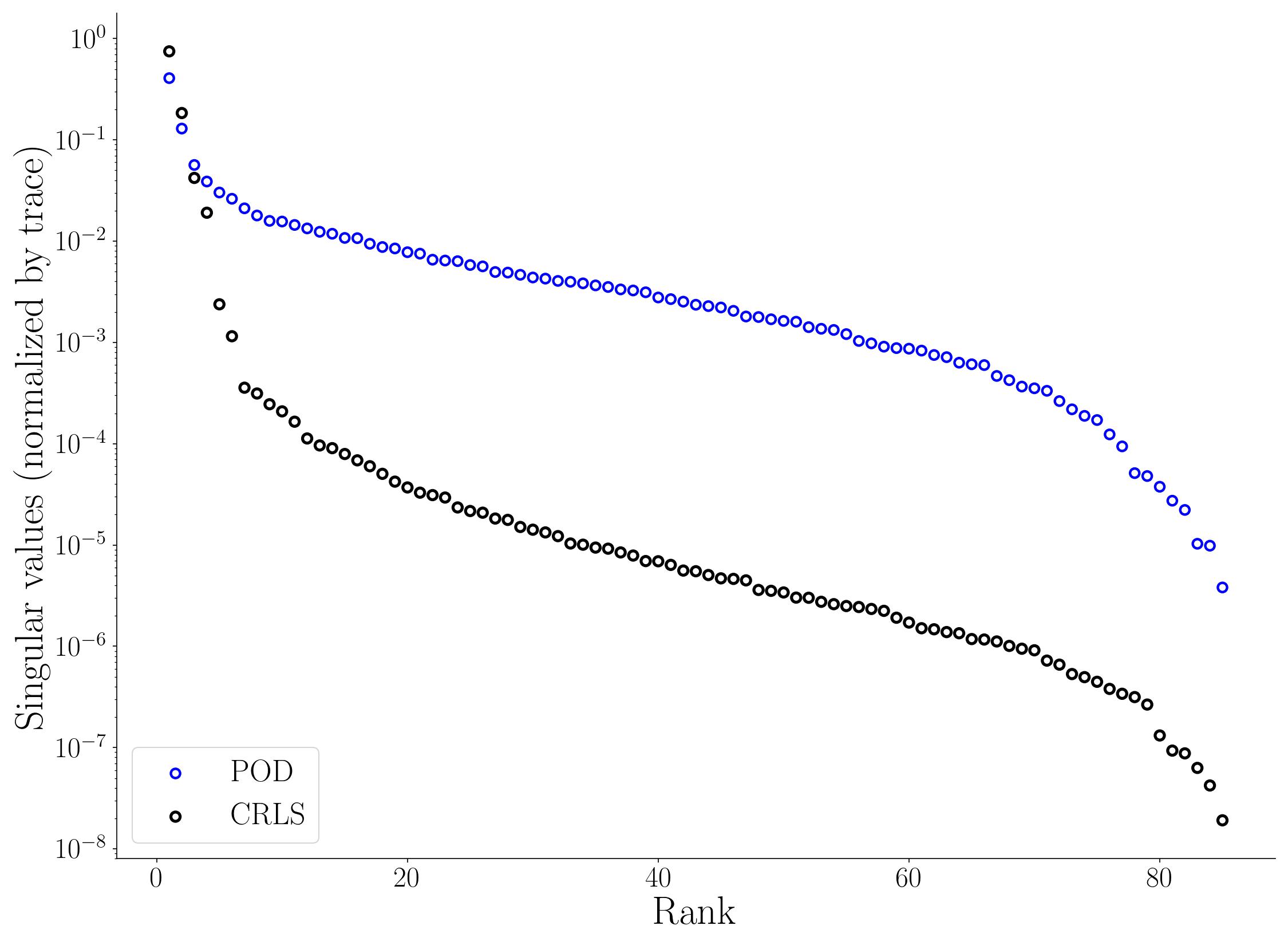}\\[-3pt]
        \includegraphics[width=1\textwidth]{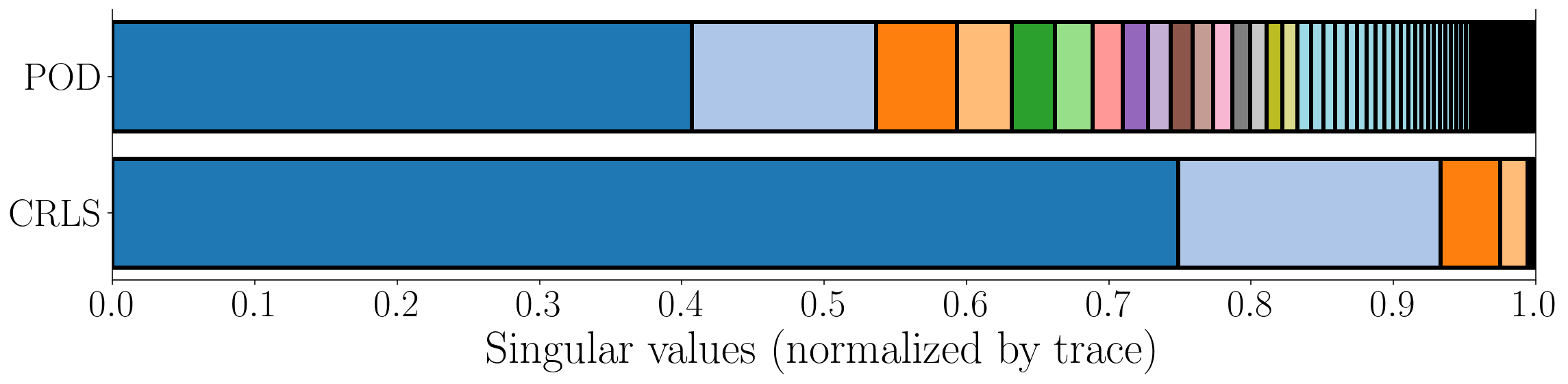}
        \label{fig:singularvalues_rae}
    \end{minipage}

    \captionsetup{justification=centering, labelfont=bf, font=small}
    \caption{Design of experiments and verification data points}
    \label{fig:DOE}
\end{figure}

	We consider two control parameters: the Mach number $M \in [0.8, 0.9]$, and the angle of attack: for which we set the range  and  the angle of attack $\alpha \in [0, 2]$. The parameter ranges are selected to ensure that the flow almost always includes a shock somewhere along the airfoil while also demonstrating substantial changes in the shock location. We generate a design of experiment, via Latin hypercube sampling~\cite{loh1996latin}, with $90$ experiments total. Of this $90$, we hold out $5$ for final validation; we split the remaining $85$ as a $90$\% - $10$\% partition for hyperparameter learning in the offline phase. The parameter scatterplot, with training and validation points highlighted, is shown in \Cref{fig:DOE}.


	\subsection{Results and discussion}
	\label{sec:results}
	We entertain three different reduced basis approaches: a traditional approach of applying POD to the snapshots without any smoothing (\texttt{POD}), POD applied to the smoothed snapshots but without regularization (\texttt{POD-c}), and the proposed approach (\texttt{CRLS}).
    The singular values plot shown in \Cref{fig:DOE} echoes the same trend observed with the $1$-D step and sigmoid function (\Cref{fig:smooth_func}) -- that is, smoothing restructures the energy distribution amongst the modes, allowing more energy to be captured by fewer modes.
 Specifically, after smoothing, we notice that the first singular value corresponds to roughly $75$\% of the energy, and the second singular value corresponds to about $18$\% of the energy -- so $93$\% of the total energy is captured by the first two modes. Overall, $99.99$\% of the total energy is captured by $48$ modes in the case of smoothened snapshots, in contrast to $81$ modes in the case of the original (nonsmooth) snapshots.
	
	\subsubsection{Airfoil surface pressure coefficients prediction}
\begin{figure}[htb!]
        \centering
        \begin{subfigure}{.5\textwidth}
        \centering
            \includegraphics[width=1\linewidth]{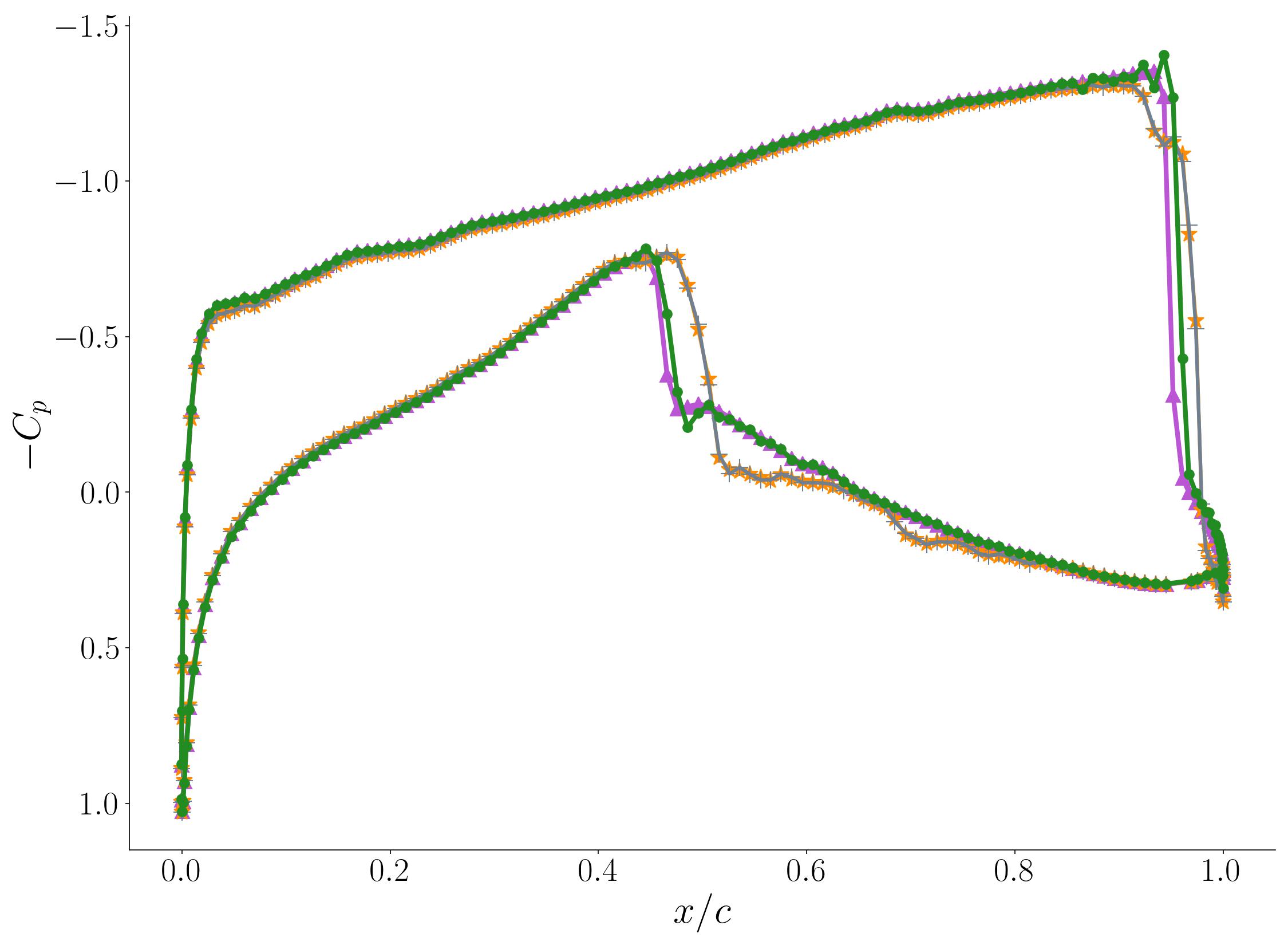}
            \caption{$M = 0.82, \alpha = 1.75$}
        \end{subfigure}%
        \begin{subfigure}{.5\textwidth}
        \centering
            \includegraphics[width=1\linewidth]{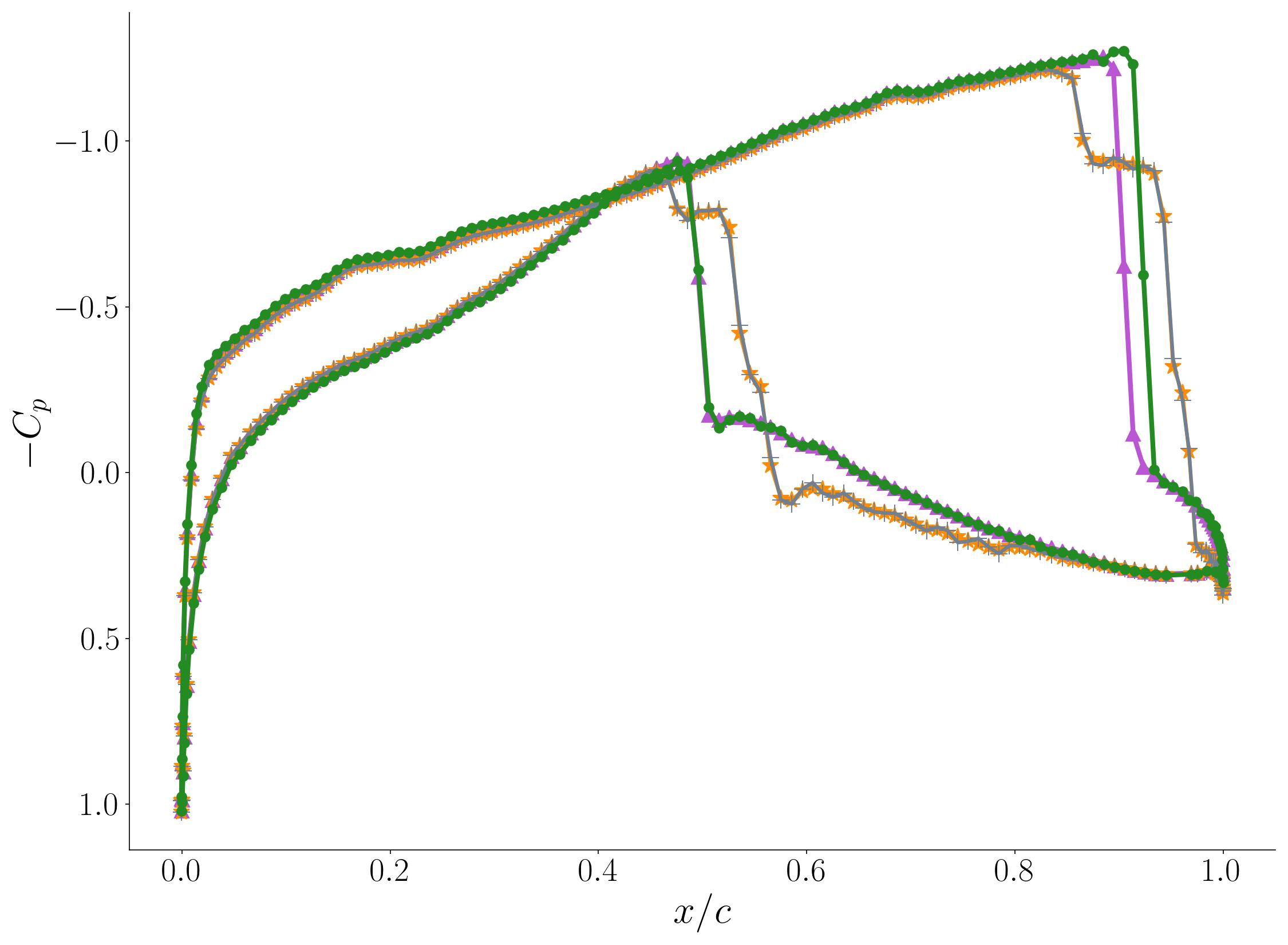}
            \caption{$M = 0.817, \alpha = 0.27$}
        \end{subfigure} \\
        \begin{subfigure}{.5\textwidth}
        \centering
            \includegraphics[width=1\linewidth]{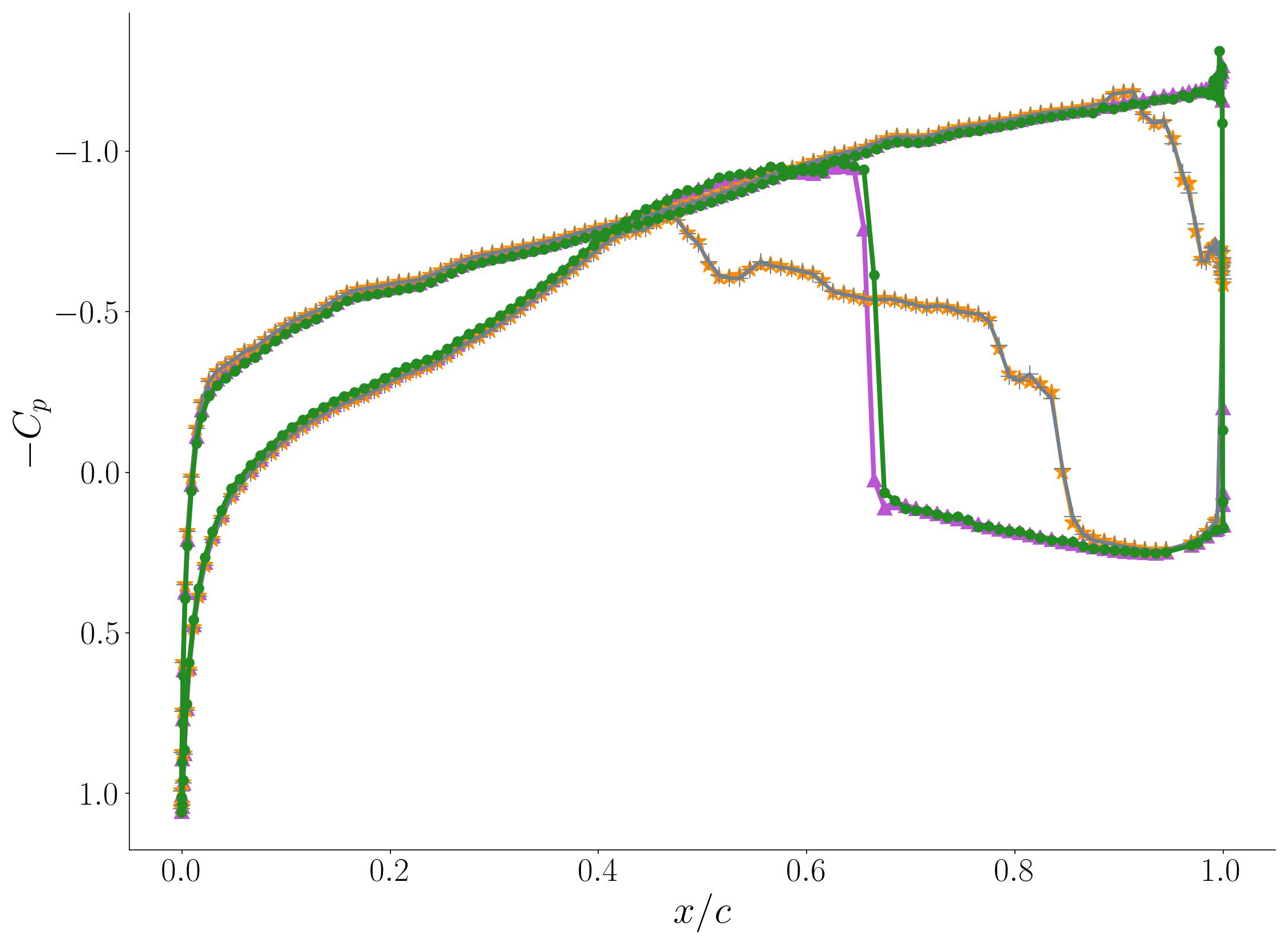}
            \caption{$M = 0.87, \alpha = 1.55$}
        \end{subfigure}%
        \begin{subfigure}{.5\textwidth}
        \centering
            \includegraphics[width=1\linewidth]{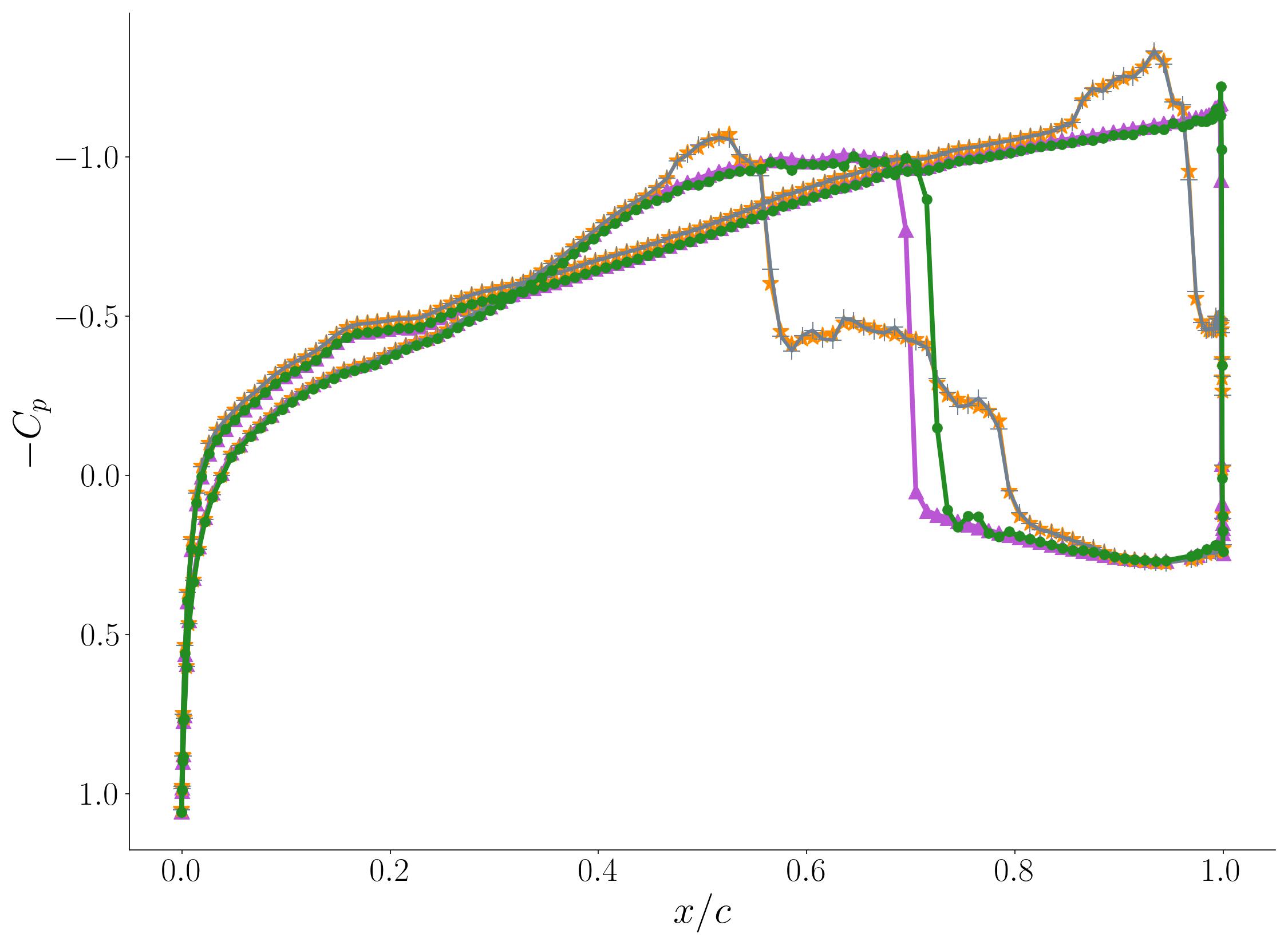}
            \caption{$M = 0.874, \alpha = 0.247$}
        \end{subfigure}\\
        \begin{subfigure}{.5\textwidth}
        \centering
            \includegraphics[width=1\linewidth]{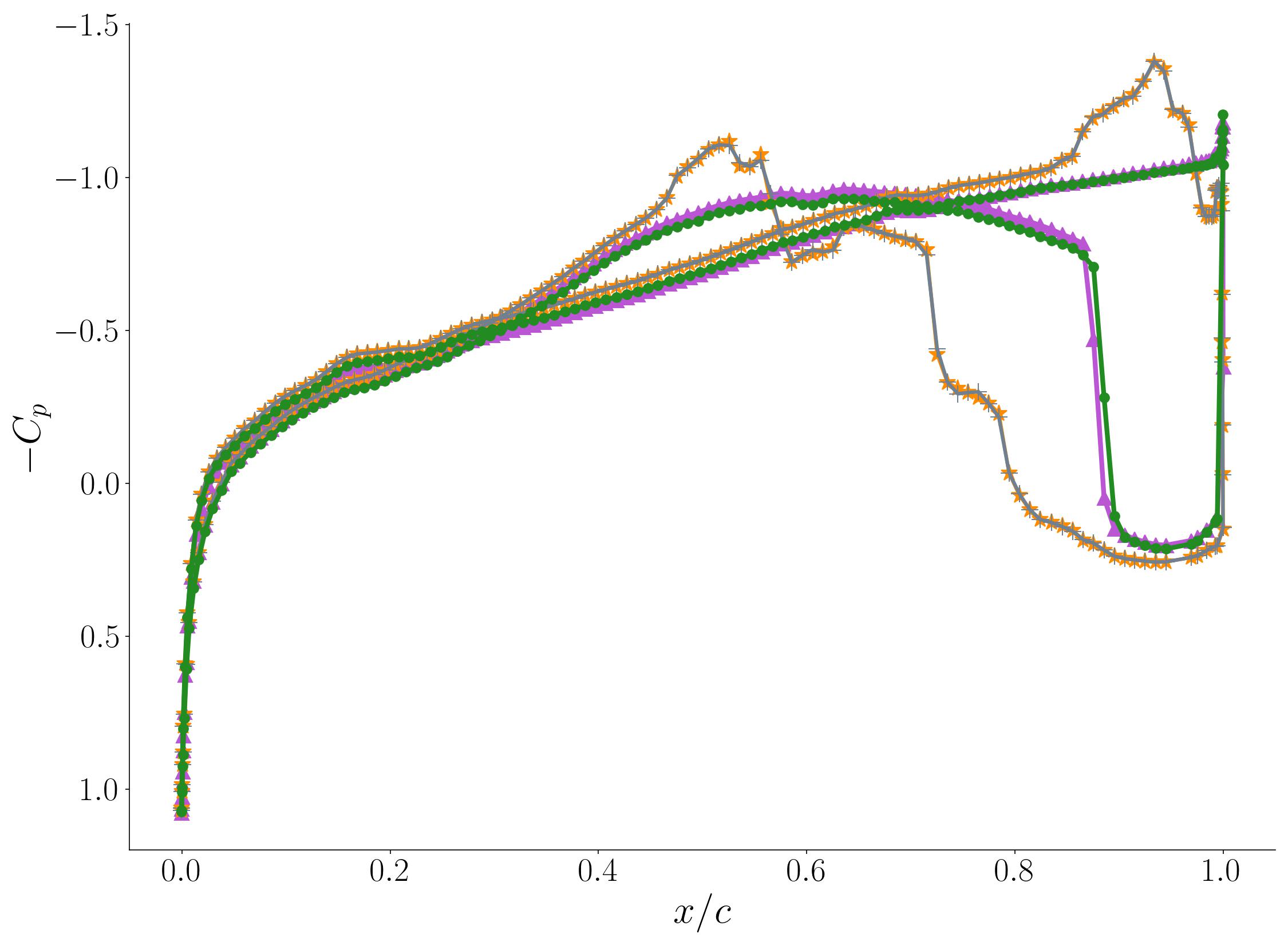}
            \caption{$M = 0.899, \alpha = 0.29$}
        \end{subfigure}%
        \begin{subfigure}{.5\textwidth}
        \centering
            \includegraphics[width=1\linewidth]{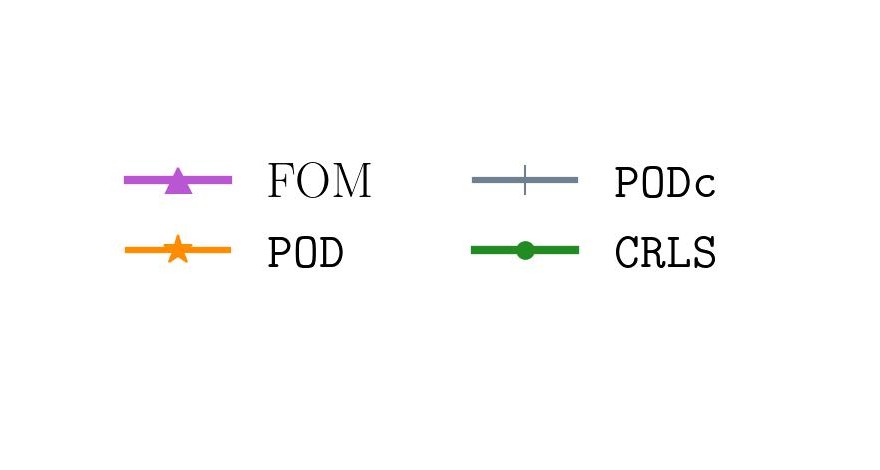}
            
        \end{subfigure}
        \caption{Comparison of pressure coefficient $(C_p)$ distribution on the airfoil surface.}
        \label{fig:airfoil_cp}
    \end{figure}

	\Cref{fig:airfoil_cp} shows the predicted pressure coefficients ($C_p$)  obtained from each of the three reduced basis approaches and the true $C_p$ obtained from solving the FOM. We retain modes that account for roughly $99.9$\% of the total energy.
    From these plots, it is quite evident that conventional approaches (without smoothing) lead to stair-stepping in the reconstructed solution, which is unsurprising. However, without regularization, even the \texttt{PODc} approach can lead to stair-stepped predictions, as illustrated in these figures; this is attributed to the ill-conditioning of the smoothing matrix $B$. On the other hand, the proposed \texttt{CRLS} method predicts physically consistent distributions that are very close to the FOM solution. Overall, the \texttt{CRLS} predictions are uniformly accurate across all the $5$ validation parameters.	
    
			

	\subsubsection{Full pressure field predictions} 	
    \begin{figure}[htb!]
        \centering
        \begin{subfigure}{.33\textwidth}
        \centering
        \includegraphics[width=1\linewidth]{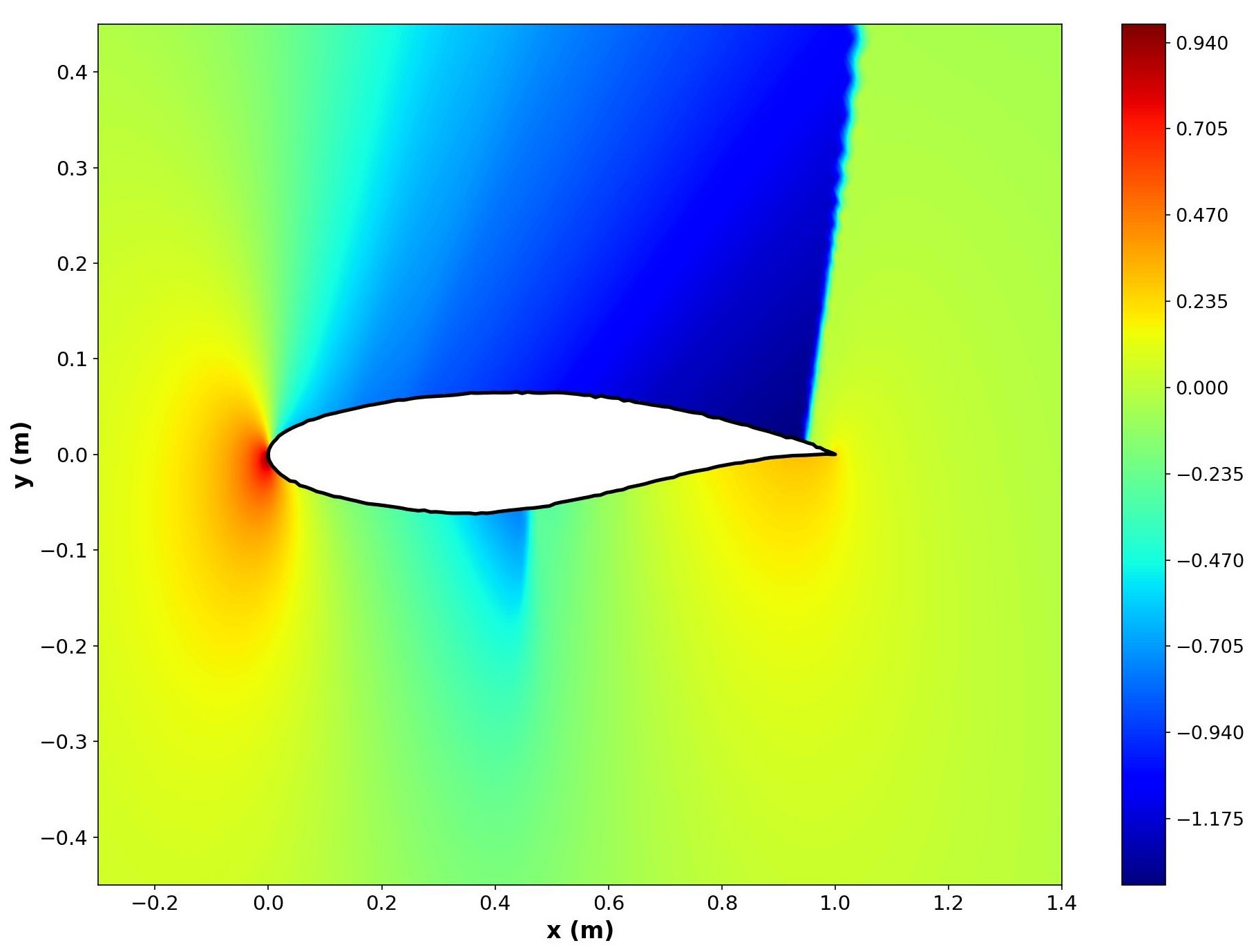}
        \end{subfigure}%
        \begin{subfigure}{.33\textwidth}  
        \centering
        \includegraphics[width=1\linewidth]{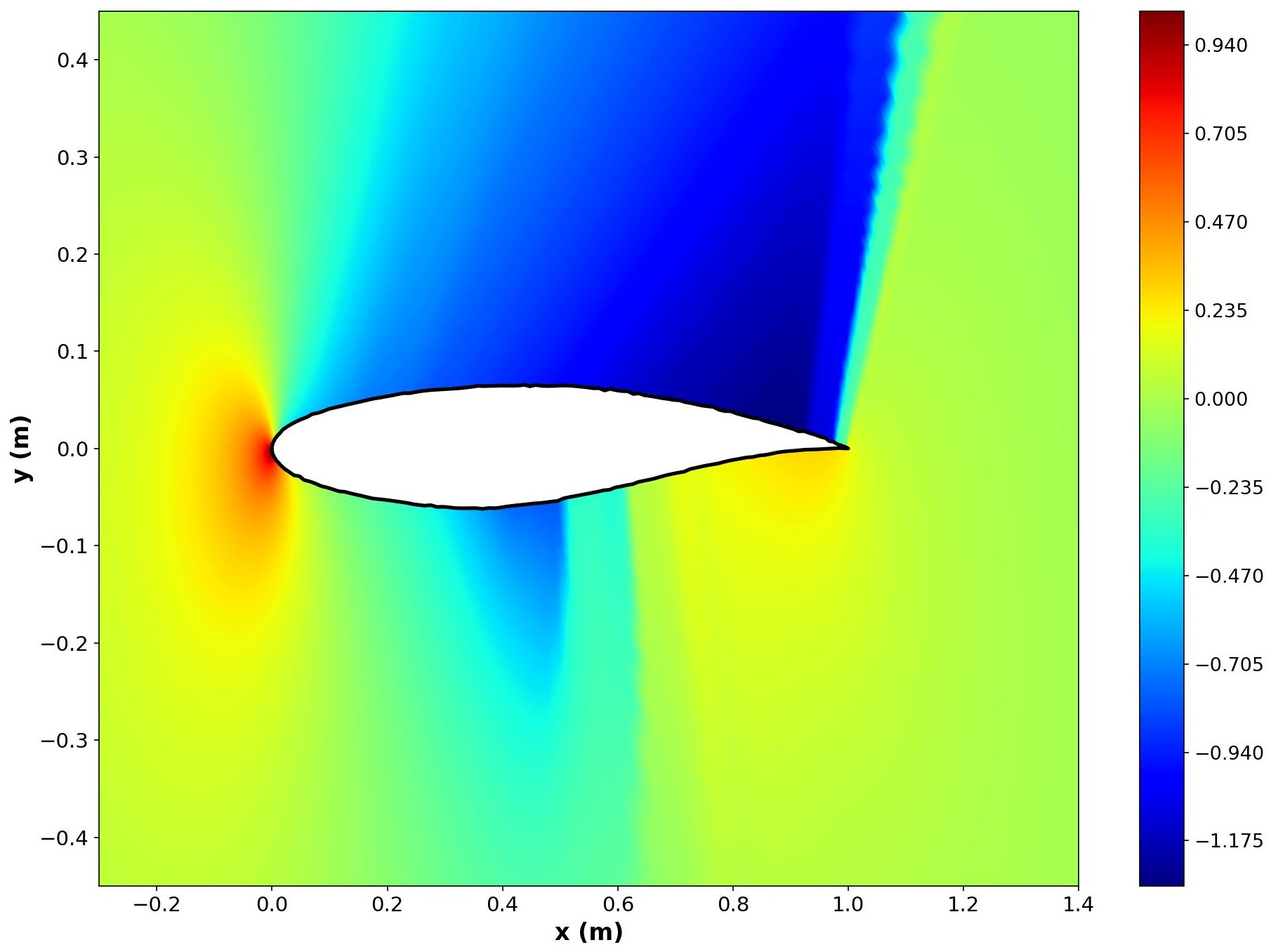}
        \end{subfigure}%
        \begin{subfigure}{.33\textwidth}
        \centering
        \includegraphics[width=1\linewidth]{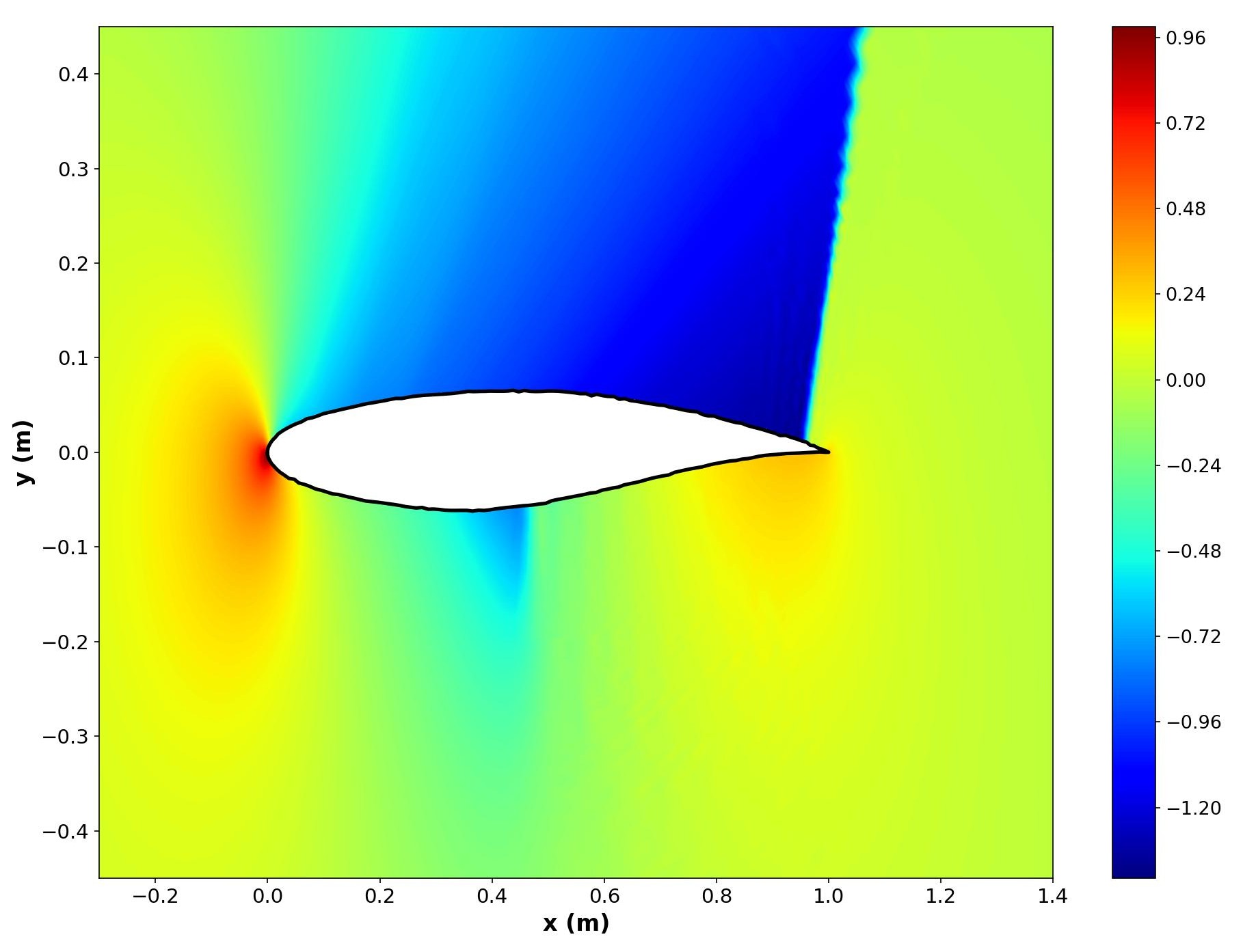}
        \end{subfigure} \\
        \begin{subfigure}{.33\textwidth}
        \centering
        \includegraphics[width=1\linewidth]{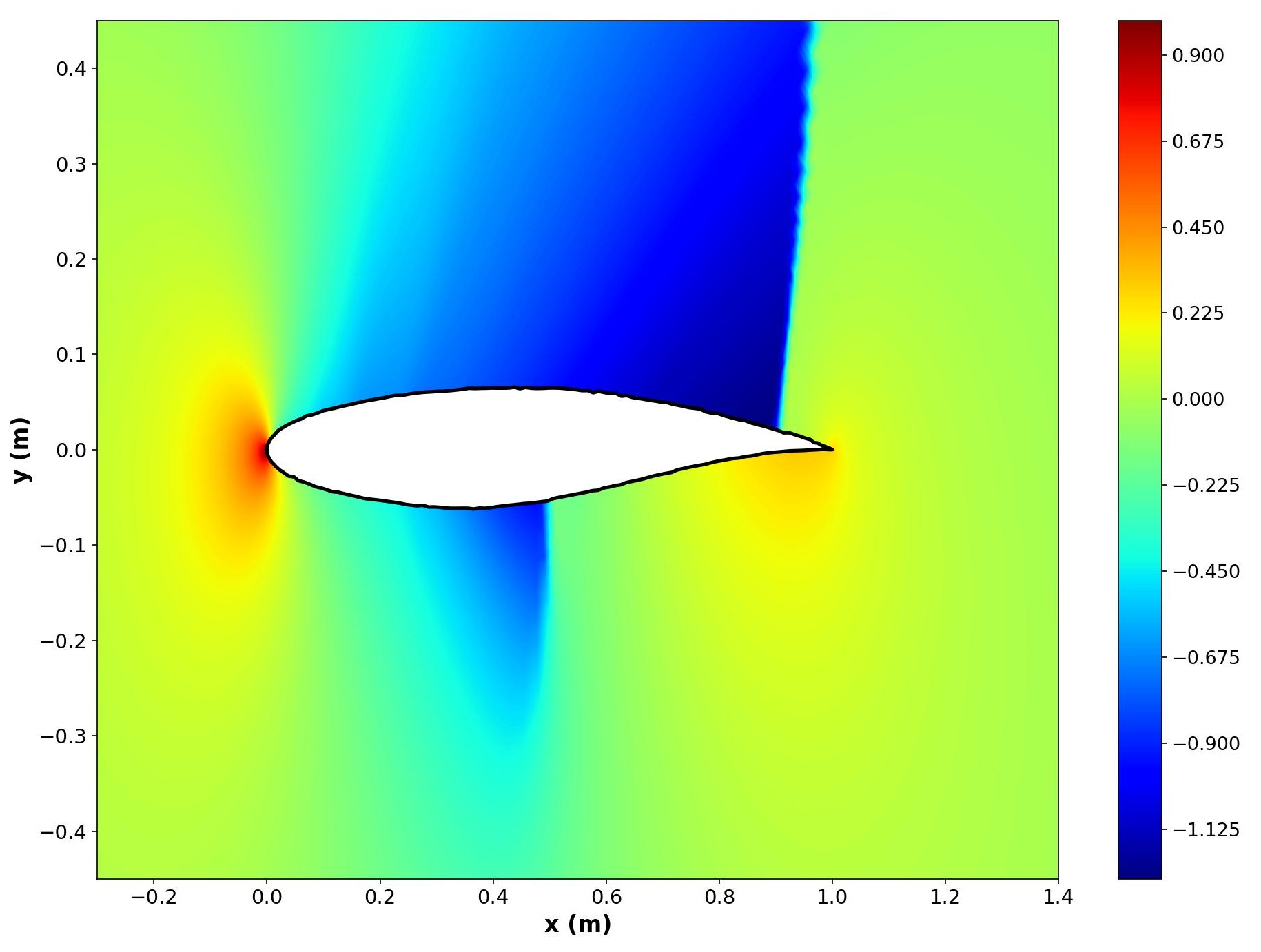}
        \end{subfigure}%
        \begin{subfigure}{.33\textwidth}  
        \centering
        \includegraphics[width=1\linewidth]{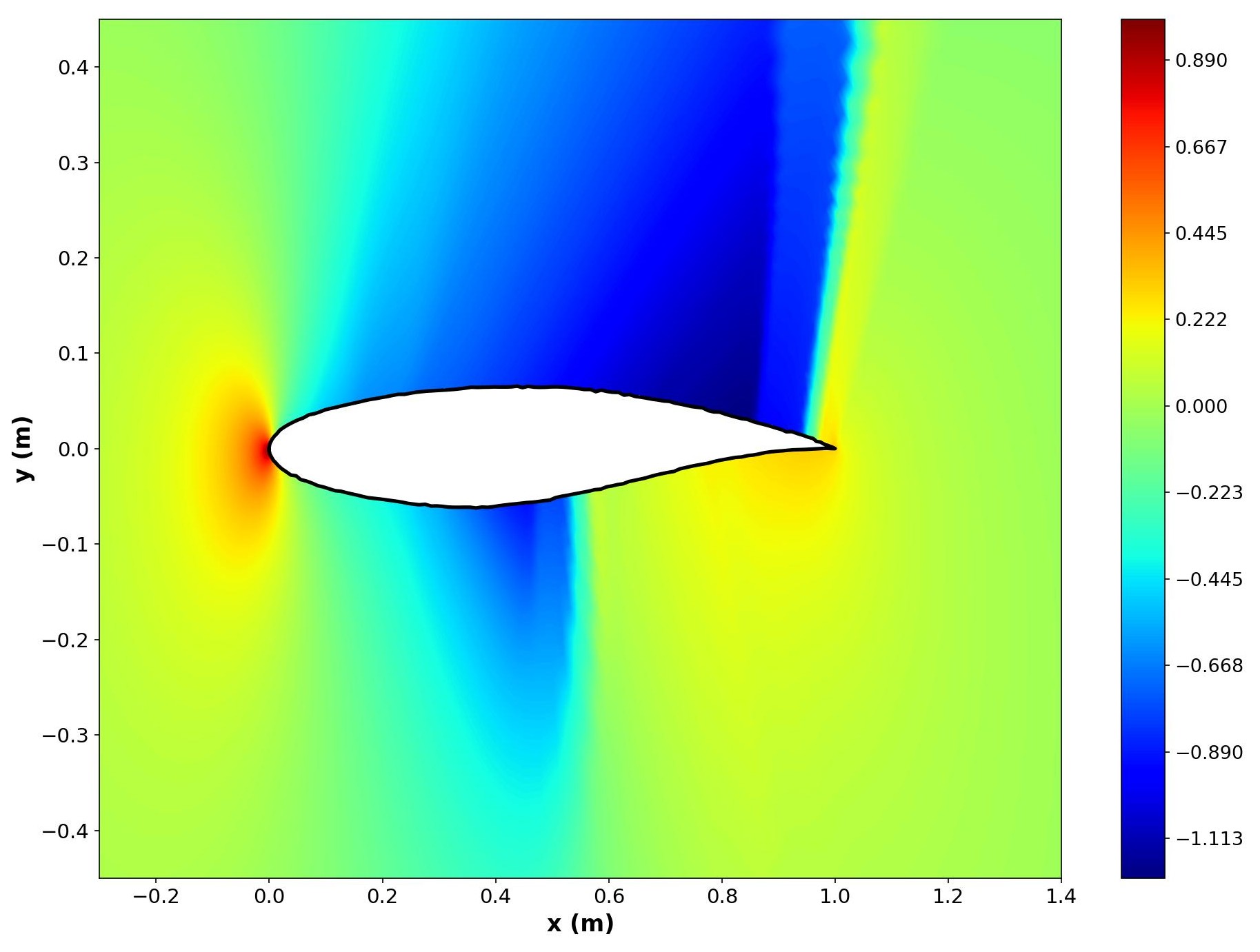}
        \end{subfigure}%
        \begin{subfigure}{.33\textwidth}
        \centering
        \includegraphics[width=1\linewidth]{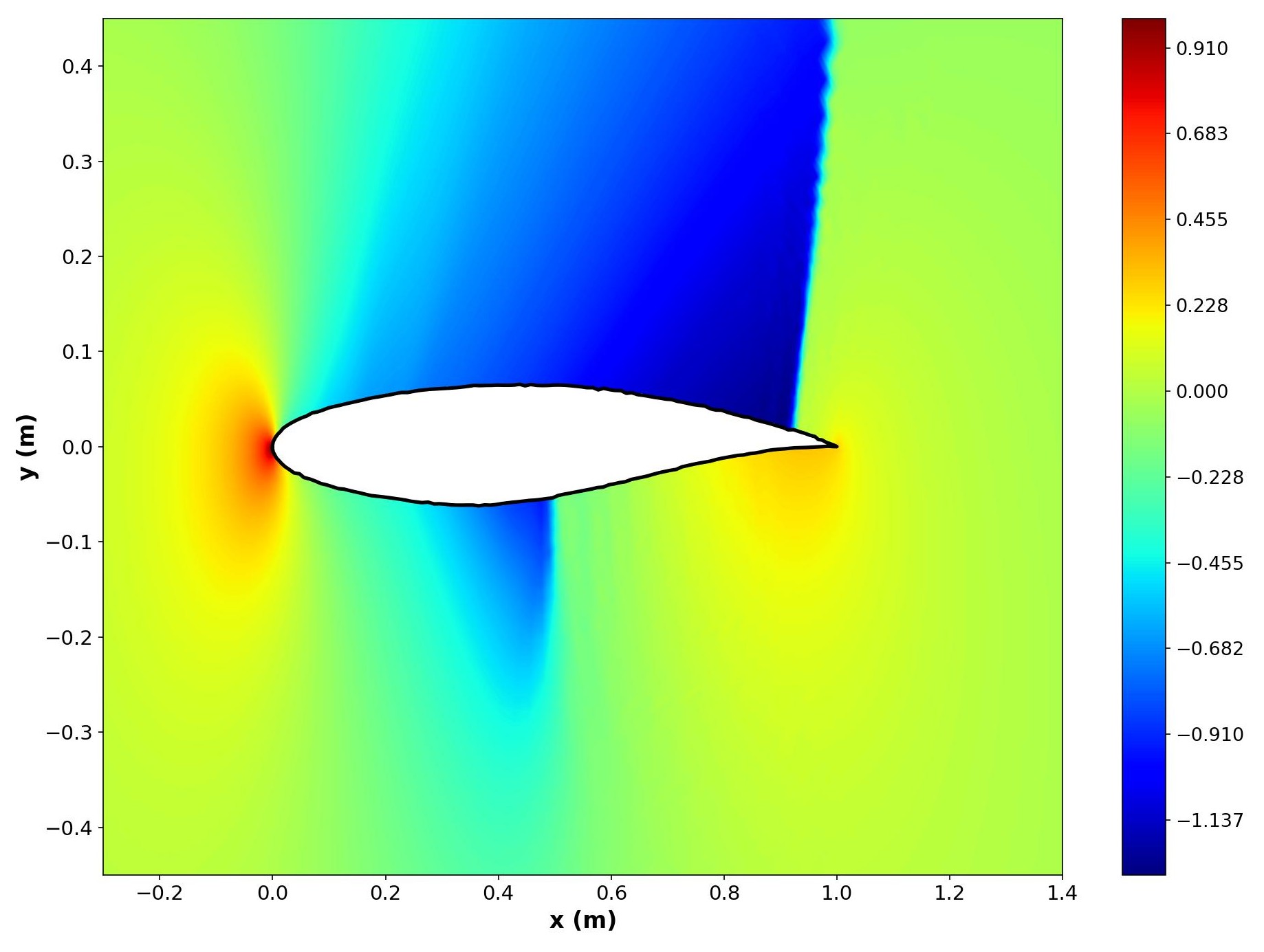}
        \end{subfigure} \\
        \begin{subfigure}{.33\textwidth}
        \centering
        \includegraphics[width=1\linewidth]{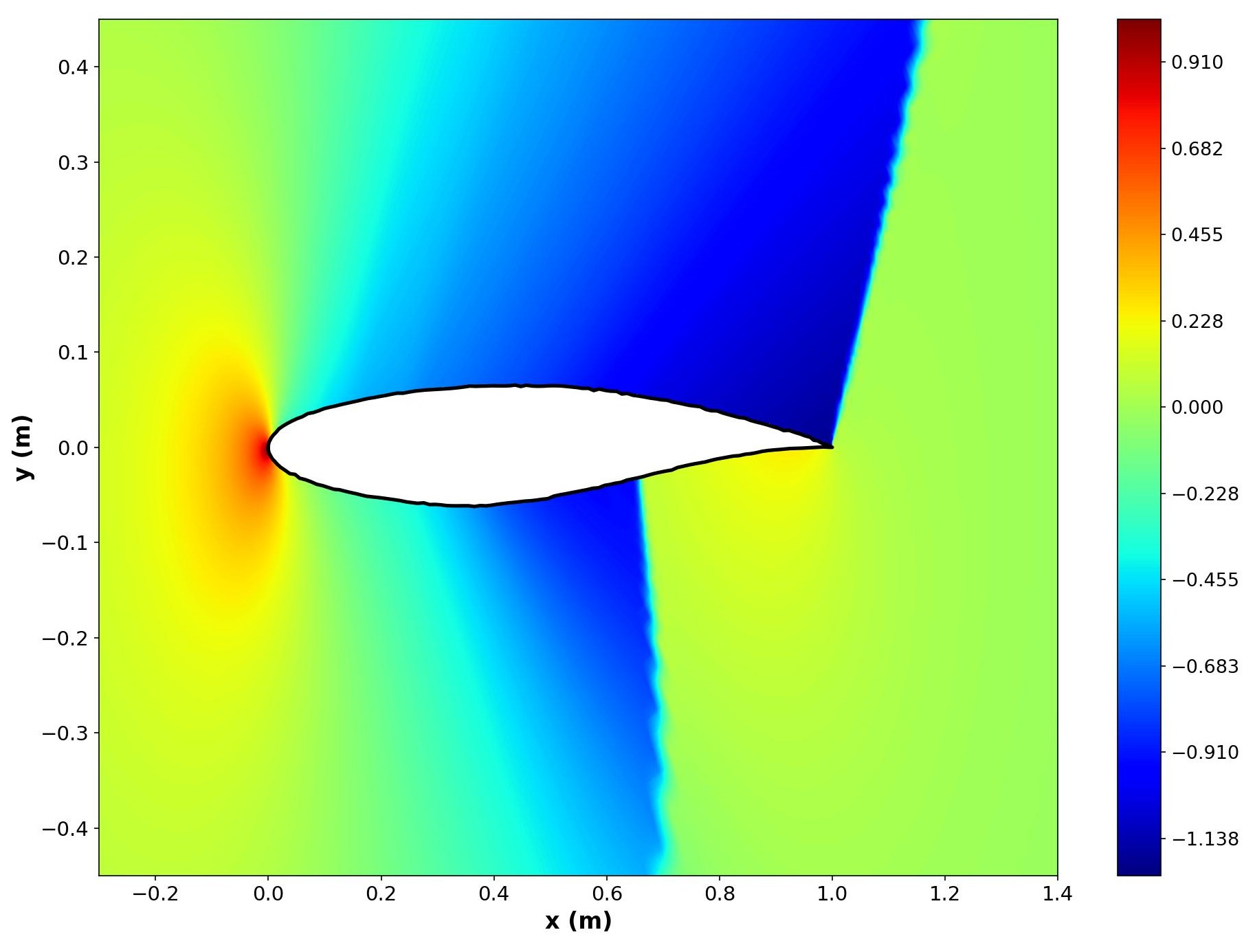}
        \end{subfigure}%
        \begin{subfigure}{.33\textwidth}  
        \centering
        \includegraphics[width=1\linewidth]{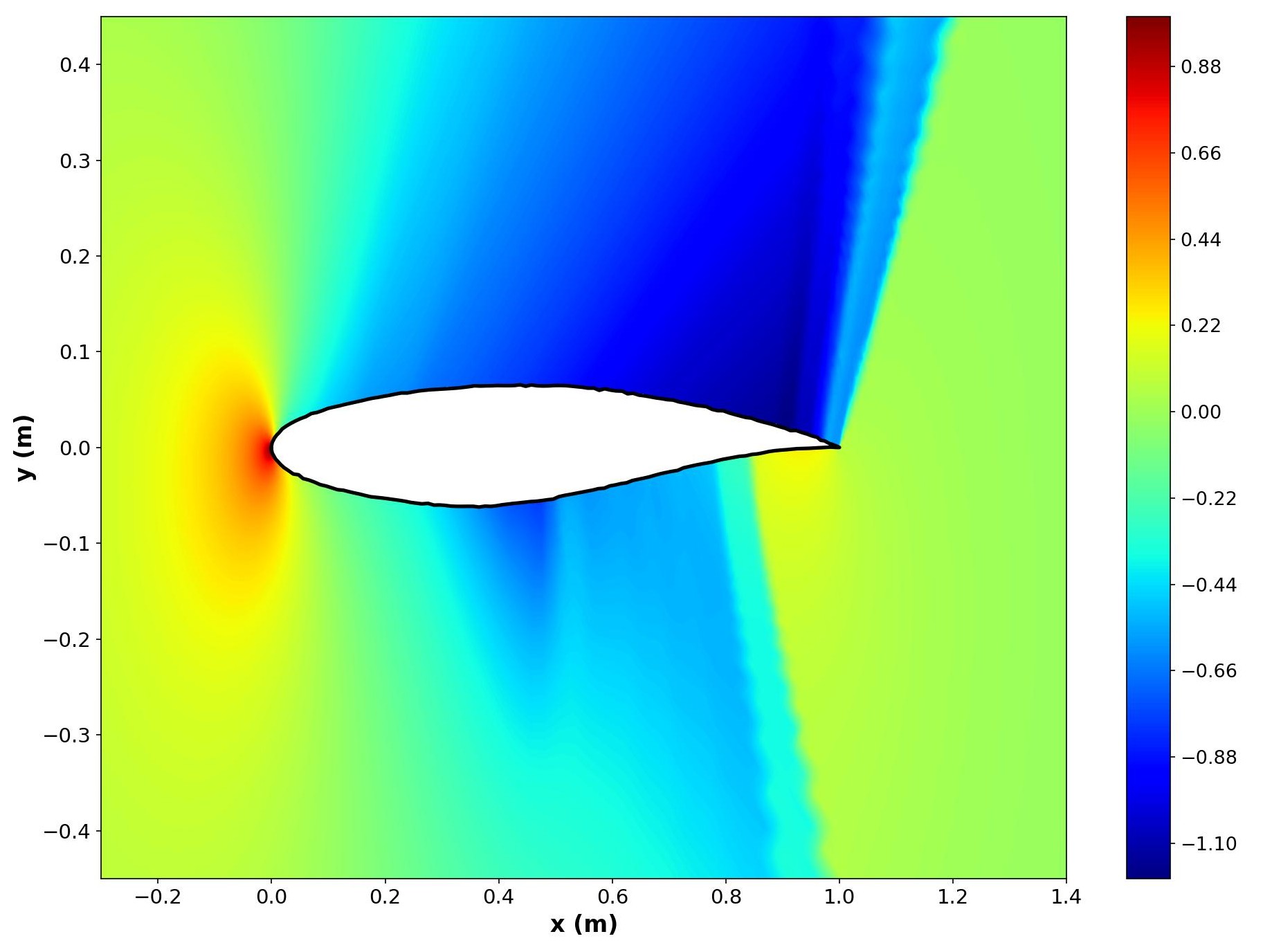}
        \end{subfigure}%
        \begin{subfigure}{.33\textwidth}
        \centering
        \includegraphics[width=1\linewidth]{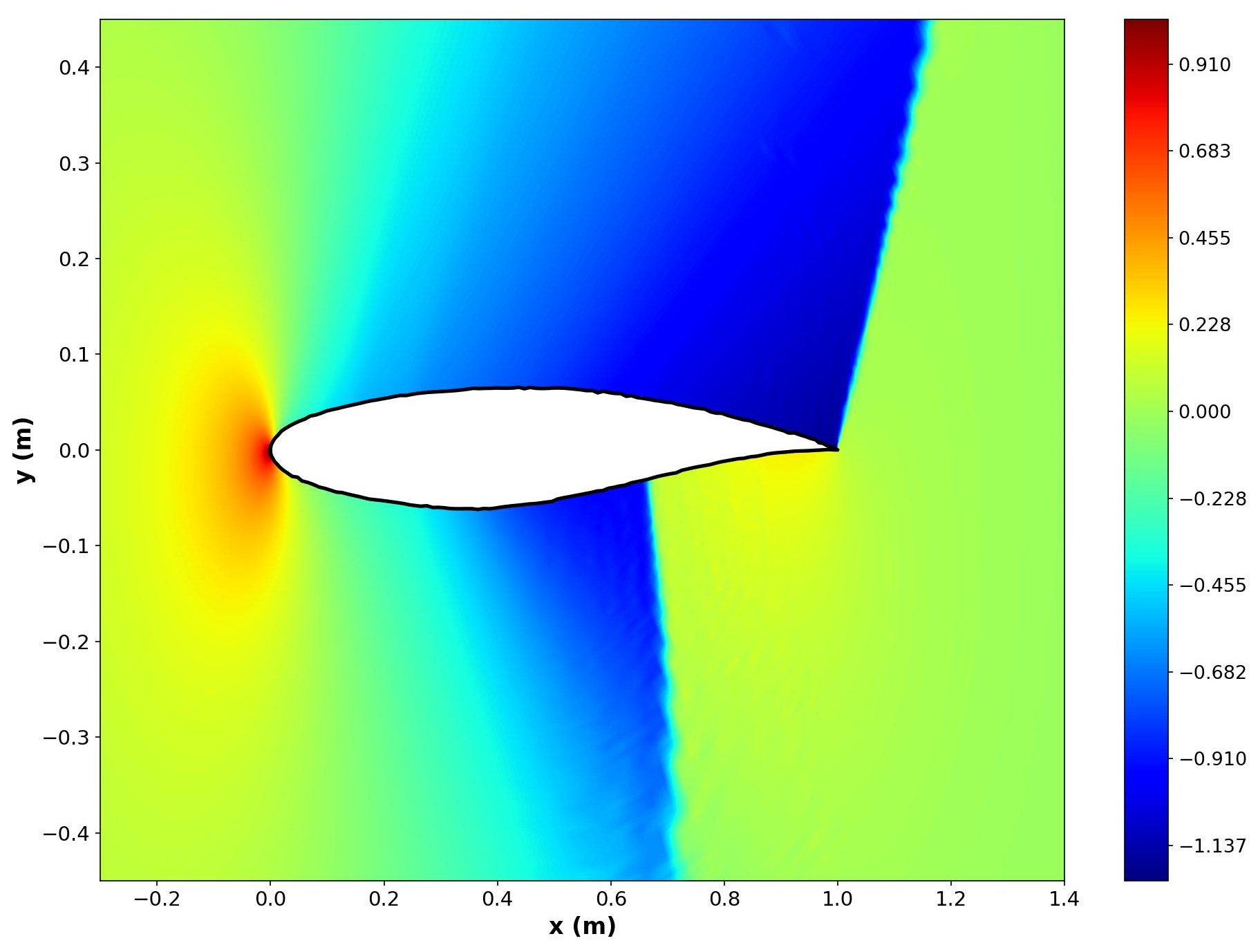}
        \end{subfigure} \\
        \begin{subfigure}{.33\textwidth}
        \centering
        \includegraphics[width=1\linewidth]{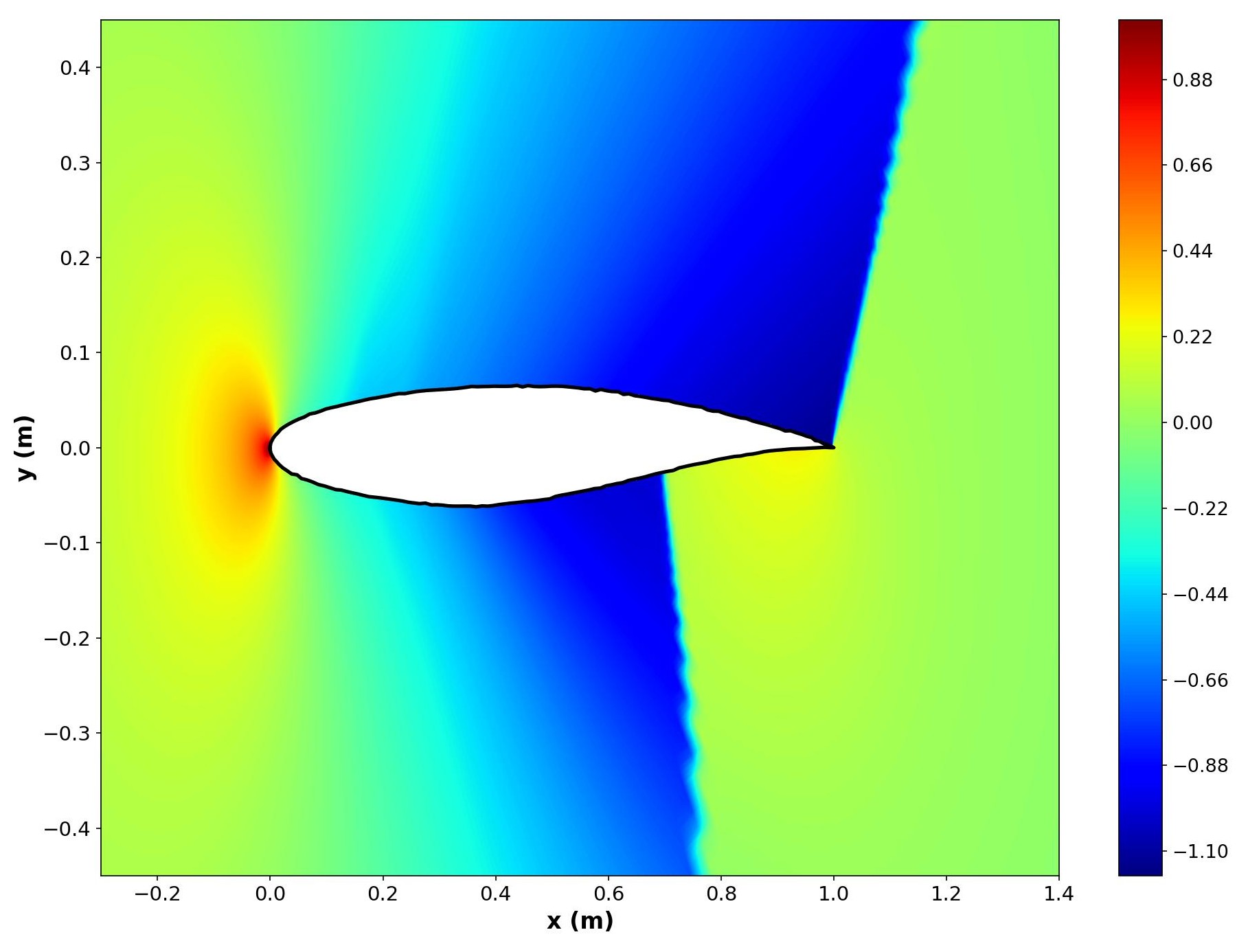}
        \end{subfigure}%
        \begin{subfigure}{.33\textwidth}  
        \centering
        \includegraphics[width=1\linewidth]{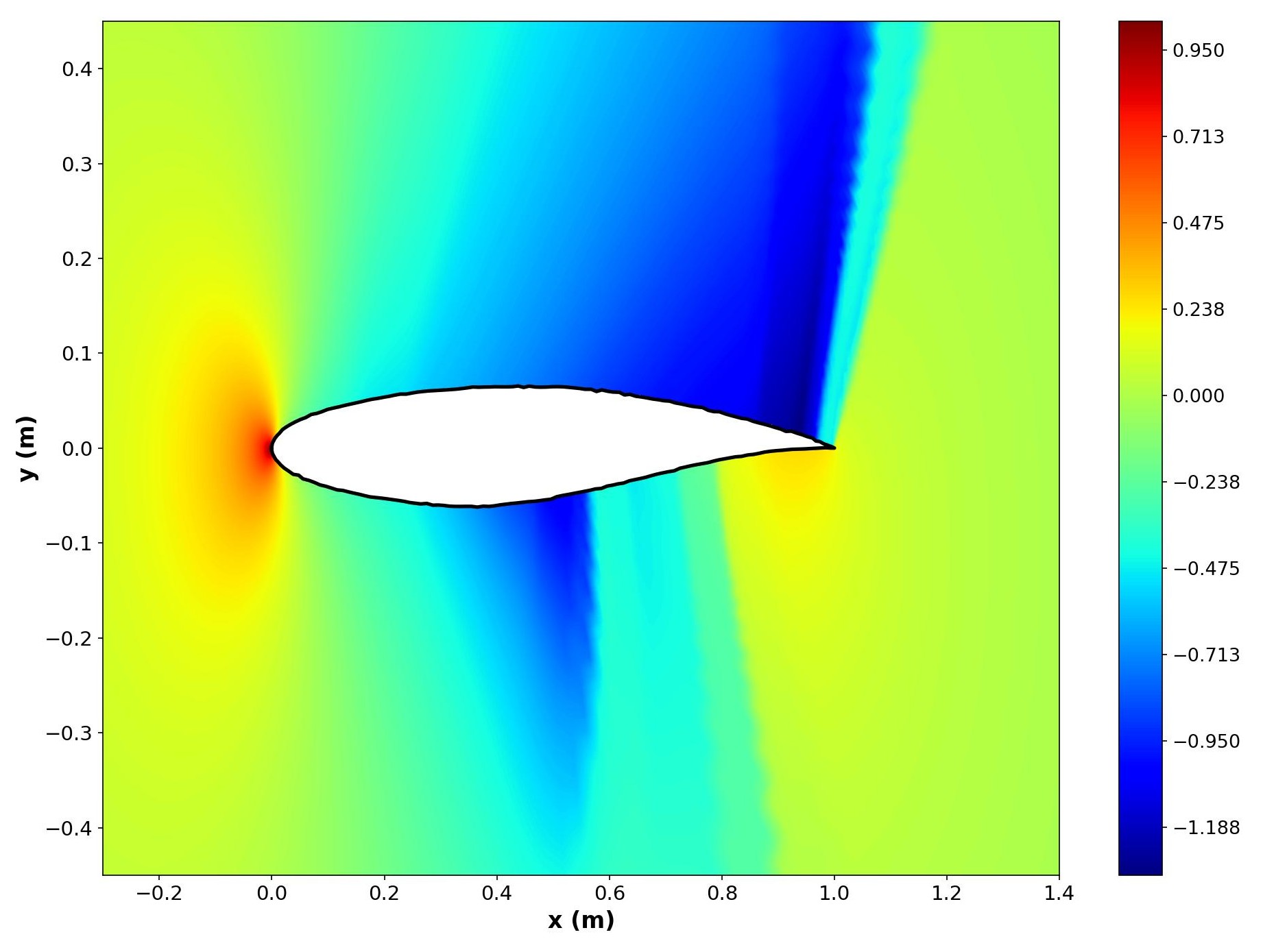}
        \end{subfigure}%
        \begin{subfigure}{.33\textwidth}
        \centering
        \includegraphics[width=1\linewidth]{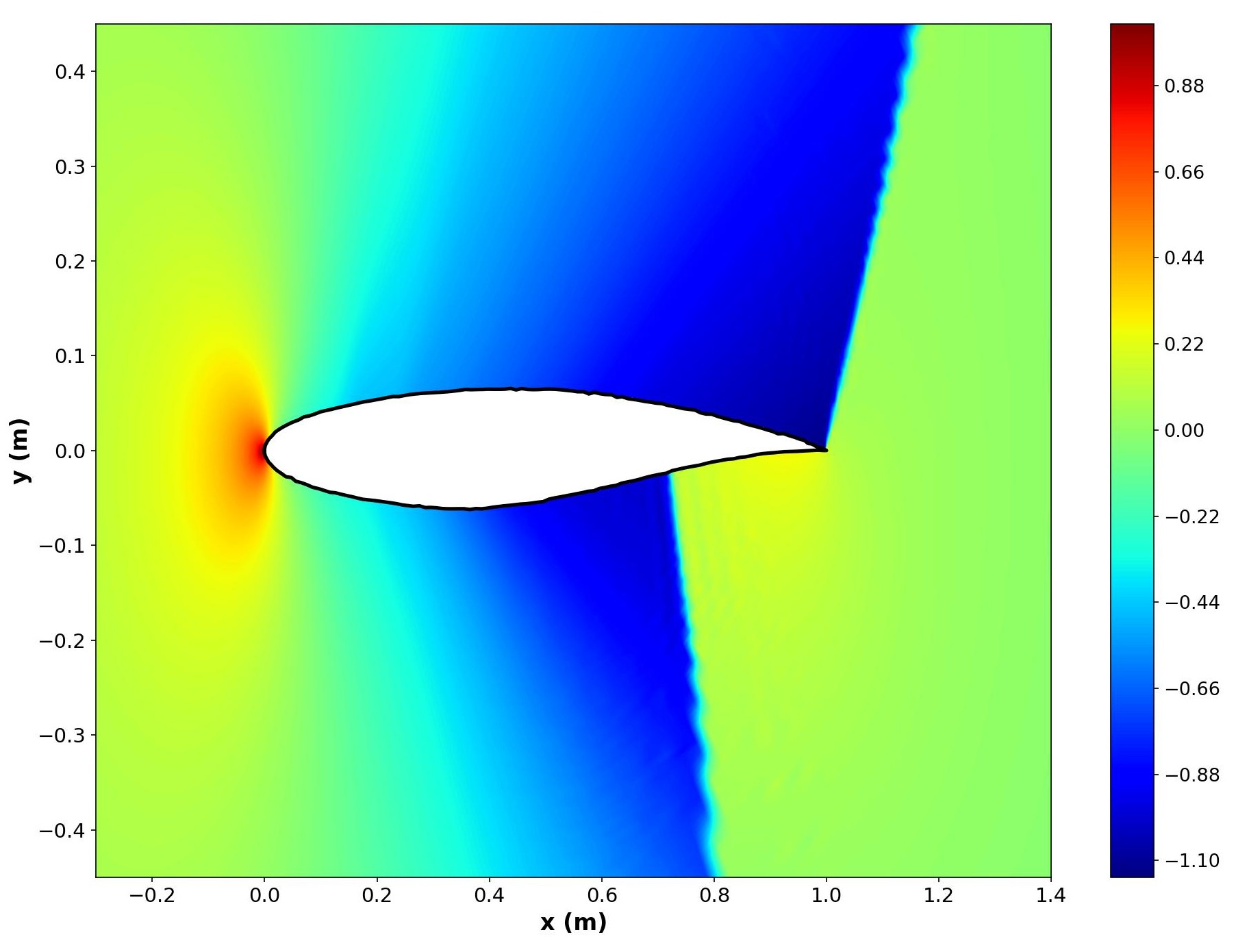}
        \end{subfigure} \\
        \begin{subfigure}{.33\textwidth}
        \centering
        \includegraphics[width=1\linewidth]{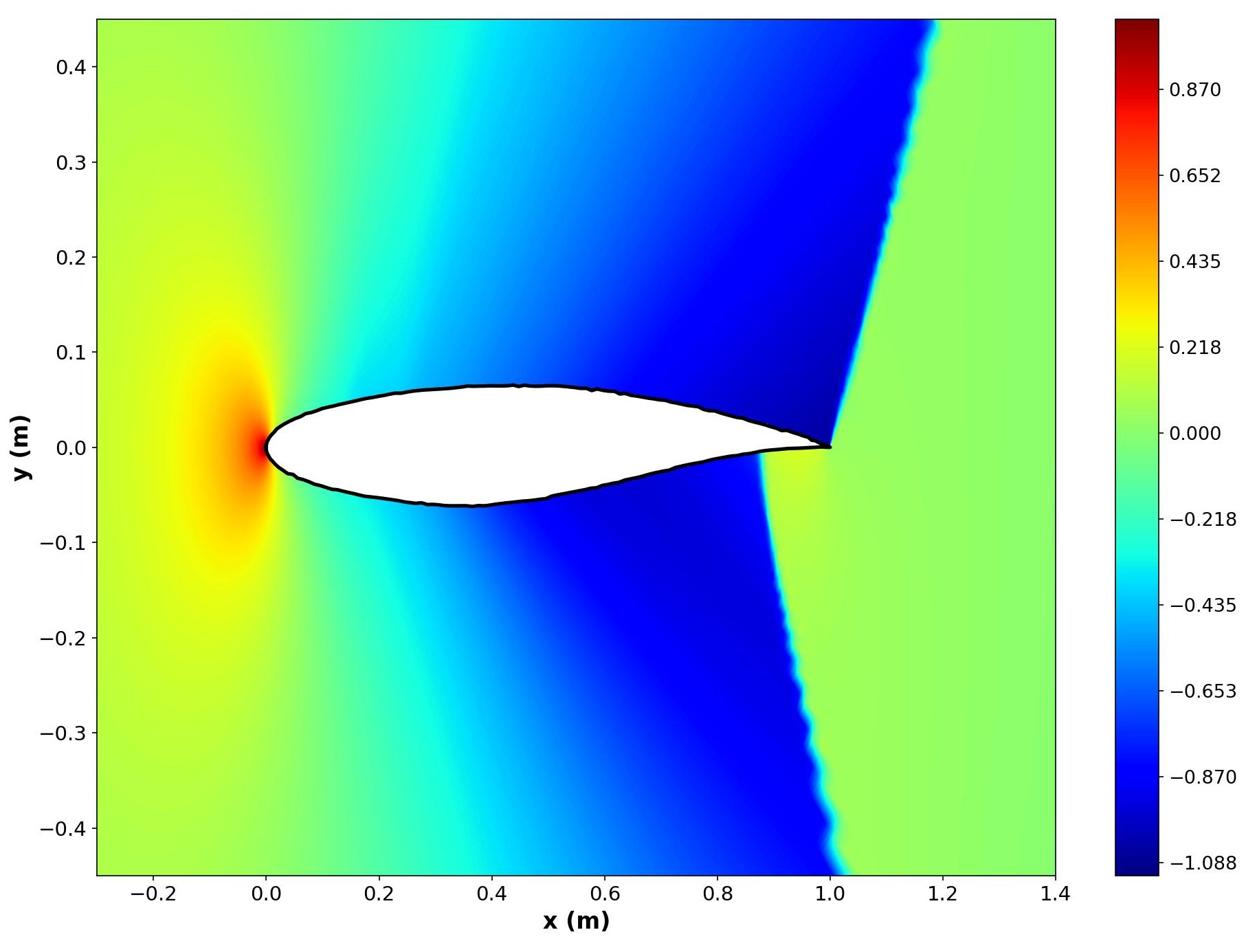}
        \caption{\texttt{FOM}}
        \end{subfigure}%
        \begin{subfigure}{.33\textwidth}  
        \centering
        \includegraphics[width=1\linewidth]{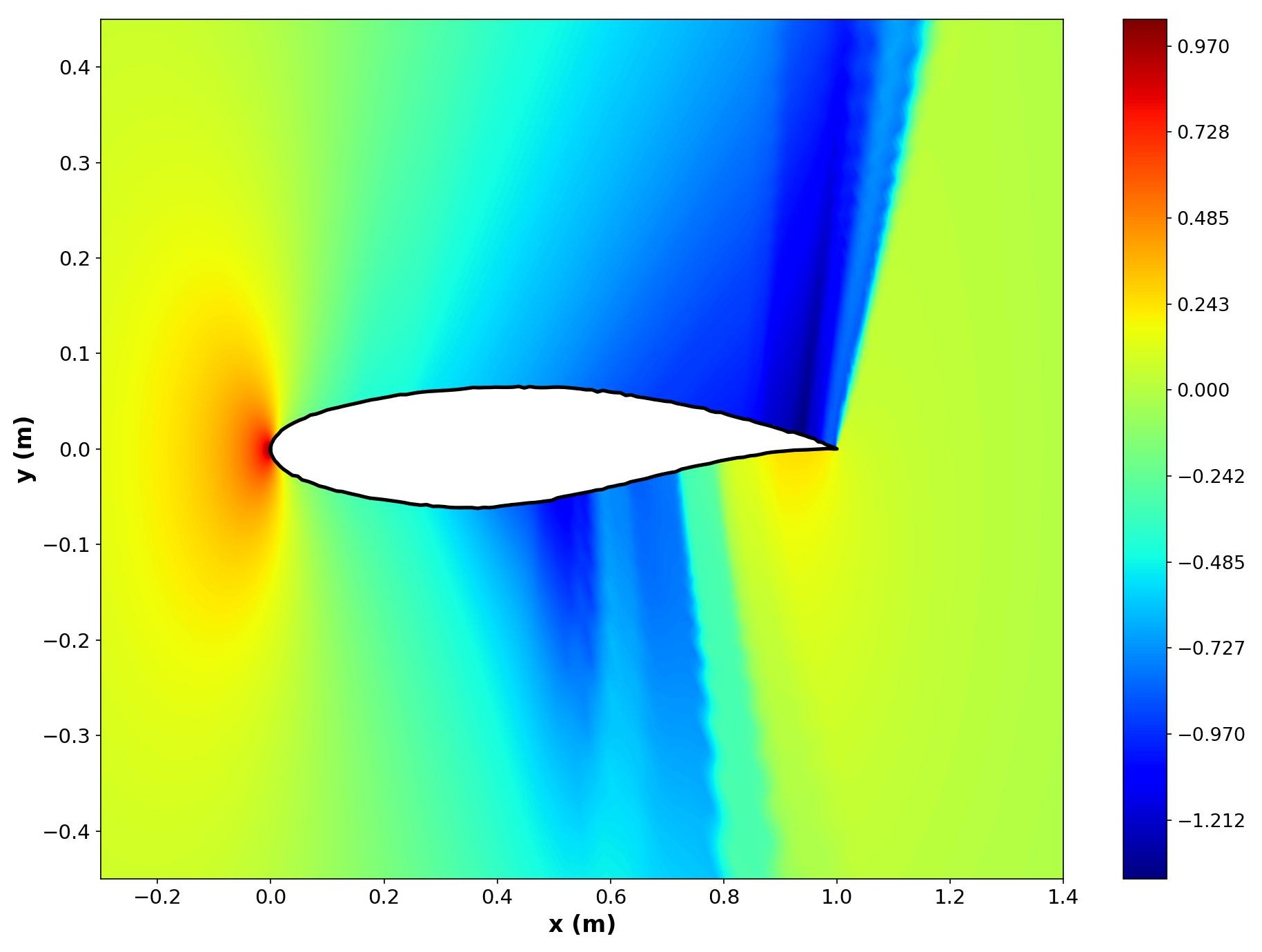}
        \caption{(Naive) \texttt{POD} }
        \end{subfigure}%
        \begin{subfigure}{.33\textwidth}
        \centering
        \includegraphics[width=1\linewidth]{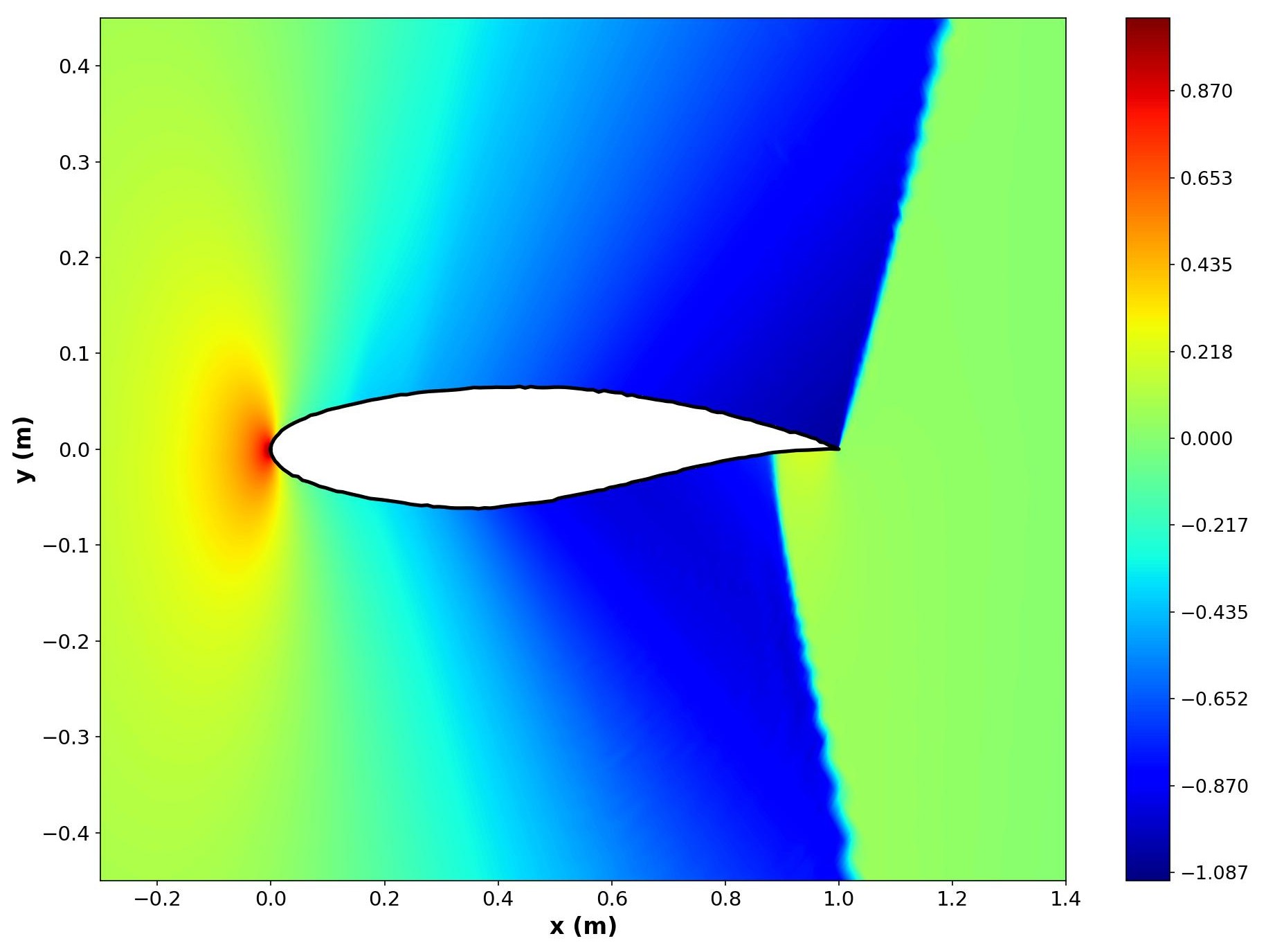}
        \caption{\texttt{CRLS} (ours)}
        \end{subfigure} 
        \label{fig:cp_contours}
        \caption{Comparison of pressure coefficient $(C_p)$ contours. Each row corresponds to parameter values in panels (a) through (e) in \Cref{fig:airfoil_cp}.}
    \end{figure}
    \Cref{fig:cp_contours} shows the predictions of the full pressure coefficient field in the flow domain for the same set of parameters in \Cref{fig:airfoil_cp}; note that we only show a close-up view of the airfoil in the figures.
    Here we include only the \texttt{PODc} in the comparison. Notice that \texttt{PODc} results in the non-physical ``double shocks'' on the upper and lower surfaces, which is consistent with the stair-stepping observed in the $C_p$ plots. On the other hand, the \texttt{CRLS} predicts physically consistent fields that are quite similar to the FOM solutions. We provide a visualization of the pointwise error fields next.
		
	
		
\begin{figure*}[!tbp]
		\centering
		\includegraphics[width=0.9\textwidth]{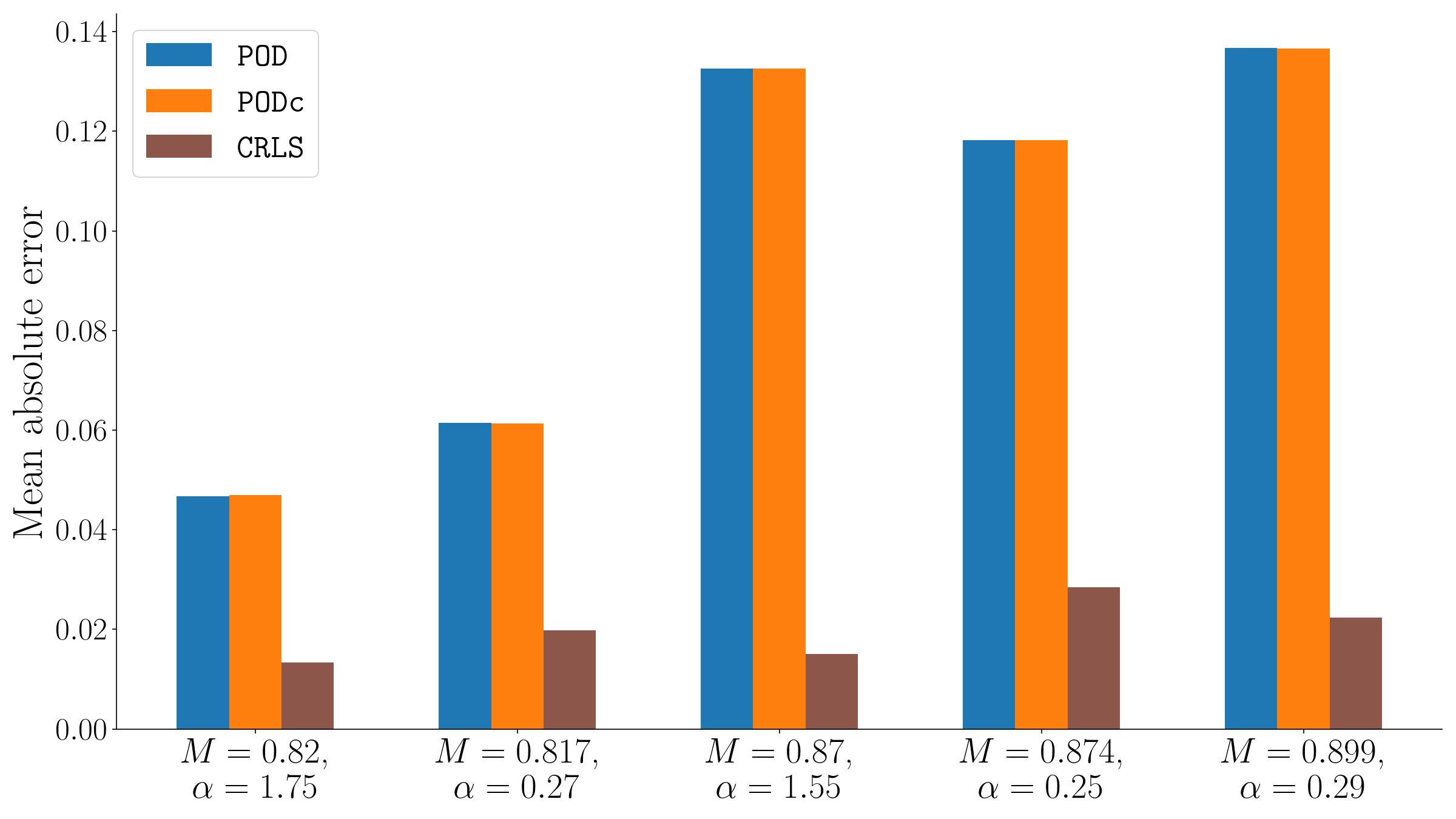}
		\captionsetup{justification=centering, labelfont=bf, font=small}
		\caption{Mean absolute error of $C_p$ on airfoil surface}
		\label{error_1d}
	\end{figure*}
\begin{figure*}[!tbp]
		\centering
		\includegraphics[width=0.9\textwidth]{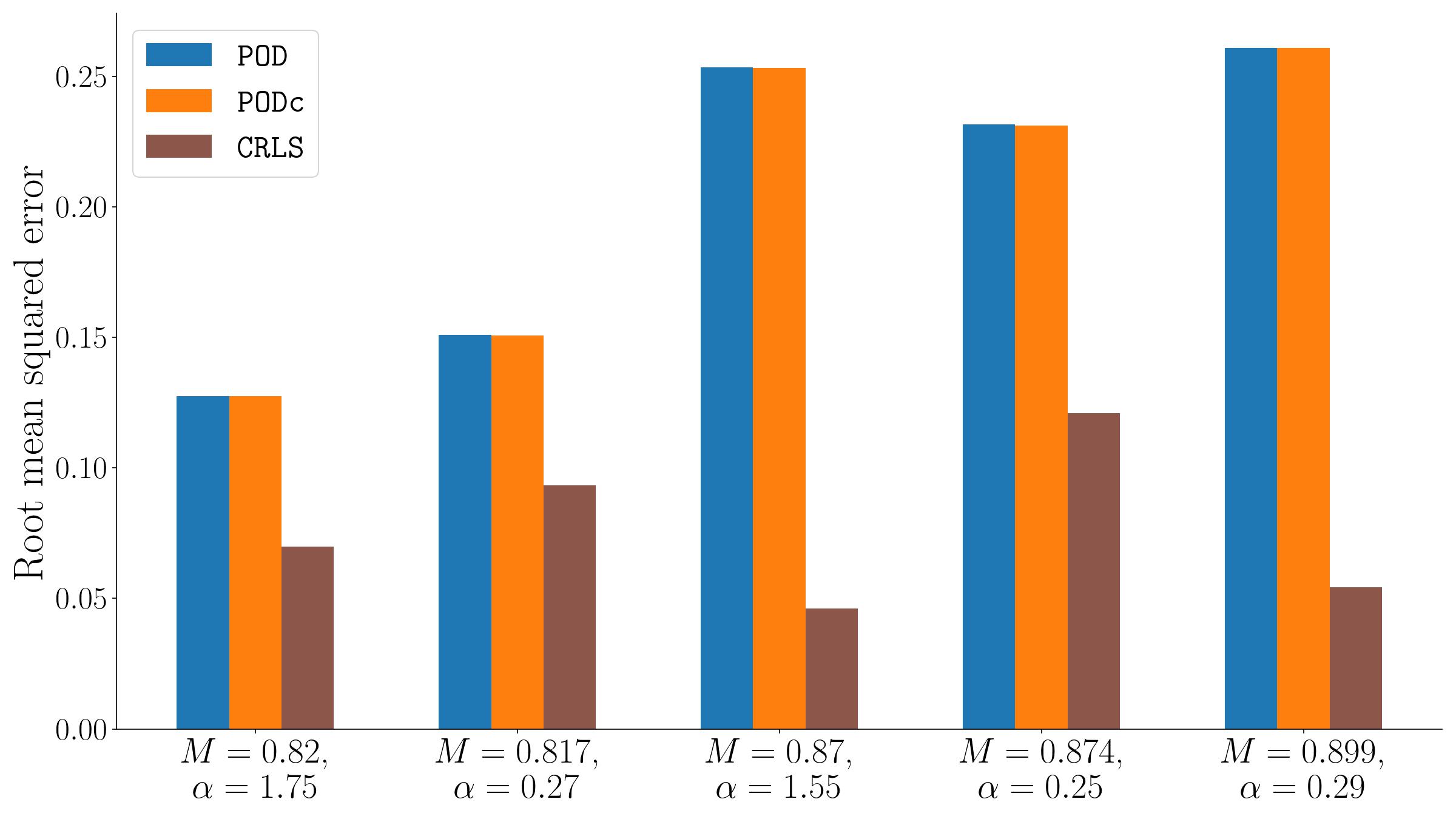}
		\captionsetup{justification=centering, labelfont=bf, font=small}
		\caption{Root mean squared error of $C_p$ on airfoil surface}
		\label{error_1d_2}
	\end{figure*}

    To quantify the prediction error, we begin by observing the mean absolute errors (MAE) and root mean squared error (RMSE) shown in figures \ref{error_1d} and \ref{error_1d_2}, respectively. For a set of $N$ spatial locations with prediction $\bs{u}$ and FOM
reference $\bs{u}^{\mathrm{FOM}}$, these are written as
\begin{equation}
	\mathrm{MAE}
	= \frac{1}{N} \|\bs{u} - \bs{u}^{\mathrm{FOM}} \|_1,
	\qquad
	\mathrm{RMSE}
	= \sqrt{ \frac{1}{N}  \|\bs{u} - \bs{u}^{\mathrm{FOM}} \|_2^2 }.
	\label{eq:mae_rmse}
\end{equation}
Whereas MAE measures the average pointwise deviation, RMSE penalizes larger local errors more. Notice that both \texttt{POD} and \texttt{PODc} give almost the same error, with \texttt{PODc} being slightly lower. Unsurprisingly, the \texttt{CRLS} gives the lowest MAE and RMSE while surpassing the other two approaches by a substantial margin.
    \begin{figure}[htb!]
        \centering
        \begin{subfigure}{.33\textwidth}
        \centering
        \includegraphics[width=1\linewidth]{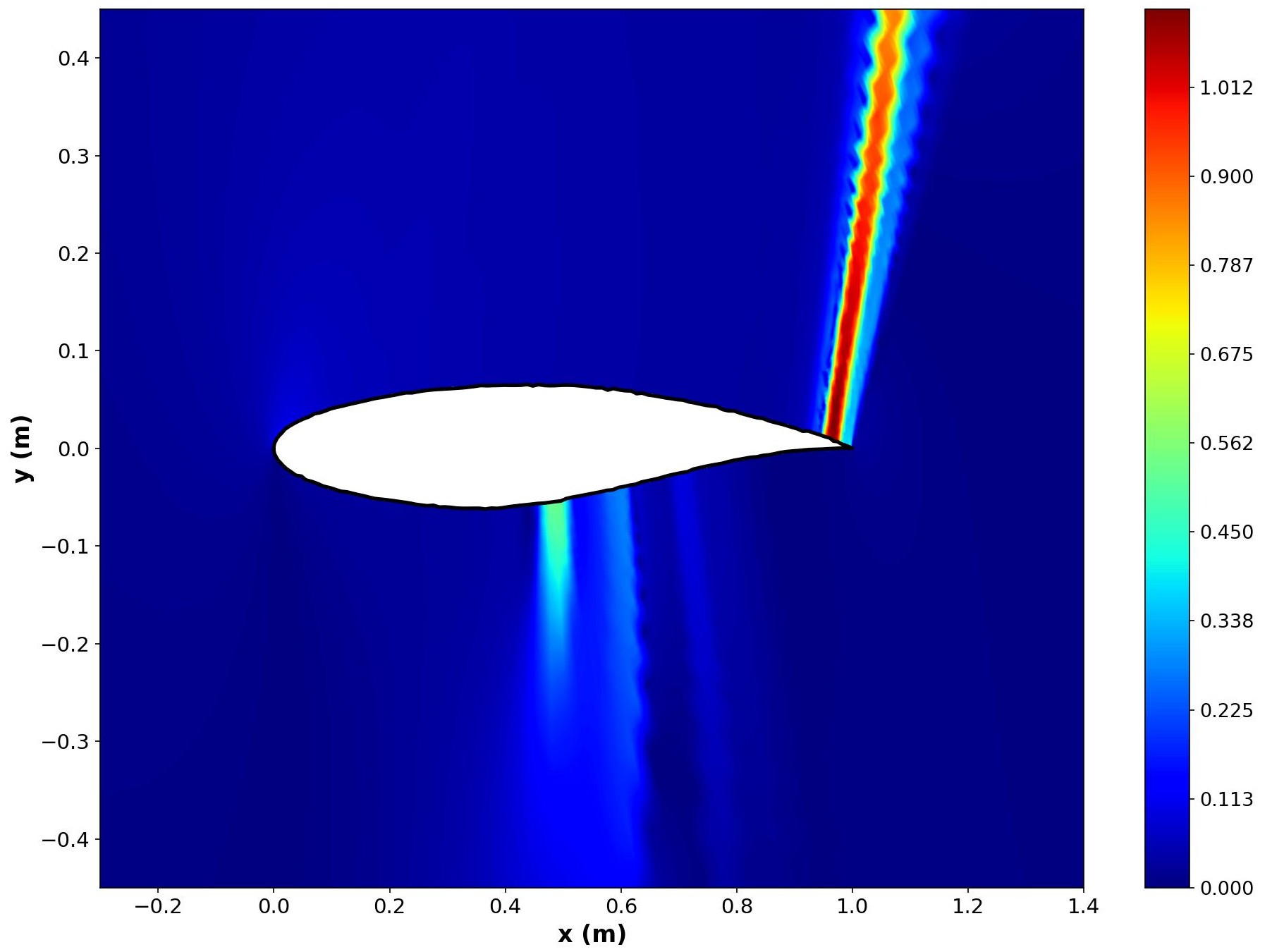}
        \end{subfigure}%
        \begin{subfigure}{.33\textwidth}  
        \centering
        \includegraphics[width=1\linewidth]{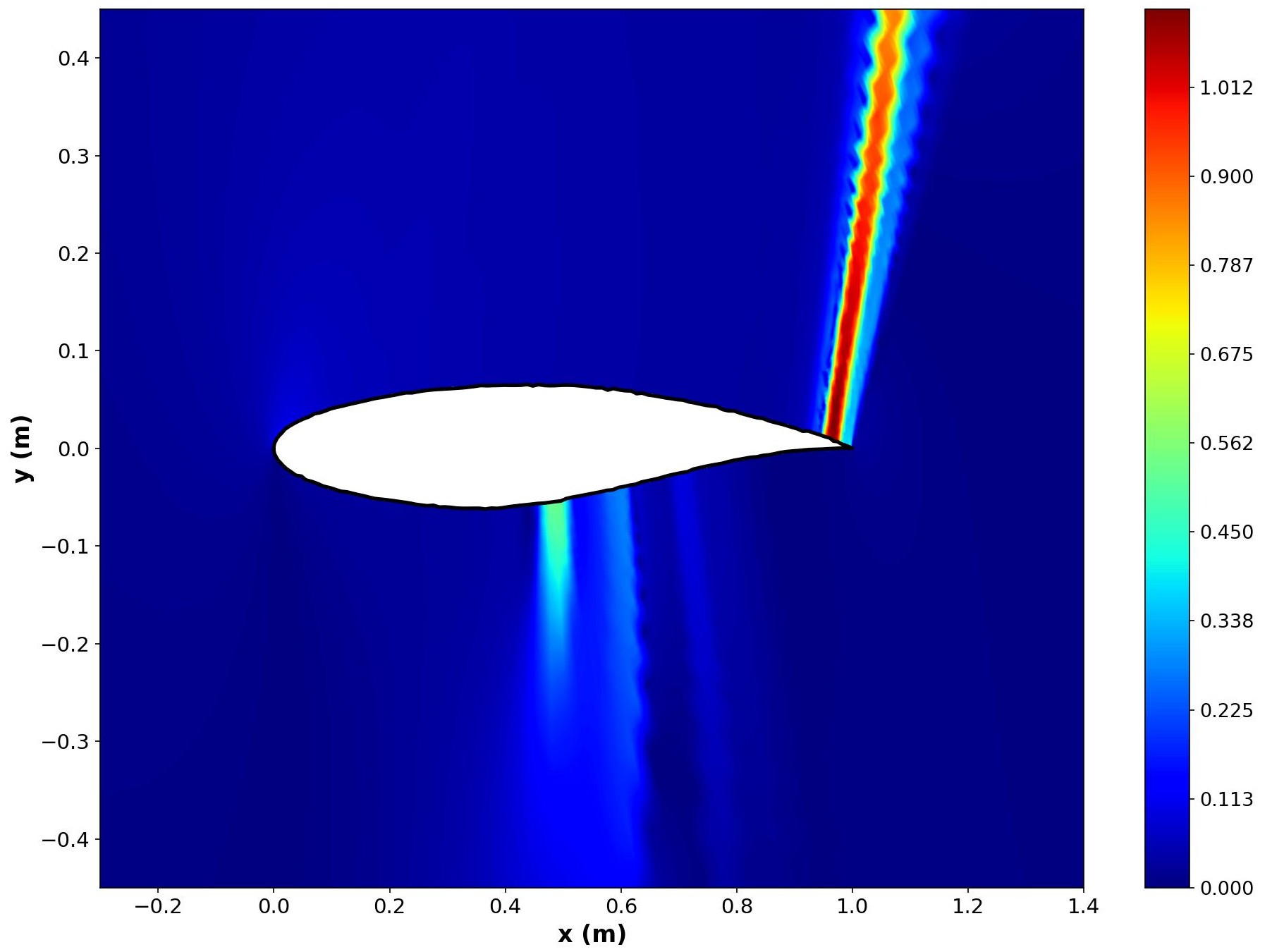}
        \end{subfigure}%
        \begin{subfigure}{.33\textwidth}
        \centering
        \includegraphics[width=1\linewidth]{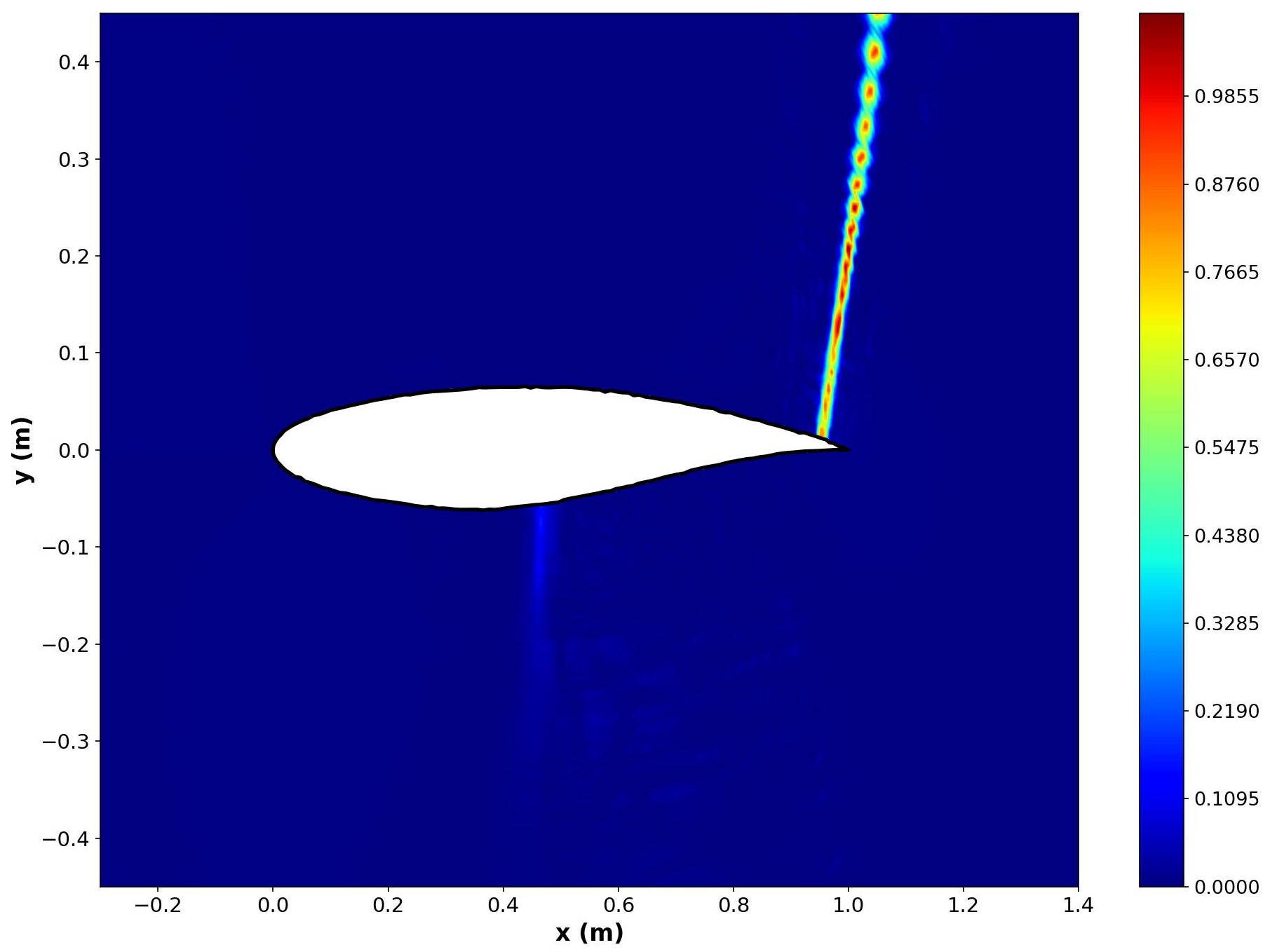}
        \end{subfigure} \\
        \begin{subfigure}{.33\textwidth}
        \centering
        \includegraphics[width=1\linewidth]{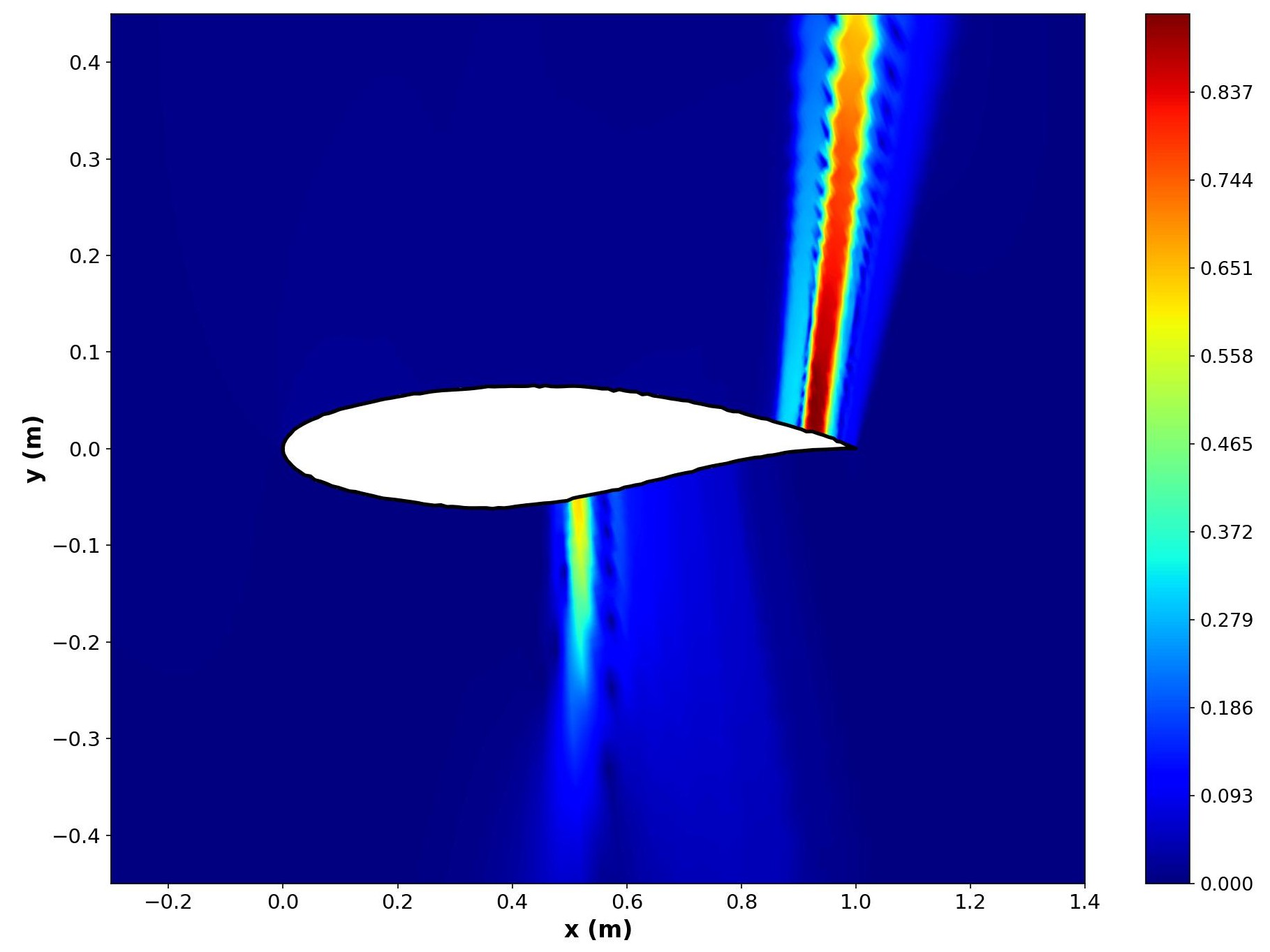}
        \end{subfigure}%
        \begin{subfigure}{.33\textwidth}  
        \centering
        \includegraphics[width=1\linewidth]{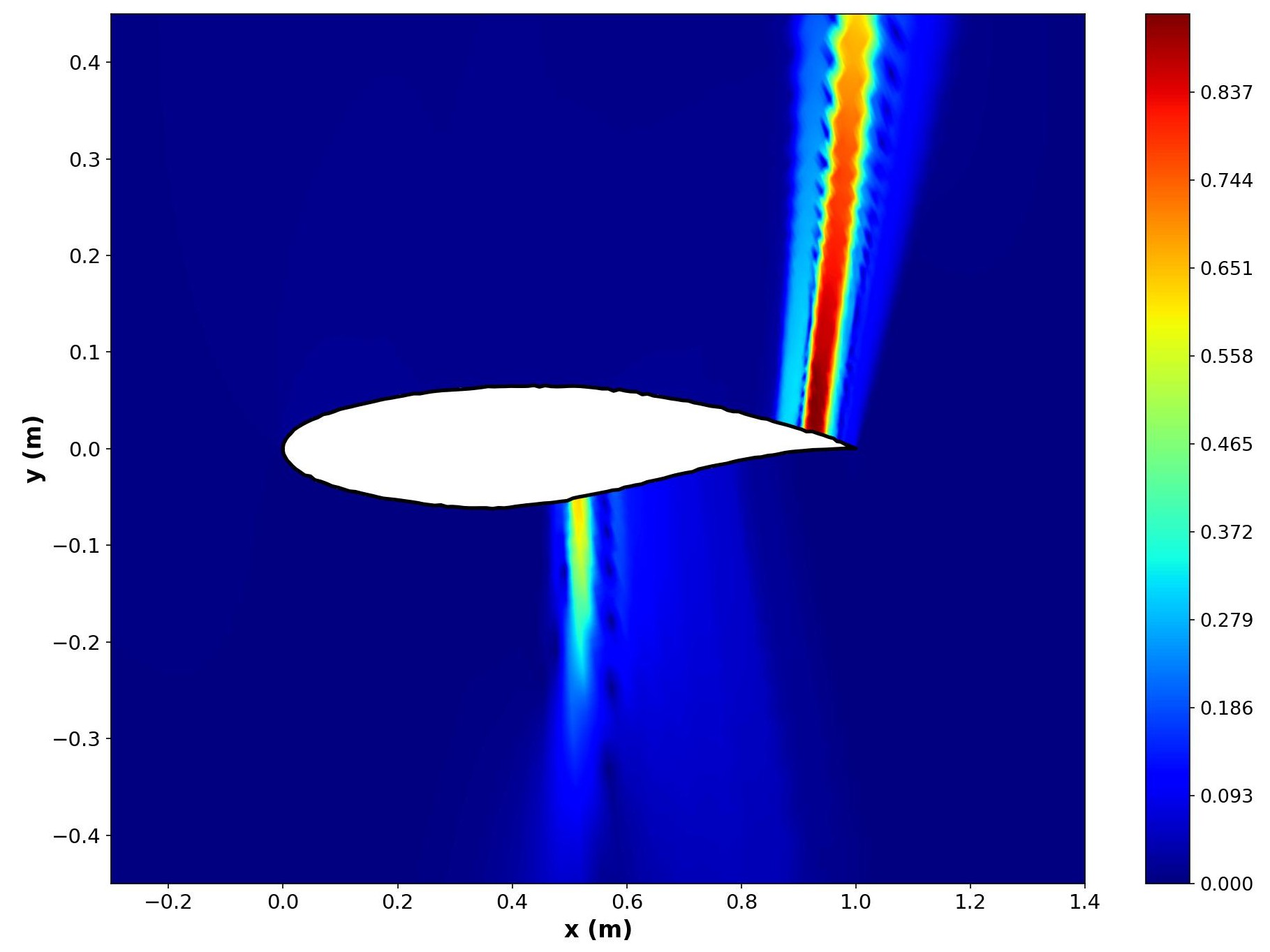}
        \end{subfigure}%
        \begin{subfigure}{.33\textwidth}
        \centering
        \includegraphics[width=1\linewidth]{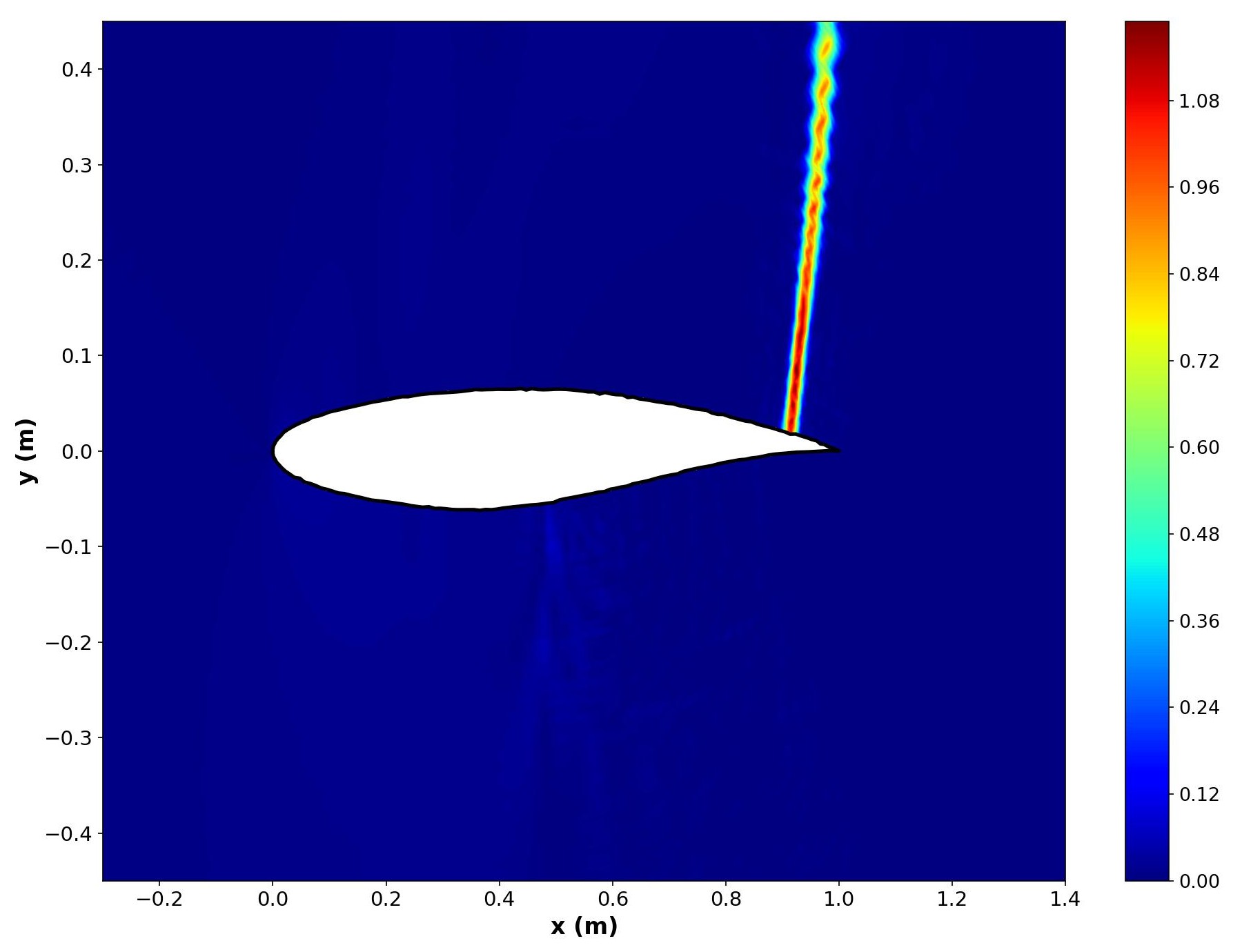}
        \end{subfigure} \\
        \begin{subfigure}{.33\textwidth}
        \centering
        \includegraphics[width=1\linewidth]{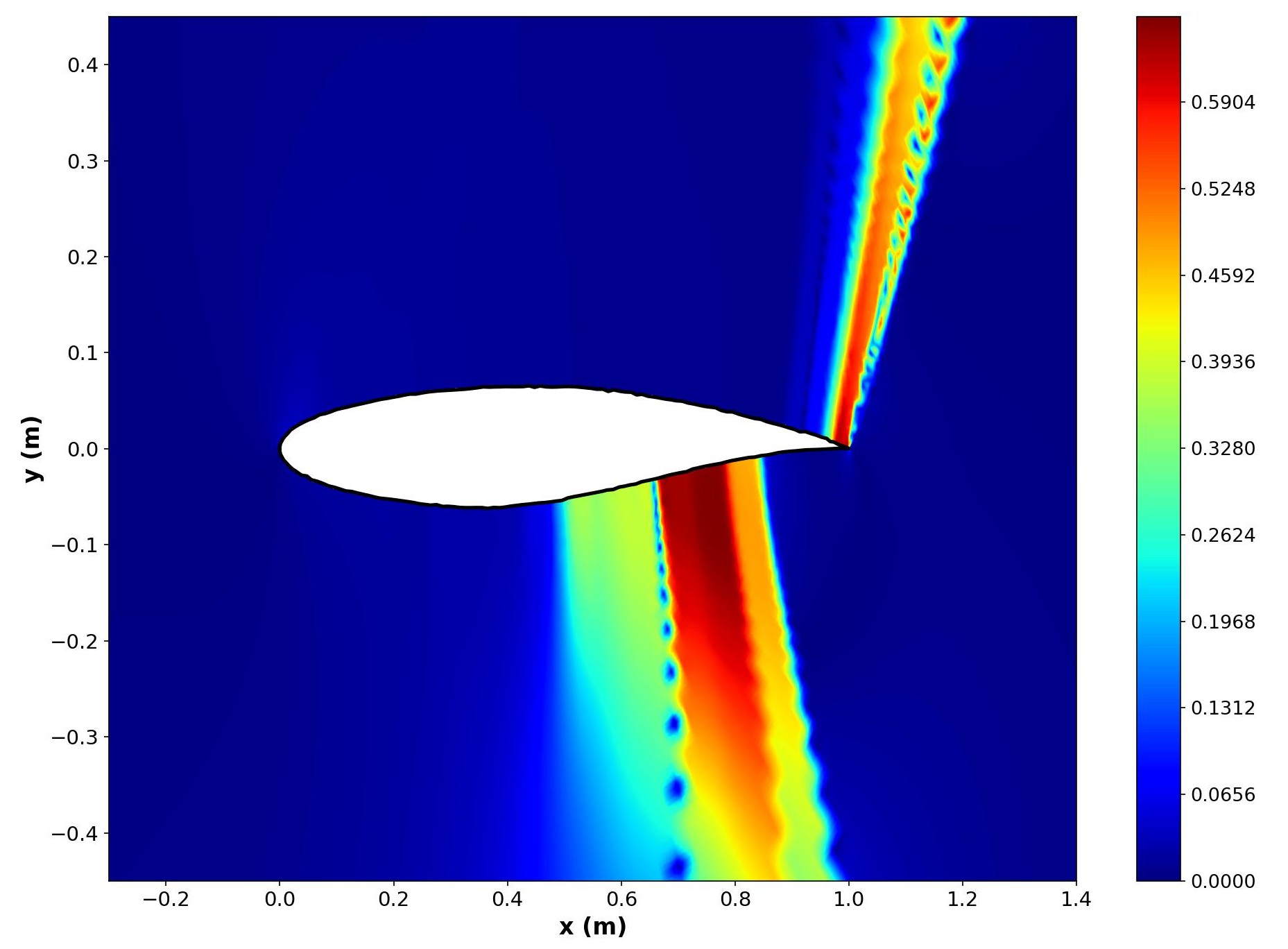}
        \end{subfigure}%
        \begin{subfigure}{.33\textwidth}  
        \centering
        \includegraphics[width=1\linewidth]{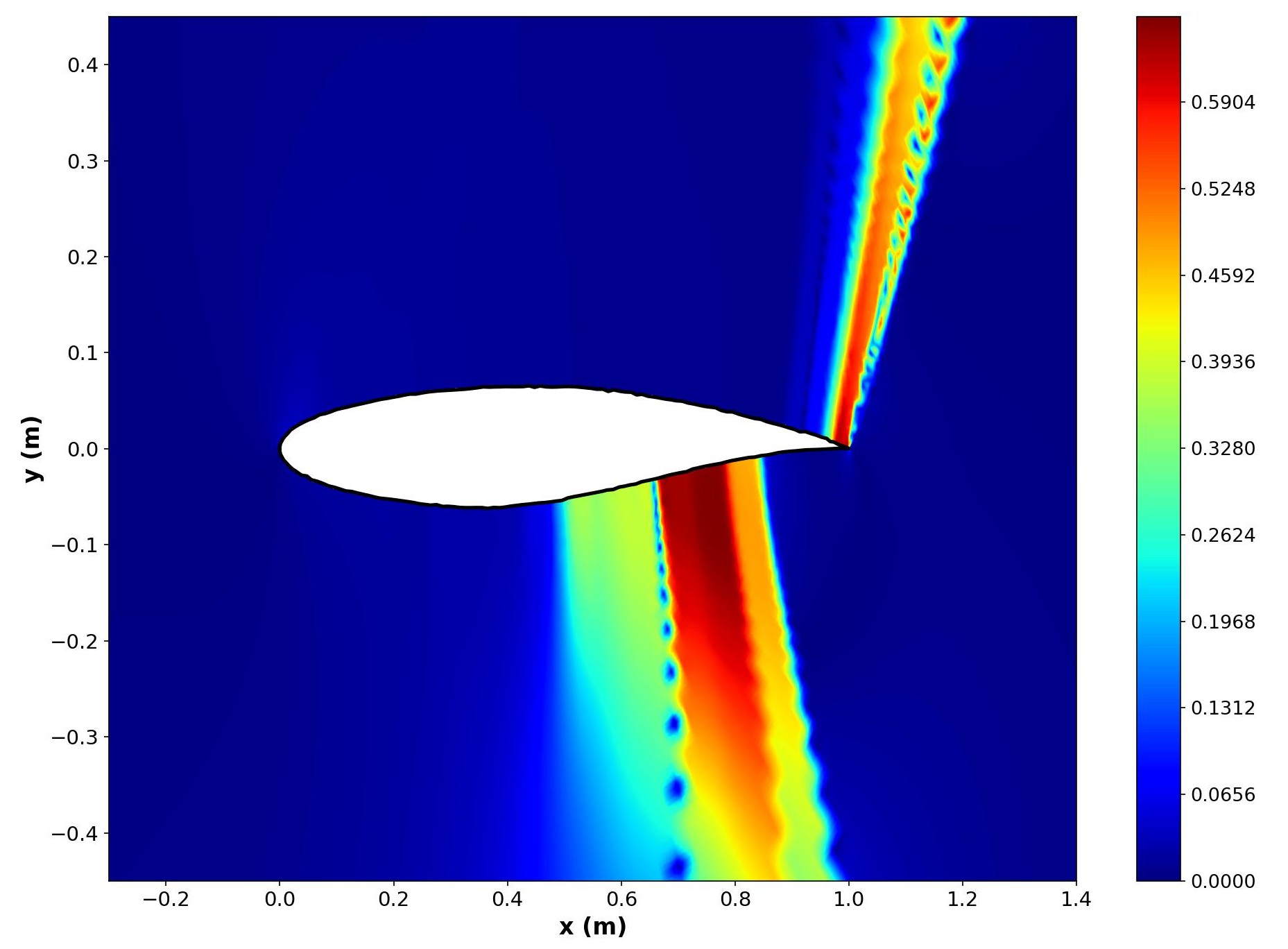}
        \end{subfigure}%
        \begin{subfigure}{.33\textwidth}
        \centering
        \includegraphics[width=1\linewidth]{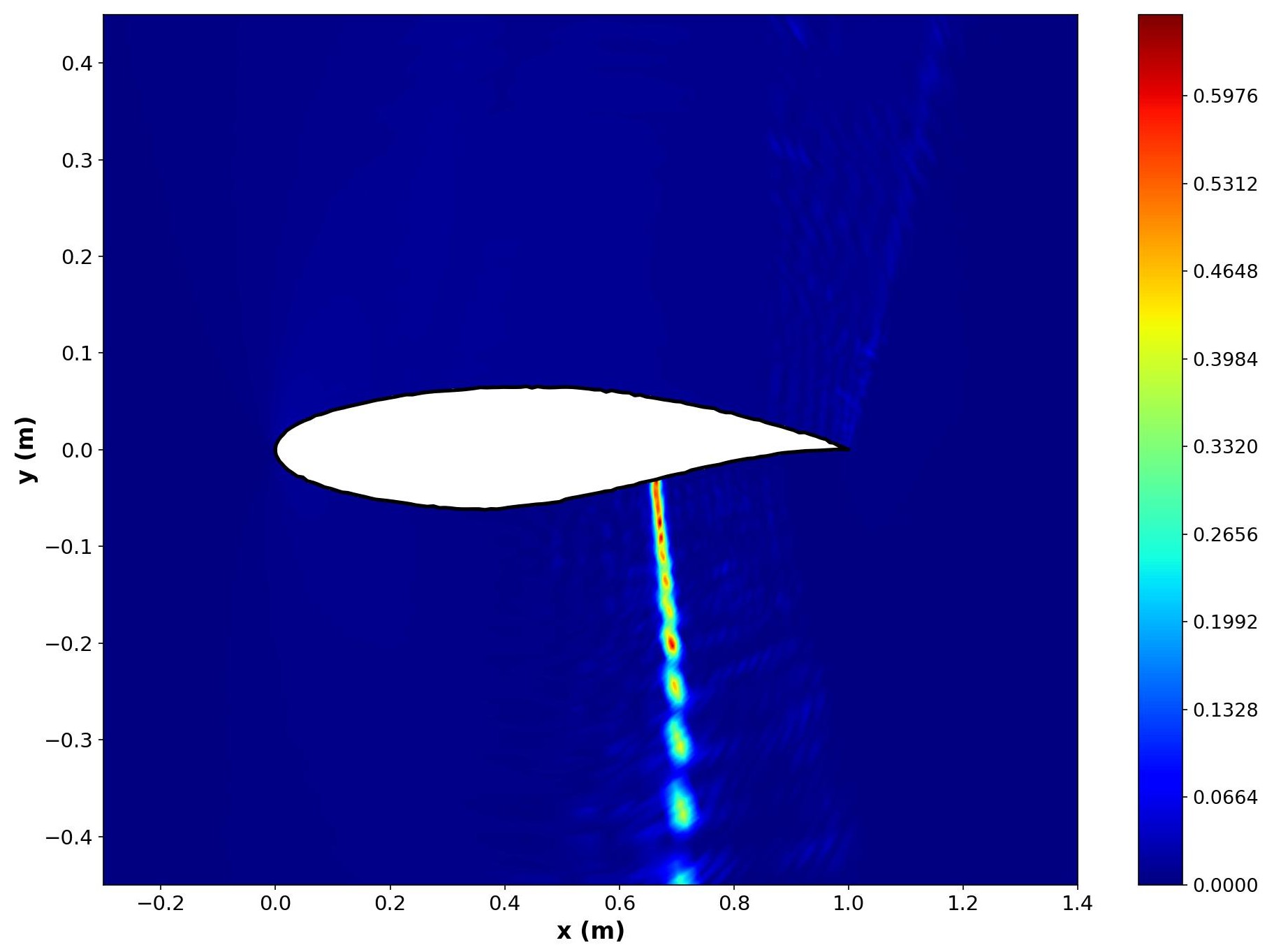}
        \end{subfigure} \\
        \begin{subfigure}{.33\textwidth}
        \centering
        \includegraphics[width=1\linewidth]{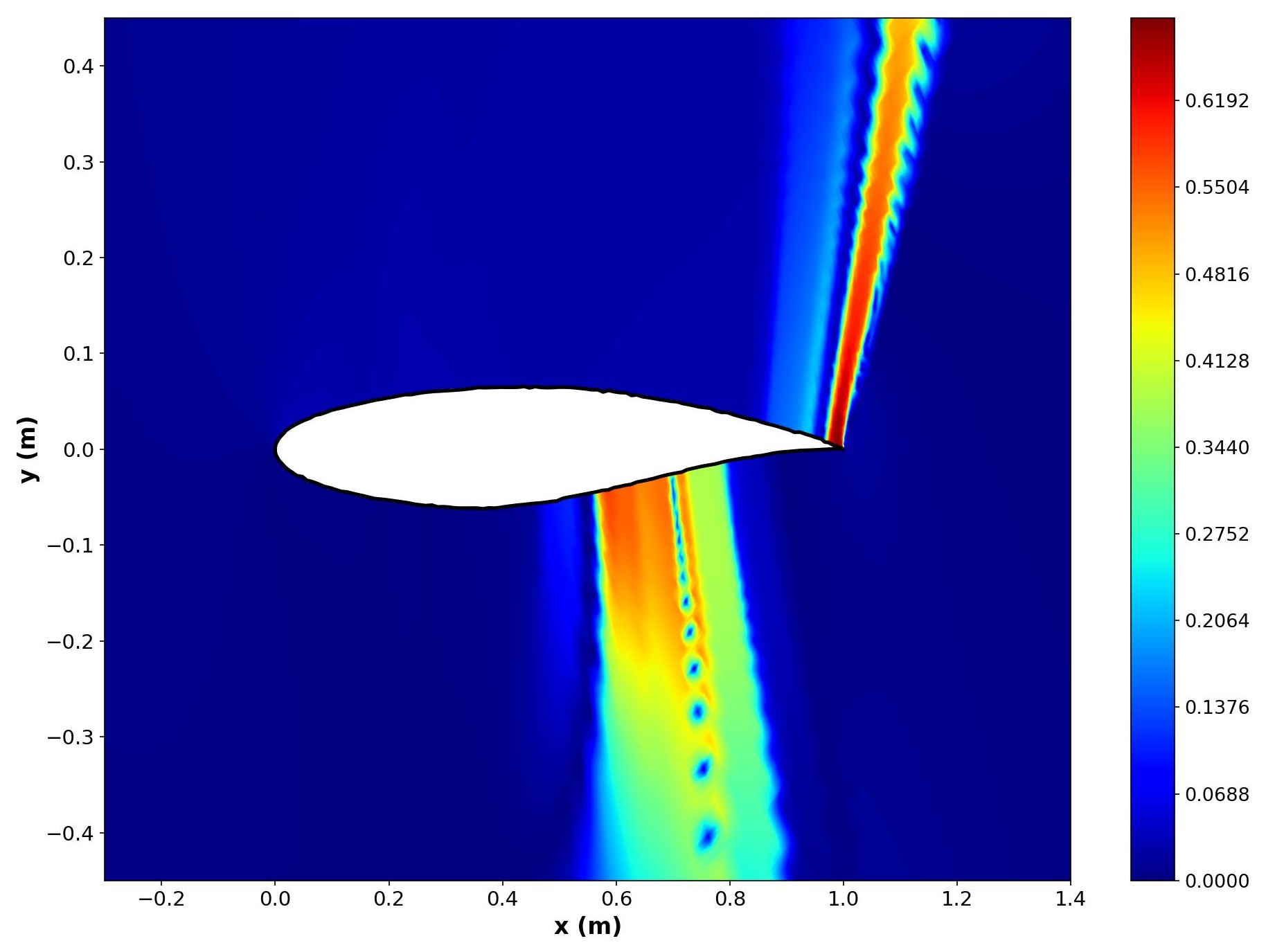}
        \end{subfigure}%
        \begin{subfigure}{.33\textwidth}  
        \centering
        \includegraphics[width=1\linewidth]{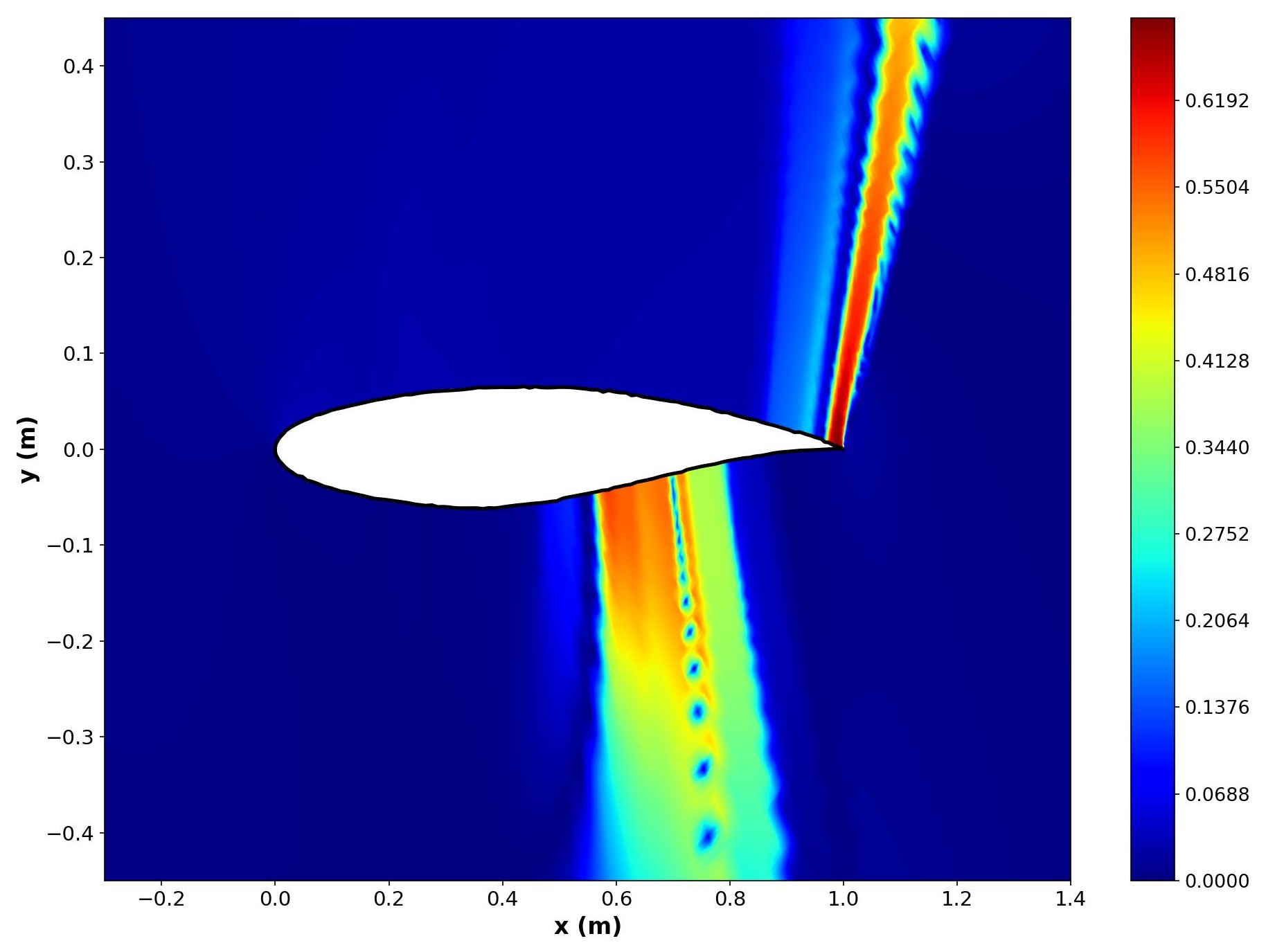}
        \end{subfigure}%
        \begin{subfigure}{.33\textwidth}
        \centering
        \includegraphics[width=1\linewidth]{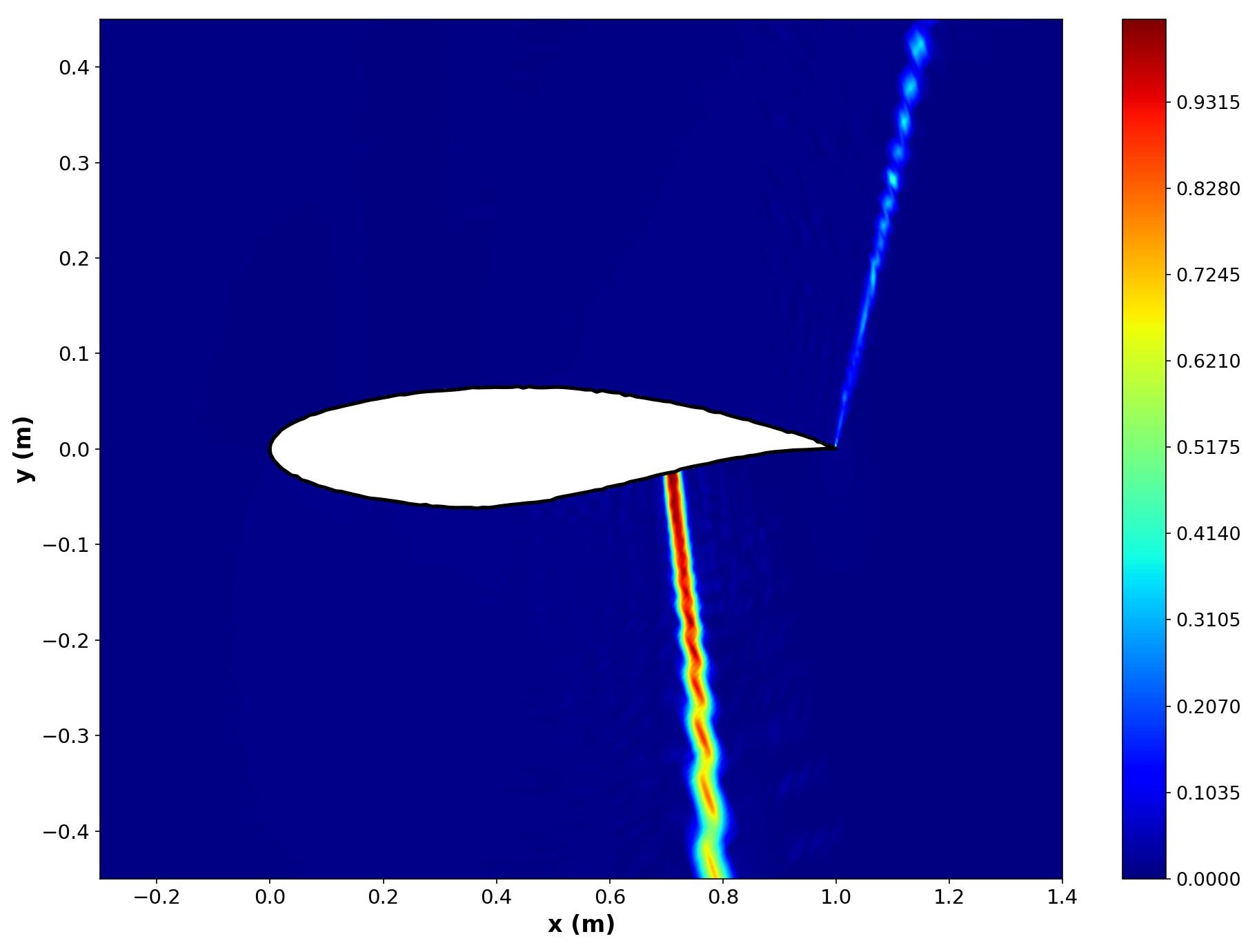}
        \end{subfigure} \\
        \begin{subfigure}{.33\textwidth}
        \centering
        \includegraphics[width=1\linewidth]{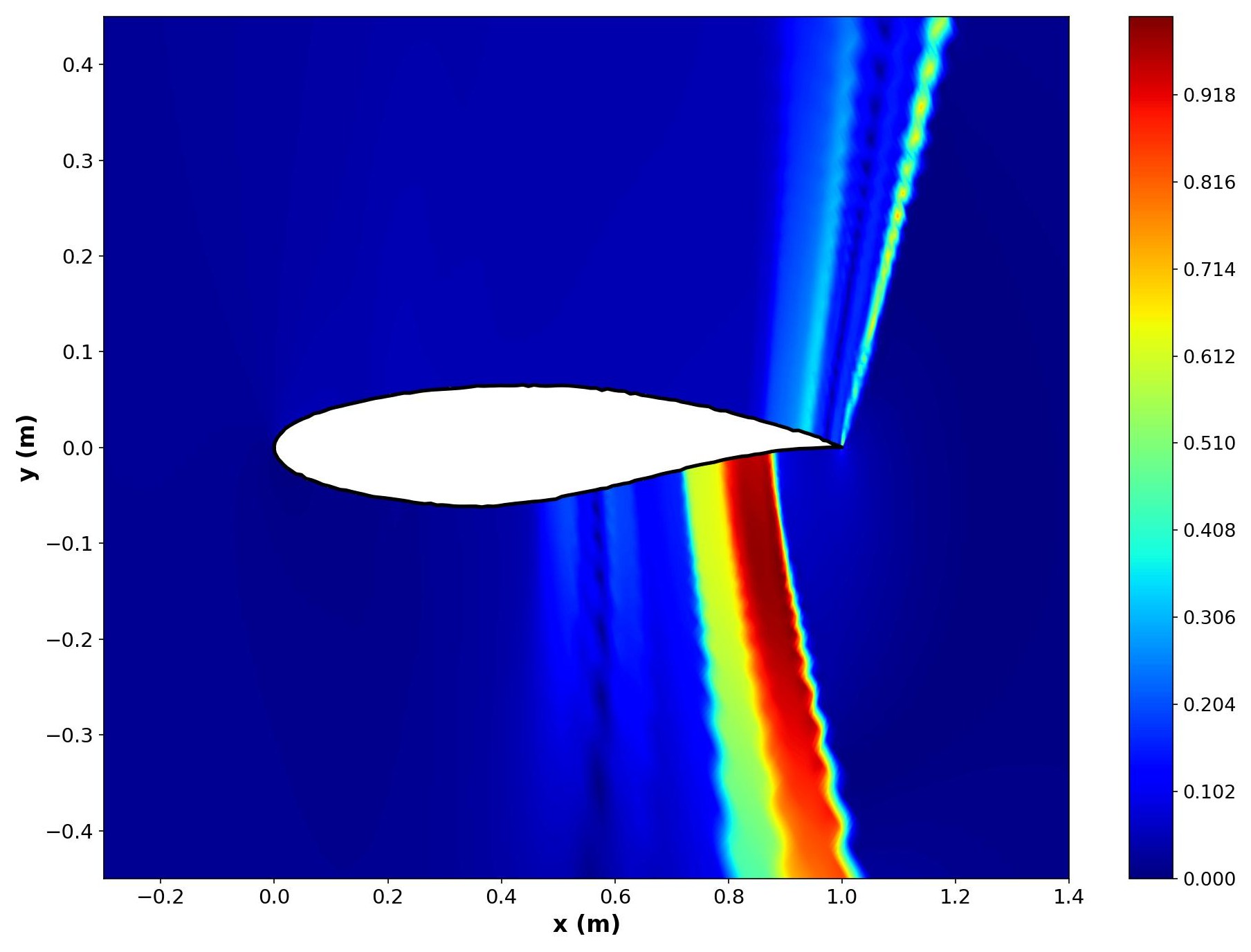}
        \caption{\texttt{FOM}}
        \end{subfigure}%
        \begin{subfigure}{.33\textwidth}  
        \centering
        \includegraphics[width=1\linewidth]{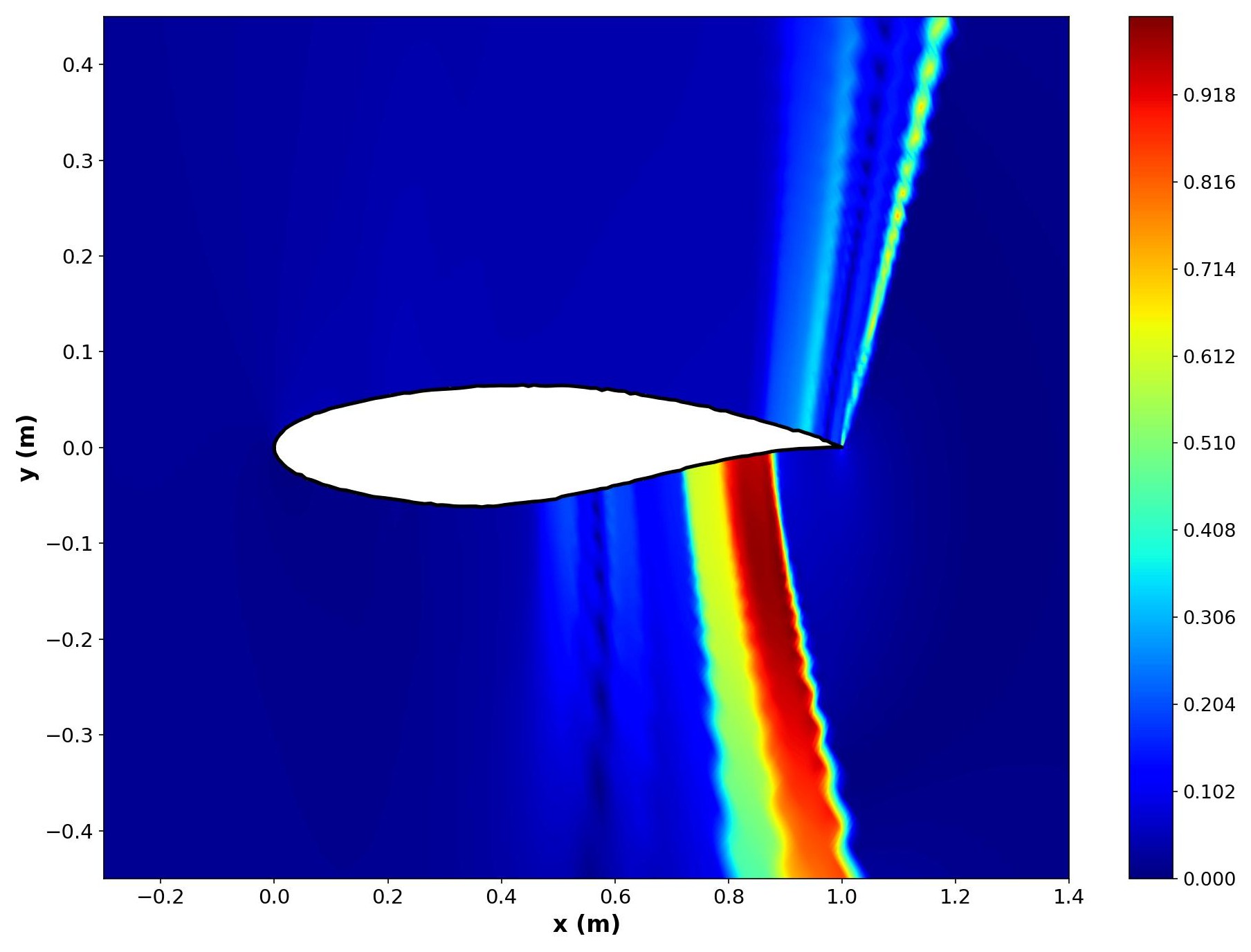}
        \caption{(Naive) \texttt{POD} }
        \end{subfigure}%
        \begin{subfigure}{.33\textwidth}
        \centering
        \includegraphics[width=1\linewidth]{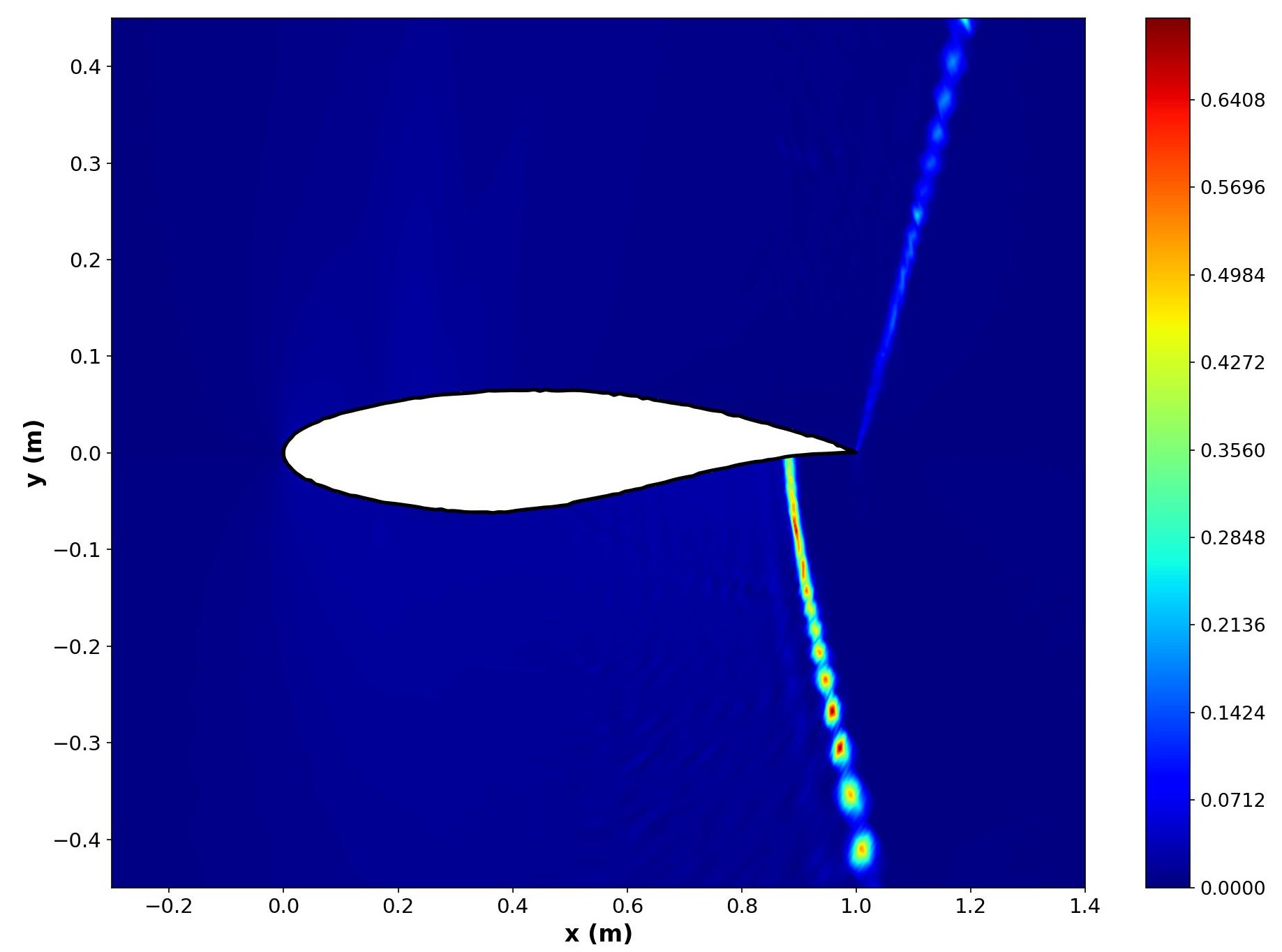}
        \caption{\texttt{CRLS} (ours)}
        \end{subfigure}         
        \caption{Comparison of the mean absolute error in pressure coefficient $(C_p)$ contours. Each row corresponds to parameter values in panels (a) through (e) in \Cref{fig:airfoil_cp}.}
        \label{fig:cp_contour_errors}
    \end{figure}

	We then compute the absolute difference between the predicted and true fields and visualize their contours, shown in \Cref{fig:cp_contour_errors}. Similar to airfoil surface pressure results, \texttt{POD} and \texttt{PODc} show similar error contours. Notably, \texttt{PODc} requires only about $48$ modes, while \texttt{POD} requires $81$ modes out of $85$, to capture the same amount of energy. From the error contours, it can also be observed that the shock locations for \texttt{POD} and \texttt{PODc} are significantly different from those for \texttt{CRLS} despite requiring far fewer modes.

	\paragraph{Implementation details.} To keep the convolution and deconvolution operations in \texttt{CRLS} tractable, we must take some measures. Specifically, the arrangement of the snapshots on a structured grid can drastically improve the computational complexity of these operations. 
    To enable this, we interpolate snapshots on unstructured grids onto a structured ``O-grid'' -- we show an illustration in \Cref{fig:ogrid}.
    
	\section{Conclusions}
	\label{sec:conclusion}
    In this work we have introduced a \texttt{CRLS} (convolutional regularized least squares) framework for building reduced-order models of flows with parameter-dependent shocks. The central idea is to exploit the fact that proper orthogonal decomposition (POD) is most effective when applied to smooth fields; we first map snapshots containing sharp gradients into a smoother space via a one-dimensional Gaussian convolution with reflect padding, construct POD bases and interpolate POD coefficients in that space, and then reconstruct sharp features by solving a regularized deconvolution problem. The convolution hyperparameters and the deconvolution regularization are selected automatically by solving targeted optimization problems, so that the entire procedure is largely data-driven and requires minimal user tuning.

A simple one-dimensional step function example illustrates the main mechanism: smoothing dramatically accelerates the decay of the singular values and allows POD to capture the dominant behavior with far fewer modes, while the deconvolution step recovers sharp transitions that are otherwise smeared or stair-stepped. We then applied \texttt{CRLS} to a more realistic and challenging setting -- steady, inviscid, transonic flow over the RAE2822 airfoil governed by the compressible Euler equations. For this configuration, \texttt{CRLS} consistently produced pressure coefficient distributions and full pressure fields that closely track the full-order model, while standard POD and a purely smoothed POD (without regularized inversion) exhibited non-physical oscillations, double shocks, and noticeable shock misplacement. Quantitatively, the proposed method yields the lowest mean absolute error and root-mean-square error across all validation cases and achieves a significant reduction in the number of POD modes required to represent a fixed fraction of snapshot energy, translating directly into lower online evaluation costs.

Beyond these specific results, \texttt{CRLS} has several practical advantages. It operates nonintrusively on snapshot data and therefore can be wrapped around existing CFD solvers without modifying their discretizations or numerics. The reflect-padded convolution and its matrix representation are compatible with structured grids obtained from interpolating unstructured meshes (such as O-grids), which keeps both the smoothing and inversion steps computationally tractable. The use of Bayesian optimization to tune the convolution kernel width and support, coupled with nearest-neighbor regularization in parameter space, makes the approach adaptable to different flow regimes and parameter ranges with minimal manual intervention.

Despite its promising performance, the present \texttt{CRLS} framework has a few limitations. First, the method relies on mapping snapshots to a structured O-grid and applying a one-dimensional Gaussian convolution with fixed orientation; while this is done to make the computational complexity of the approach tractable, more innovative methods to perform the convolution and deconvolution can further enhance the method. Second, \texttt{CRLS} uses radial basis function interpolation and a nearest-neighbor reference in parameter space; both ingredients assume reasonably dense sampling and are not designed for aggressive extrapolation beyond the training envelope. Finally, the current implementation does not explicitly enforce physical constraints such as conservation or stability of long-time integrations, and the impact of noisy or inconsistent snapshot data on the regularized inversion has not been systematically quantified; addressing these issues will be important for deploying \texttt{CRLS} in broader aerodynamic design and control workflows.

The present study focuses on steady, inviscid, two-dimensional transonic aerodynamics; however, the methodology is more general. Future work will extend \texttt{CRLS} to viscous and turbulent flows, unsteady problems, and three-dimensional configurations of interest. It will also be valuable to explore alternative kernels and anisotropic or directionally aligned convolutions, richer regularization strategies that encode physical constraints, and combinations of \texttt{CRLS} with hyper-reduction or data-driven closure models. Finally, embedding \texttt{CRLS}-based surrogates into optimization and uncertainty quantification loops, for example in reliability-based design or multiobjective aerodynamic optimization, offers a promising avenue to obtain shock-resolving reduced-order models that remain both accurate and computationally affordable at scale.


	\clearpage

	\bibliographystyle{plainnat}
	\bibliography{references2}

\begin{appendices}

    	\begin{figure*}[htb!]
		\centering
		\includegraphics[width=0.55\textwidth]{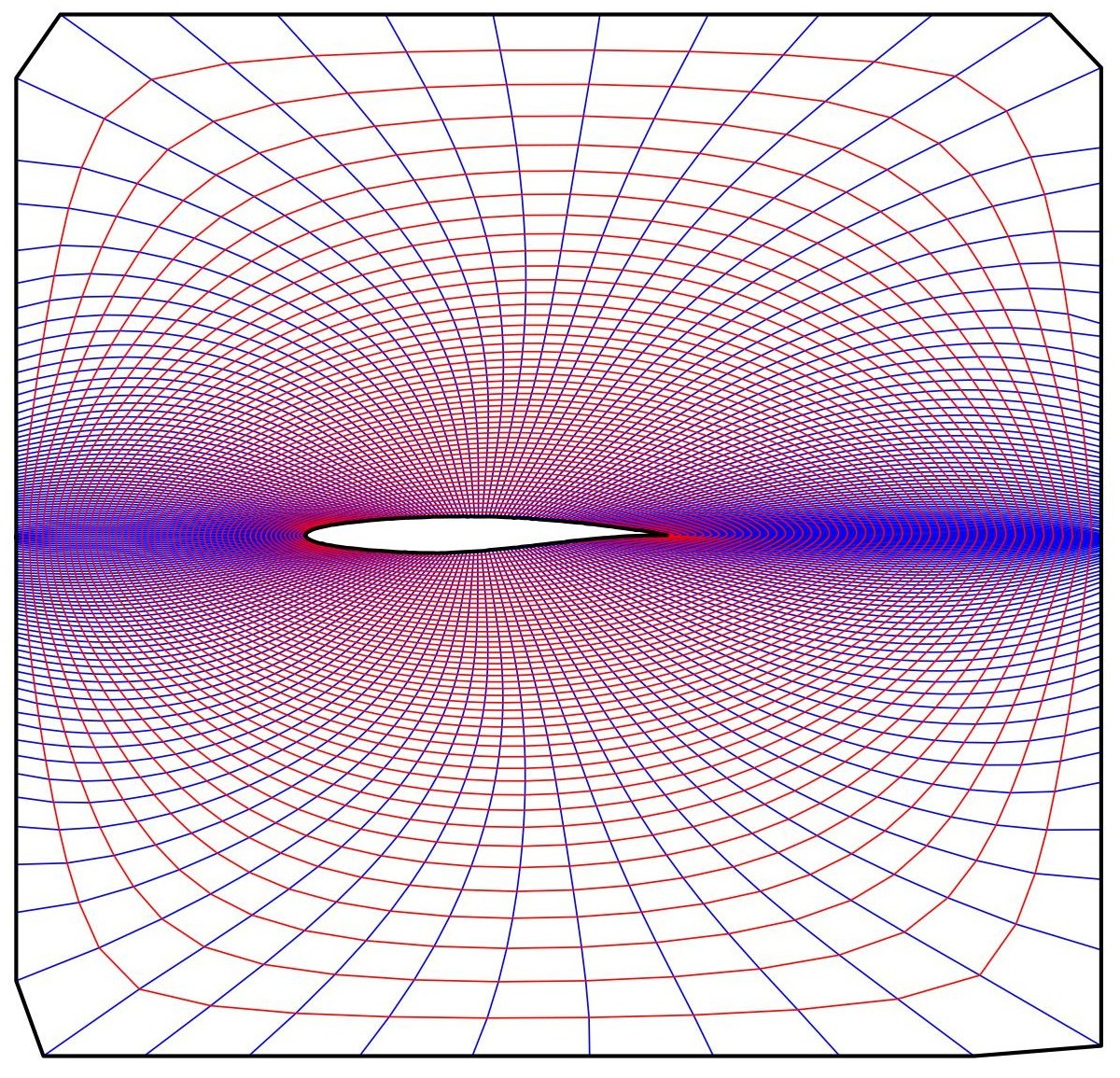}
		\captionsetup{justification=centering, labelfont=bf, font=small}
		\caption{zoomed view of O-grid mesh}
		\label{fig:ogrid}
	\end{figure*}
    
	\section{Example of kernel matrix} \label{Appendix_1}

\noindent To better understand this, consider a snapshot $\bm{u}$ of $n=5$ elements. Then, one can obtain the convolved output $\bm{u^s} = {B} \bm{u}$ in matrix form as;
	\begin{equation}\label{2.2.7}
		\begin{pmatrix}
			\bm{u}^s[0] \\
			\bm{u}^s[1] \\
			\bm{u}^s[2] \\
			\bm{u}^s[3] \\
			\bm{u}^s[4]
		\end{pmatrix}
		=
		\begin{pmatrix}
			\bm{k}[2] & \bm{k}[1] + \bm{k}[3] & \bm{k}[0] + \bm{k}[4] & 0 & 0 \\
			\bm{k}[1] & \bm{k}[0] + \bm{k}[2] & \bm{k}[3] & \bm{k}[4] & 0 \\
			\bm{k}[0] & \bm{k}[1] & \bm{k}[2] & \bm{k}[3] & \bm{k}[4] \\
			0 & \bm{k}[0] & \bm{k}[1] & \bm{k}[2] + \bm{k}[4] & \bm{k}[3] \\
			0 & 0 & \bm{k}[0] + \bm{k}[4] & \bm{k}[1] + \bm{k}[3] & \bm{k}[2]
		\end{pmatrix}.
		\begin{pmatrix}
			\bm{u}[0] \\
			\bm{u}[1] \\
			\bm{u}[2] \\
			\bm{u}[3] \\
			\bm{u}[4]
		\end{pmatrix}
	\end{equation}
	
	The main idea is to convolve the given snapshot data matrix $U$, construct ROM for the convolved (smoothened) snapshots $\bm{u^s}$  and invert the predicted convolved snapshot to obtain the unsmoothened prediction.

\end{appendices}

\end{document}